\documentclass[sigconf]{acmart}
\usepackage{siunitx}
\usepackage{xcolor}
\usepackage{soul}
\usepackage{float}

\usepackage{multirow}

\AtBeginDocument{%
  }

\copyrightyear{2026}
\acmYear{2026}
\setcopyright{cc}
\setcctype{by}
\acmConference[CHI '26]{Proceedings of the 2026 CHI Conference on Human Factors in Computing Systems}{April 13--17, 2026}{Barcelona, Spain}
\acmBooktitle{Proceedings of the 2026 CHI Conference on Human Factors in Computing Systems (CHI '26), April 13--17, 2026, Barcelona, Spain}
\acmPrice{}
\acmDOI{10.1145/3772318.3791964}
\acmISBN{979-8-4007-2278-3/2026/04}

\acmSubmissionID{9635}

\begin{document}

\title[Design for Effective but Sustainable Virtual Reality Exposure Studies]{Responsible Trauma Research: Designing Effective and Sustainable Virtual Reality Exposure Studies}

\author{Annalisa Degenhard}
\orcid{0000-0002-3190-0496}
\email{annalisa.degenhard@uni-ulm.de}
\affiliation{%
  \institution{Ulm University}
  \city{Ulm}
  \country{Germany}
}

\author{Sophia Ppali}
\orcid{0000-0001-5630-1679}
\affiliation{%
  \institution{CYENS Centre of Excellence}
  \city{Nicosia}
  \country{Cyprus}}
\email{s.ppali@cyens.org.cy}

\author{Fotis Liarokapis}
\orcid{0000-0003-3617-2261}
\affiliation{%
  \institution{CYENS Centre of Excellence}
  \city{Nicosia}
  \country{Cyprus}}
\email{f.liarokapis@cyens.org.cy}

\author{Enrico Rukzio}
\orcid{0000-0002-4213-2226}
\affiliation{%
  \institution{Ulm University}
  \city{Ulm}
  \country{Germany}}
\email{enrico.rukzio@uni-ulm.de}

\author{Jennifer Spohrs}
\authornote{Both authors contributed equally to this research.}
\orcid{0000-0001-8886-0395}
\affiliation{
\department{Department of Psychiatry, Psychotherapy and Psychotraumatology}\institution{Military Medical Centre}
\city{Ulm}
\country{Germany}}
\affiliation{
\department{Department for Child and Adolescent Psychiatry and Psychotherapy}\institution{Ulm University Medical Centre}
\city{Ulm}
\country{Germany}}
\email{jennifer.spohrs@uni-ulm.de}

\author{Stefan Tschoeke}
\authornotemark[1]
\orcid{0000-0002-8612-7752}
\affiliation{
\department{Clinic for Psychiatry and  Psychotherapy I (Weissenau)}\institution{Ulm University}
\city{Ulm}
\country{Germany}}
\affiliation{\institution{Centre for Psychiatry S\"{u}dw\"{u}rttemberg}
\city{Ravensburg-Weissenau}
\country{Germany}}
\email{stefan.tschoeke@zfp-zentrum.de}

\renewcommand{\shortauthors}{Degenhard et al.}

\begin{abstract}
Virtual reality exposure therapy (VRET) enables controlled exposure to trauma-related stimuli to facilitate memory access and emotional processing. However, the field remains underexplored for complex post-traumatic stress disorder (C-PTSD). Unlike single-trauma PTSD, C-PTSD requires highly individualized triggers that are difficult to identify and implement safely. We conducted a feasibility study with 11 patients, two trauma therapists, and a VR developer to explore integrating VRET into C-PTSD treatment while safeguarding all stakeholders. Initial findings indicate that simple objects can be just as effective as complex scenes, therapeutic success does not correlate with VR presence levels, and the design process itself became integral to therapy rather than preparatory. However, involving developers in therapy sessions led to considerable emotional stress and role confusion, which required a cautious approach. Based on these insights, we provide methodological recommendations for safe and patient-centered VRET studies that balance therapeutic effectiveness with stakeholder safety across the research process.
\end{abstract}

\ccsdesc[500]{Human-centered computing~User studies}
\ccsdesc[500]{Human-centered computing~Empirical studies in HCI}

\keywords{C-PTSD, exposure, virtual reality}

\begin{teaserfigure}
\centering
  \includegraphics[width=0.9\textwidth]{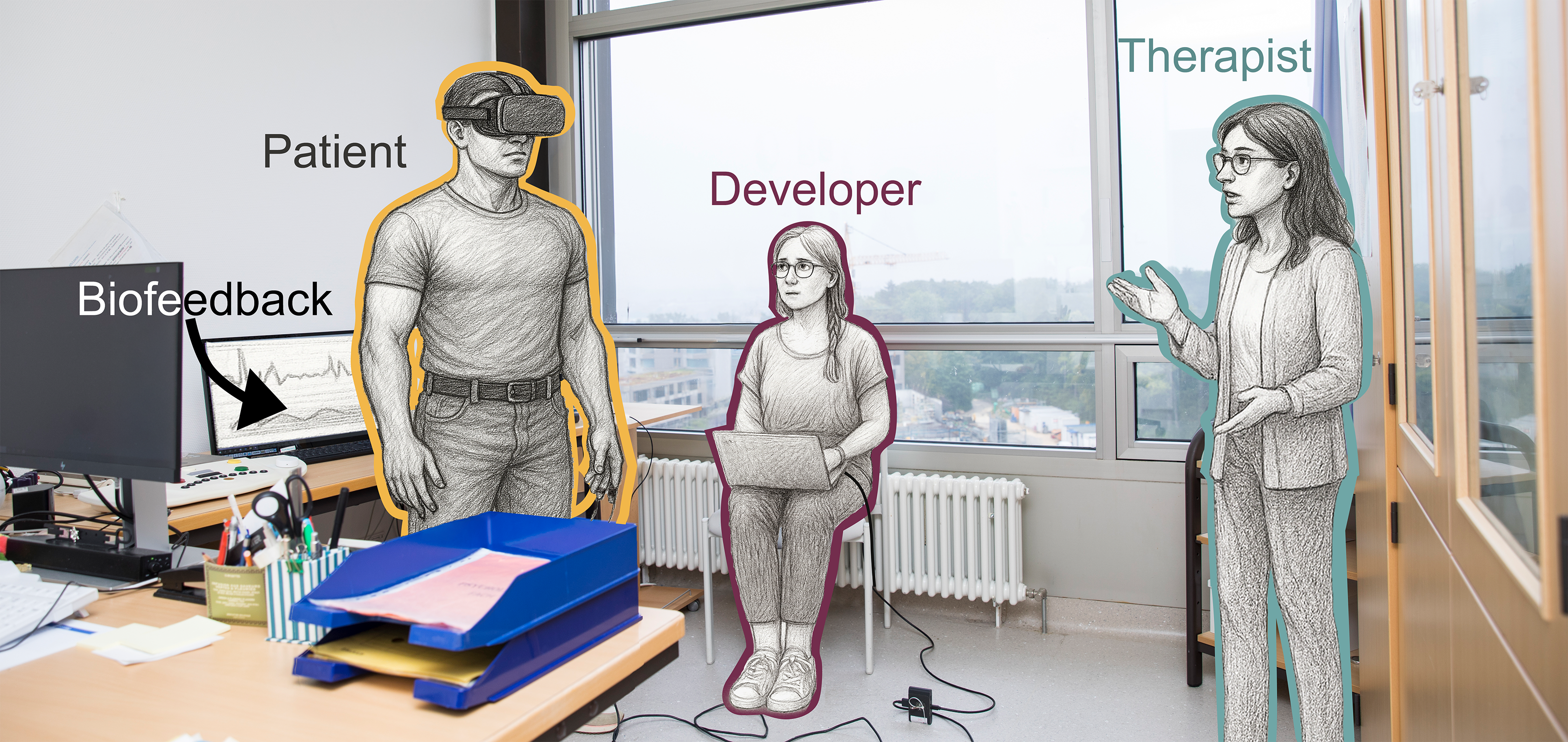}
  \caption{A typical setting of a virtual reality exposure session with a developer involved.}
  \Description{The image shows a therapy room with a patient standing in the middle of the room wearing a virtual reality headset. He wears physiological sensors that are connected to a computer behind him, screening the collected data. A developer sits in the background with a laptop on her lap, connected to the virtual reality headset. On the right of the room is a therapist who is talking to the patient.}
  \label{fig:teaser}
\end{teaserfigure}


\maketitle
\textbf{Content warning:} This paper discusses and visualizes trauma-related material. Text and figures include
references to traumatic memories (e.g., violence, abuse,
war-related events, and hospital/clinical settings). While presented for scientific
purposes, some readers may find this distressing.

\section{{Introduction}}

Virtual reality (VR) technologies are increasingly used in trauma-focused exposure therapy. Their immersive qualities allow controlled presentation of trauma-related stimuli, offering new ways to support memory retrieval and emotional processing~\cite{rizzoVirtualRealityExposure2014, Botella2015Virtual, Loucks2019You}. While clinical studies show promising results, research has paid little attention to how Virtual Reality Exposure Therapy (VRET) scenarios are actually designed and implemented with vulnerable patients \cite{knaust_virtual_2020}. Most work evaluates system effectiveness rather than addressing the methodological and ethical complexities of running such studies safely \cite{rizzo_clinical_2017}. Previous work demonstrates that VRET can be effective for specific trauma types, such as combat-related post-traumatic stress disorder (PTSD)~\cite{Beidel2017Trauma, Botella2015Virtual, rizzoVirtualRealityExposure2014}, but it offers limited guidance on how scenarios can be designed and adapted for individuals with a history of multiple traumas, such as childhood abuse \cite{Botella2015Virtual}.
\noindent This gap becomes critical in the context of complex post-traumatic stress disorder (C-PTSD). Unlike PTSD, C-PTSD develops through prolonged or repeated traumatic events (e.g. torture, genocide campaigns, prolonged domestic violence, repeated childhood sexual or physical abuse), in a way that affects a person's emotional regulation, negative self-concept, and in sustaining relationships ~\cite{cloitreDistinguishingPTSDComplex2014,wilsonPTSDComplexPTSD2004, cloitreDistinguishingPTSDComplex2014}. 

The key point is that traumatic stimuli in C-PTSD are numerous, highly individualized and fragmented, and are only transferable to other patients to a limited extent~\cite{wilsonPTSDComplexPTSD2004}. 

\noindent As a result, C-PTSD is particularly challenging for VRET implementation. In the absence of established guidelines~\cite{Botella2015Virtual, Gonalves2012Efficacy}, scenario development usually follows a trial-and-error model: designers build exposures, test them with patients, observe responses, and adjust. For mono-traumatized individuals, this process can be manageable, since triggers often share commonalities. For C-PTSD, however, trial and error is both inadequate and risks therapy dropouts. To be viable, VRET must allow highly individualized and adjustable exposures, developed through close collaboration between therapists and developers~\cite{harrington_deconstructing_2019, mullins_patient-centeredness_2014}. Yet, implementation processes are rarely documented in detail~\cite{Gonalves2012Efficacy}, leaving little guidance on how to conduct such collaborations safely, protect stakeholders, or integrate design practices into therapy.
Against this backdrop, our work asks:
\begin{itemize}
    \item [\textbf{RQ1}] How can VRET studies for C-PTSD explore challenges and needs while protecting stakeholders?
    \item [\textbf{RQ2}] What design processes are most effective for creating exposure scenarios for patients with C-PTSD?
    \item [\textbf{RQ3}] What challenges arise when involving technical developers directly in trauma therapy sessions?
\end{itemize}

\noindent To explore these questions, we conducted a feasibility study with eleven patients (five with military trauma, six with childhood trauma), two experienced trauma therapists, and a VR developer.

With the overarching aim to promote sustainable research on VRET for C-PTSD, our objective was to identify the challenges and needs of patients, therapists, and developers at different stages of VRET, and to understand how studies can create safe conditions for meaningful insights. Our approach was necessarily exploratory. We tested different collaboration models, varied levels of patient involvement in design, and experimented with triggers ranging from simple objects to complex environments.
Although it was a first exploratory feasibility study with a limited sample size, a maximum of five sessions per patient, and no control group, some patients achieved breakthroughs that they had not reached in years of therapy; for instance, retrieving avoided memories or a reduction of trauma-related agoraphobia to a bearable level. At the same time, random triggers, communication barriers, and emotional stress of the developer~\cite{williamson_secondary_2020} challenged both acceptance and study integrity.

\noindent Through our study, we provide an examination of VRET adaptation for C-PTSD. We contribute lessons learned for designing individualized scenarios, highlight the therapeutic value of collaborative design, and show that surprisingly simple, low-cost implementations can support the recollection of avoided memories when adapted to patients’ needs. In some cases, findings indicated that less may even be more effective in overcoming avoidance and retrieving emotional memories. We also identify risks to non-clinical staff and offer practical insights for safeguarding all stakeholders. Our findings challenge assumptions about VR presence and provide a foundation for safer, more effective study design to explore the field of VRET for C-PTSD.\\

\noindent \textbf{Contribution statement}~\cite{wobbrock_research_2016}: This work contributes an \textbf{empirical study} with ($N=11$) VRET for C-PTSD and provides \textbf{methodological insights} into the challenges of such studies.

\section{{Background}}
In fear learning, the acquisition and maintenance of PTSD symptoms are linked to excessive consolidation and failures of extinction of conditioned fear responses. Behavioral patterns, physiological reactions, and emotional responses may resurface when cues resemble the original threatening situation~\cite{ledoux2016using,dunsmoorLaboratoryModelsPosttraumatic2022,shalev2024neurobiology,cushing2024Metacognition}.

A core symptom is hyperarousal, characterized by a persistent sense of threat and impaired emotion regulation. Recent neuroscientific models attribute this to hyperactivity in an evolutionarily developed threat-detection and defense system \cite{ledoux2016using,shalev2024neurobiology,LANIUS2017109,CESARI2023175}. Further symptoms include cognitive impairments, executive dysfunction, and difficulties integrating trauma-related information into context. This can reduce the ability to judge risky social situations and to differentiate threat from safety, increasing vulnerability to revictimization \cite{shalev2024neurobiology,messman2003Childhood}.

Individuals often avoid trauma-related confrontation due to intense negative sensations. Nonetheless, controlled repeated confrontation and emotional processing can reduce reaction intensity~\cite{foaEmotionalProcessingFear1986} and lower the risk of uncontrolled triggers. The goal is therefore to overcome avoidance and foster engagement. Numerous studies show that higher self-efficacy supports trauma processing, resilience, and therapeutic success ~\cite{bandura1989regulation,BENIGHT20041129}.

\subsection{Trauma-focused psychotherapy and accessibility of treatment for C-PTSD}
Trauma-focused exposure psychotherapy is widely regarded as the most effective treatment for PTSD following both single and repeated traumatic events\cite{HOPPEN2024112,Billings.2025}. The approach centers on retrieving avoided traumatic memories to practice handling them~\cite{brewin2025post,cloitre2011treatment} and to support autobiographical integration~\cite{berntsen2007trauma,smeets2010Autobiographical}. Overcoming avoidance behavior is of central importance here because it is a symptom of the disorder, yet it hinders exposure therapy and an improvement in symptoms. This is especially difficult in cases of C-PTSD, which is mainly caused by repeated interpersonal trauma, such as childhood sexual abuse~\cite{Herman.1992,Billings.2025}.
 
Therefore, a core aim is to help patients reduce avoidance and regain a sense of control by explaining symptoms, teaching coping strategies, and creating safe conditions for confronting triggers. Therapy typically involves four stages. In the \emph{psychoeducation} phase, patients learn about PTSD symptoms, including common concerns such as ``Why am I restless?'' or ``Why can’t I sleep''. Controllability in trigger situations is central to this approach: controlled exposure is expected to reduce emotional reactivity over time. After initial skills training, in which patients practice self-regulation techniques such as diaphragmatic breathing or cognitive reframing, later exposure sessions revisit traumatic memories to demonstrate their tolerability. This rests on inhibitory learning theory, which holds that repeated disconfirmation of threat expectations reduces fear responses~\cite{CRASKE201410}. The scenario itself is not therapeutic; its value lies in enabling therapeutic dialogue and facilitating engagement and emotional processing~\cite{BENIGHT20041129}. In the debriefing phase, patients integrate these experiences with the therapist.
Retrieving traumatic memories is often difficult, particularly for patients with high levels of dissociation. Approaches include retrospective accounts (e.g., narrative exposure therapy\cite{neunerNarrativeExpositionstherapieNET2021}), post-hoc observations, and deliberately provoking flashbacks~\cite{vanderkolkDissociationFragmentaryNature1995}. The efficacy of these approaches, however, is challenged by the greater levels of dissociation and more complex, individualized triggers associated with C-PTSD~\cite{maercker2022complex,karatzias2017ptsd}. Psychopharmacological treatments are available, but are not recommended as the primary treatment option~\cite{leichsenring2024Borderline}, making psychotherapy with trauma exposure the most promising intervention and the first choice of guidelines~\cite{12schafer2019s3}. Visualization-based approaches have shown potential in the treatment of both single-event and multiple-event PTSD through individualized exposure therapy~\cite{chuDissociativeSymptomsRelation1990,HOPPEN2024112}. However, they also face practical limitations: creating tailored visualizations can be time-consuming, costly, and difficult to adapt to the fragmented nature of C-PTSD memories~\cite{Nester29072022}. Here, immersive technologies such as VR present both opportunity and risk.

VRET was found to facilitate the development of flexible and economical alternatives to in vivo exposure. However, the economic feasibility was bound to the existence of respective media or the reuse of costly, specially designed media. The rise of generative artificial intelligence (GAI) has rendered the effective generation of customized media increasingly feasible, making a supply of individualized VRET conceivable for the first time.

\subsection{Challenges in exposure therapy}
VR exposure for psychotherapy offers a unique way for a controlled and adaptable visualization of traumatic memories; on the other hand, several challenges function as barriers to engagement, risking ineffective therapy and drop-outs. Key challenges on the patient side include:

\begin{itemize}
    \item\emph{Intense recall}: Retrieving avoided memories can provoke severe emotional responses and ego disorganization, including anxiety, despair, or impulses toward self-destructive behavior~\cite{herman1987recovery}. Panic attacks may also occur~\cite{lyssenkoDissociationPsychiatricDisorders2018}, and loss of control may result from absorption~\cite{murrayAbsorptionDissociationLocus2007,glicksohnExplorationsVirtualReality1997,banosPsychologicalVariablesReality1999}.
    \item\emph{Unintentional flashbacks}: Individuals with C-PTSD are at heightened risk of being triggered unpredictably. If exposure exceeds tolerance, hyperarousal can occur~\cite{fordTraumaMemoryProcessing2018}.
    \item\emph{Avoidance}: Avoidant behaviors may reduce engagement in memory retrieval and self-disclosure, and in severe cases lead to dropout~\cite{charltonWaysCopingPsychological1996}.
    \item\emph{Acceptance} Success depends on patients’ acceptance of both the therapy and the technology. Low acceptance risks disengagement and dropout~\cite{kimDevelopmentHealthInformation2012,ICHE6R21997}.
\end{itemize}

\noindent For exposure therapy to be effective, traumatic stimuli must be confronted directly~\cite{bohusDialektischBehavioraleTherapie2024}. Yet high levels of avoidance in PTSD can make this difficult~\cite{charltonWaysCopingPsychological1996}. Approaches based on imagination alone are often unreliable, as therapists cannot control whether patients actually engage with the intended stimuli~\cite{vincelli1999imagination,difedeInnovativeUseVirtual2002}. Here, VR provides a potential advantage: by visualizing stimuli, it can partially circumvent avoidance and allow therapists to calibrate intensity~\cite{meggelenComputerBasedInterventionElements2019,meggelenRandomizedControlledTrial2022,Haft2025-ko}.

\subsection{VR exposure therapy for C-PTSD}
Graded VRET has proven effective across multiple phobias, including acrophobia \cite{emmelkampVirtualRealityTreatment2001,rothbaumVirtualRealityGraded1995,hodgesVirtualEnvironmentsTreating1995}, fear of flying \cite{muhlbergerRepeatedExposureFlight2001}, claustrophobia \cite{botellaVirtualRealityTreatment1998}, arachnophobia \cite{carlinVirtualRealityTactile1997}, and social anxiety \cite{andersonVirtualRealityExposure2003,harrisBriefVirtualReality2003}. By simulating in vivo exposure scenarios with fixed grading options, these applications activate fear structures and enable emotional processing. This approach extends to trauma-related fears, where VR simulations facilitate the development of coping mechanisms \cite{Carl2019-jk}. Beyond general fear activation, some applications specifically promote traumatic memory recollection by simulating known trauma aspects. \citet{difedeInnovativeUseVirtual2002,difedeVirtualRealityExposure2007}, for instance, retrieved previously suppressed memories through graded VR simulation of the World Trade Center incident, while \citet{josmanBusWorldAnalogPilot2008,josmanBusWorldDesigningVirtual2006} simulated an Israeli bus bombing across four scenarios eliciting varying discomfort levels. Both projects visualized suspected traumatic events and implemented grading through controlled incident staging. Building on this work, Rizzo et al. developed military trauma exposure software enabling in situ VR adaptations for individualized dynamic grading, involving veterans throughout design. Their findings indicate high VRET feasibility for combat-related PTSD, since traumatic experiences often share simulatable features across patients \cite{rizzo2006usercentered,rizzo2021fromcombat}. As an alternative, \citet{stevens2021clinician} discuss 360° videos for personalized exposure, offering an economical option—though these remain sensory-rich and non-adjustable in situ.

Findings showed that VRET for combat-related PTSD can be highly feasible, since traumatic experiences often share recognizable features that can be simulated across patients~\cite{rizzo2006usercentered,rizzo2021fromcombat}. At the same time,~\citet{stevens2021clinician} discuss the aptitude of 360° videos for personal exposure scenarios, offering an economic perspective on individualized exposure therapy. Nevertheless, such videos are sensory-rich and can not be adjusted in situ.

Across this body of work, literature consistently emphasizes the need for adjustable, customizable environments that reliably activate trauma-related fears while maintaining safety and predictability \cite{Haft2025-ko,Sherrill2020-mn,Sherrill2022-vl}. \citet{Sherrill2019-io,Sherrill2025-fo} specifically recommend flexible adjustment of sensory intensity and scenario complexity, while stressing that VR exposure itself does not engender therapeutic effects - the clinician does \cite{Sherrill2025-fo}. Yet VRET still faces high dropout rates \cite{benbowMetaanalyticExaminationAttrition2019} and fear relapse \cite{vervliet2013fear}. To address these challenges, both VR design and therapeutic workflow integration must prioritize engagement, increasing patient autonomy and self-efficacy to reduce avoidance and dropout \cite{Sherrill2025-fo}.

For C-PTSD, designing effective VRET that avoids dropouts is more complex \cite{Vermetten2025zu,rizzo2021fromcombat}. Triggers are more idiosyncratic, dissociation more pronounced, and recall often follows trial-and-error patterns \cite{brewin2025post,hyland2020relationship}. Unsuccessful attempts are likely to increase dropout risk. VRET design should therefore minimize failed attempts through patient-centered, experience-based approaches. Given lower resilience in C-PTSD, exposure scenario complexity should be significantly reduced, and the design process itself may elicit reactions and should be considered part of therapy.

Building on proven benefits of immersive technologies for trauma therapy, human-computer-interaction (HCI) research has explored more trauma-informed VR exposure designs tailored to individual characteristics such as communication style \cite{randazzoIfSomeoneDownvoted2023}. Guidelines for trauma-informed design aim to create safer, more empowering spaces for trauma survivors \cite{randazzo2023TraumaInformedDesign,scott2023SocialMedia}. Multi-Modal-Motion-Assisted Memory Desensitization and Reconsolidation (3MDR) exemplifies individualized exposure design \cite{gelderen2018Innovative}, enabling effective therapy for patients unresponsive to previous treatments \cite{Vermetten2025zu}. These findings demonstrate that individualized approaches achieve therapeutic success beyond generic designs. However, 3MDR's complexity may be too high for more severe trauma such as C-PTSD \cite{Vermetten2025zu}.

\noindent Taken together, VR offers a medium capable of bypassing avoidance and enabling personalized exposures, but its very strength, intense and immersive confrontation, also raises the risk of excessive demands that could cause therapy dropouts. Feasibility, hence, depends on an experience-based process and exposure design.

\section{{Methodology}}

We conducted a feasibility study \cite{bowen2009we} to explore whether VR exposure could be integrated into trauma-focused therapy for C-PTSD and accepted by patients. Two experienced therapists embedded graded VRET into routine care for eleven inpatients (six with childhood trauma, five with military trauma). Because no guidelines exist for VRET with C-PTSD. The design of the exposure was oriented around guidelines for VRET for PTSD with the objective of ensuring customizability and gradability, whilst final decision-making power was placed firmly with the therapists~\cite{Haft2025-ko}. However, the study still had to deal with an unusually large uncertainty factor. This factor was associated with risks not only for patients but also for other stakeholders. Questions emerged about how to explore the design of process and scenarios while protecting participants against accidental triggers and other risks of an exploratory process. Involving a VR developer in therapy sessions entailed additional risks, due to the lack of guidelines for trauma studies with such configurations and the developer's lack of experience with trauma therapy. To capture these dynamics, we treated the process itself as data. Therapists and the developer wrote structured reflective accounts after sessions \cite{chang2016autoethnography}, which we combined with clinical notes, design artifacts, psychometrics, and biofeedback. Rather than testing efficacy, our aim was to understand how design and communication shaped safety and value. We analyzed the corpus using reflexive thematic analysis focused on critical incidents~\cite{simmons_critical_2017}, which allowed us to surface methodological and ethical challenges of running VRET in practice.

\subsection{Study approach}
\begin{figure*}
    \centering
    \includegraphics[width=\linewidth]{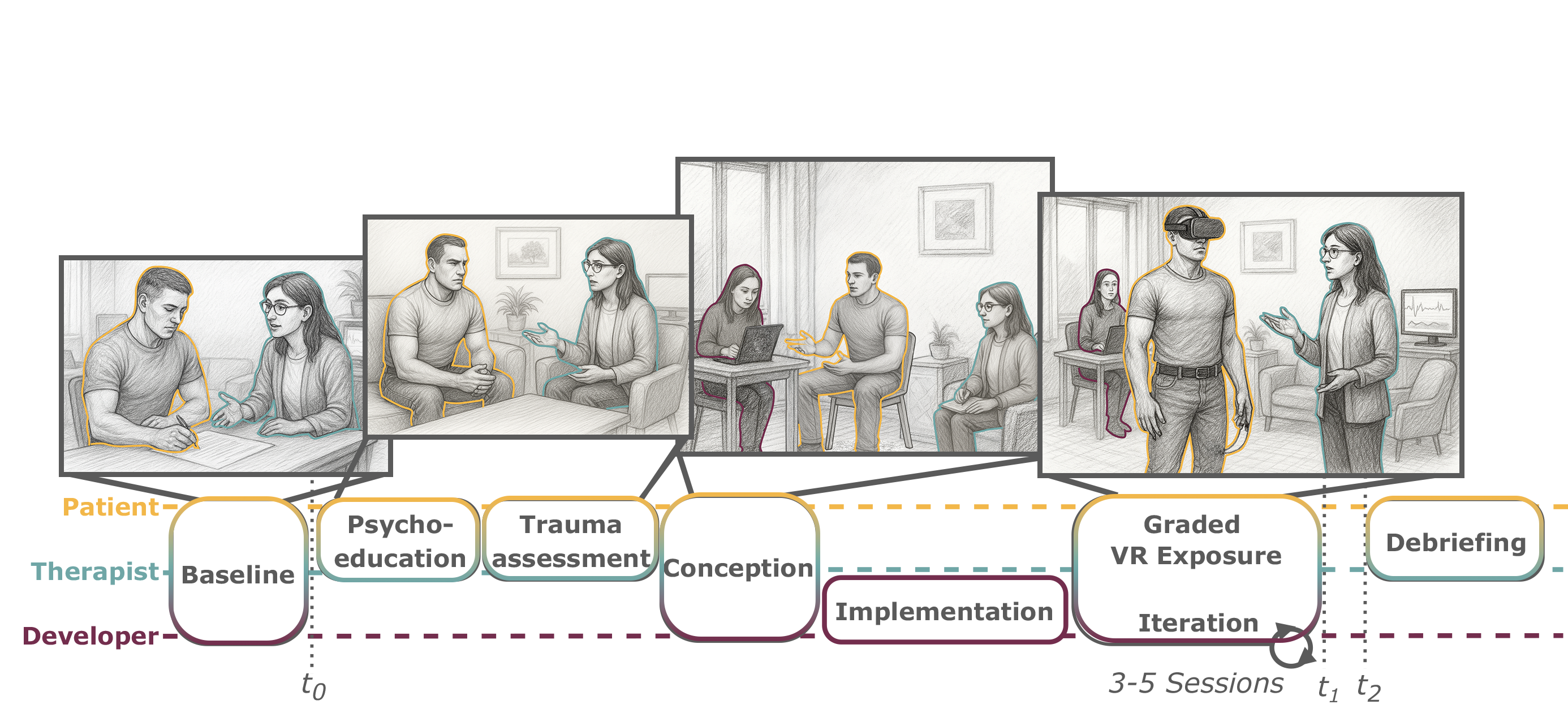}
    \caption{Procedure of our study showing which stakeholders were involved in the different processes of the study.}
    \Description{The illustration shows the study's procedure on a timeline separated by the stakeholders patient, therapist and developer. The processes are shown with exemplary images. The processes include baseline, psycho-education, trauma assessment, conception, implementation, graded exposure and iteration, and debriefing. The color of the outline and size of the processes show which stakeholders were involved in the process.}
    \label{fig:procedure}
\end{figure*}
The study followed the structure of a full trauma-focused therapy process, adapted to integrate VR exposure. We did not include a control group. All patients moved through four main phases: \emph{psychoeducation}, \emph{conception}, \emph{implementation} and \emph{exposure}, and \emph{debriefing}.

\smallskip
\noindent\emph{Psychoeducation.} At the start of therapy, patients were introduced to the nature of PTSD and C-PTSD and how trauma therapy works. They learned that therapy would not erase traumatic experiences but could provide strategies to cope with them. A central aim was to help patients recognize that they could control trigger situations and gradually make them more bearable. To support this, patients received skills training in affect regulation, including techniques such as diaphragmatic breathing, which they could use to regain control during emotional re-experiencing. In this phase, patients also set a realistic therapy goal to work toward, in order to avoid frustration from slow or incremental progress. 

\smallskip
\noindent\emph{Conception.} After psychoeducation, each patient worked with the therapist to identify the core of their trauma and the related core fear by revisiting formative experiences \cite{neunerNarrativeExpositionstherapieNET2021}. Once a suspected trigger scenario was identified, the developer was introduced. Depending on the patient’s resilience, in some cases, the developer joined the therapist and patient to refine and understand the trauma scenario. The patient first described the scenario, and the developer then asked follow-up questions or showed online images to clarify details. When the developer did not participate, the therapist or patient produced a written or visual summary of the scenario, often supplemented with photos or reference images, which was then sent to the developer for implementation. Inputs varied from short summaries of therapy sessions to photos or written descriptions provided by the patient.

\smallskip
\noindent\emph{Implementation}. The developer created VR scenes based on the described requirements, keeping implementation times short to explore the economic feasibility of the process. Level of detail was prioritized according to what patients and therapists emphasized as relevant. The scenes were built using Unity 3D\footnote{\url{https://unity.com/}, accessed: 02.09.25} and existing assets from the Unity Asset Store\footnote{\url{https://assetstore.unity.com/}, accessed: 02.09.25} and BlenderKit\footnote{\url{https://www.blenderkit.com/}, accessed: 02.09.25}. In rare cases, specific triggers had to be custom-built (e.g. Sabrina). Due to the restricted physical space and to avoid the uncontrolled effects due to erroneous locomotion, most VR scenarios were designed for stationary use where patients were asked to explore the virtual environment visually. The used gradation mechanisms and scenarios are summarized in~Table \ref{tab:gradation} and~Table \ref{tab:scenarios}. \\

\smallskip
\noindent\emph{Exposure}. Each patient then participated in three to five VR exposure sessions. The sessions were conducted in regular therapy rooms with a physical space of $2-3\mathrm{m}^2$, and lasted 45 to 60 minutes, with exposure durations of 15 to 45 minutes. Each session typically consisted of a preliminary discussion, exposure and iteration, and a post-discussion. 
Therapists stayed in verbal contact with the patients during the whole exposure and conducted the sessions in line with current guidelines on cognitive behavioural therapy trauma exposure treatment.
The Vive Pro Eye headset\footnote{\url{https://www.vive.com/us/support/vive-pro-eye/}, accessed: 02.09.25} was used via a wired connection to either a Laptop or Desktop PC. The headset has a resolution of 1440x1600 pixels per eye, a 68° field of view, and 6 degrees of freedom. Skin conductance sensors were attached to the passive hand of the patients and wired to a Nexus10 Device\footnote{\url{https://www.mindmedia.com/de/produkte/nexus-10-mkii/}, accessed: 05.09.25}. The data was streamed through a wired or Bluetooth connection to a separate computer. Data was screened using the BioTrace Software by Mindmedia\footnote{\url{https://mindmedia.com/de/products/biotrace}, accessed: 20.11.2025}. Patients could decide whether they preferred to sit or stand. During exposure, the therapists stood near the patients and talked to them from outside of VR.~Figure \ref{fig:teaser} illustrates an typical setting. Content, interaction and pacing tailored to individual resilience. When the developer was present, she took over the control of the simulation and bioscreening. Simulations could be adjusted dynamically, enabling flexible gradations and rapid iteration. All change requests were either instructed or previously approved by the therapist. When the developer was absent, therapists set up and controlled the VR system and screening themselves. They used simplified applications that gave them full control but reduced adaptability. Between sessions, scenarios were iterated based on patient responses and feedback.\\

\begin{figure*}[t]
    \centering
    \includegraphics[width=\linewidth]{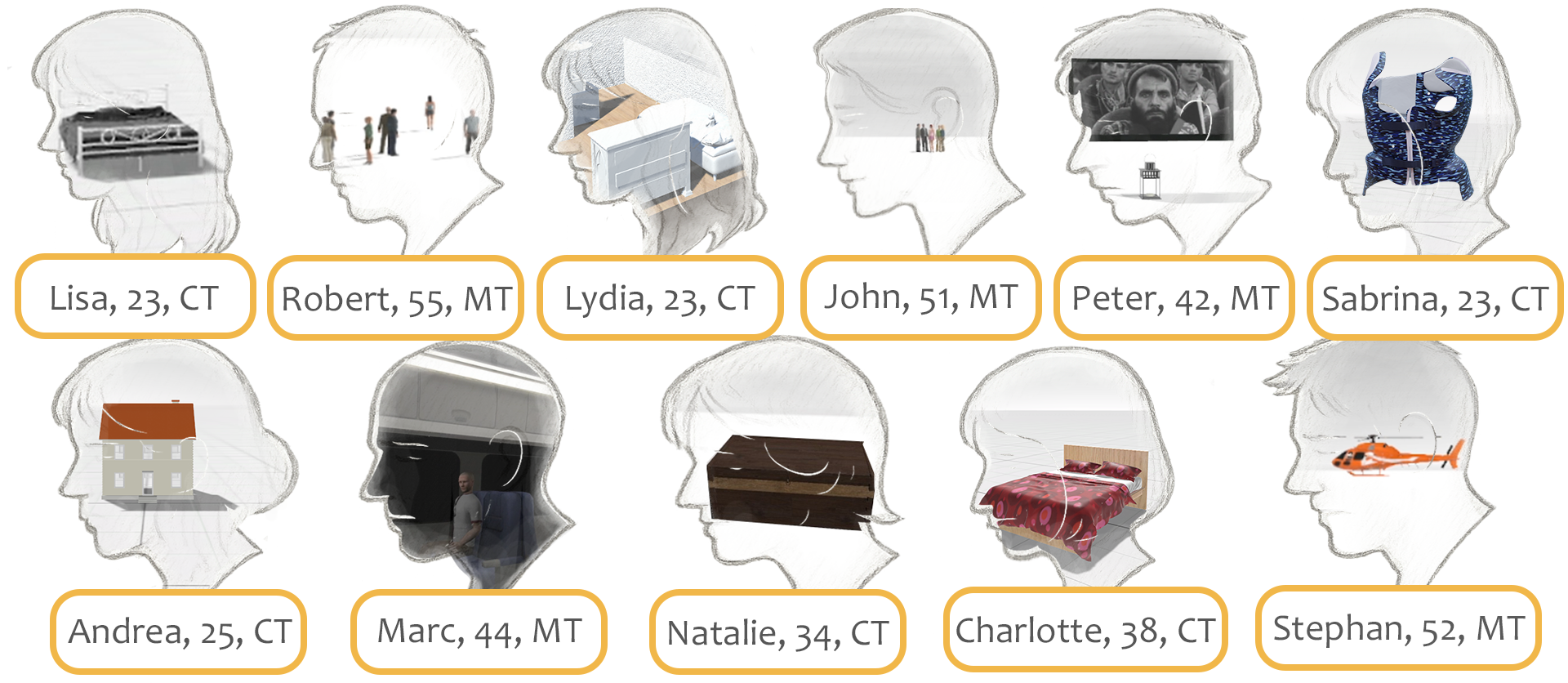}
    \caption{The patients who participated in our study, including pseudonym, age, gender, trauma type (f = female, m = male, CT = childhood trauma, MT = military trauma), and scenarios that they visited.}
    \Description{The figure shows eleven heads that represent the patients. Each head is filled with an image of the virtual reality scenario they visited. Each patient has a label including name and age.}
    \label{fig:participants}
\end{figure*}

\smallskip
\noindent\emph{Debriefing}. Therapy concluded with a debriefing phase, in which therapists and patients reflected on the VR exposure sessions, processed the emotional experiences, and prepared patients for discharge.\\

\smallskip
\noindent Because patients were highly vulnerable, each session required careful coordination of multiple tasks: attaching physiological sensors and managing assessment, running the VR simulation and adjusting exposure levels, delivering therapy, administering questionnaires, and documenting key events. To balance safety, feasibility, and effectiveness, different constellations of roles were tested. In some cases, the developer’s involvement allowed for richer exploration but introduced emotional risks. In others, therapist-only operation provided safer boundaries but limited flexibility.

\subsection{Participants}
    \begin{table*}[t]
    \centering
    \begin{tabular}{p{4cm} c c c c c c c c c}\hline 
        Construct & \multicolumn{3}{c}{Childhood/ Sexual} & \multicolumn{3}{c}{Military} & \multicolumn{3}{c}{Overall}\\ \cline{2-10}
        & Pre & Post & Delta & Pre & Post & Delta & Pre & Post & Delta \\ \hline
        Posttraumatic Symptom Scale - 10 items (PTSS-10) & \multicolumn{9}{c}{}\\ 
        \hspace*{3mm} Mean & $20.40$ & $19.40$ & $0.00$ &$19.20$ & $18.70$ & $-0.60$& $19.80$ & $18.70$ & $-0.60$\\  
        \hspace*{3mm} SD & $6.11$ & $9.17$ & $2.83$& $1.30$ & $2.12$ & $1.94$ &$4.21$ & $3.80$ & $2.50$\\  
        Borderline Symptom List (BSL) & \multicolumn{9}{c}{}\\
        \hspace*{3mm} Score  & $224.67$ & $180.20$ & $-12.60$& $206.60$ & $177.40$ & $-18.60$& $216.45$& $178.80$ & $-15.60$\\  
        \hspace*{3mm} SD  & $56.55$ & $75.74$ & $24.53$ & $34.51$ & $57.98$ & $22.86$ & $48.65$ & $67.46$& $23.90$ \\ 
        \hspace*{3mm} Percentile rank & $67$ & $48$ & $-18.6$ &$56$ & $43$ & $-12.6$& $62$ & $46$ & $-15.6$\\  
        Childhood Trauma Questionnaire (CTQ-SF) & \multicolumn{9}{c}{}\\ 
        \hspace*{3mm} Mean  & $78.80$ & - & - & $27.80$ & - & - & $53.30$& - & - \\  
        \hspace*{3mm} SD  & $22.69$ & - & - & $2.93$ & - & - & $32.02$ & - & - \\  
        Impact of Event Scale - revised (IES-R) & \multicolumn{9}{c}{}\\
        \hspace*{6mm} Mean  & $1.10$ & $0.84$ & $-0.26$& $1.80$& $1.48$& $-0.32$ & $1.45$ & $1.16$ & $-0.29$\\  
        \hspace*{6mm} SD  & $0.95$ & $1.30$ & $0.86$& $0.53$& $0.42$& $0.79$ & $0.81$ & $1.01$ & $0.78$\\  \hline
        \multicolumn{10}{l}{$N=11$ ($n=6$ with childhood trauma, $n=5$ with military trauma).}\\ 
    \end{tabular}
    \caption{Psychometric diagnostics of the patients}
    \label{tab:diagnostics}
\end{table*}

The feasibility study involved two experienced trauma therapists, one VR developer, and eleven patients clinically diagnosed with C-PTSD. Each group contributed different perspectives that were necessary to understand the feasibility and challenges of VRET for C-PTSD. Patients experienced the therapy directly, therapists guided and safeguarded the clinical process, and the developer created and adapted the VR scenarios while also reflecting on her role in the sessions. The ethics vote was given by the Ethics Committee Ulm \#216/23. The feasibility study was registered with the German 
Register for Clinical Studies (DRKS00032739). The study was commissioned by the Bundeswehr Medical Academy under the number \textit{SoFo 51K3-S-32 2426}.

\noindent\textbf{Therapists.} Two trauma specialists with extensive clinical experience participated in the study. Both had over a decade of practice in psychotherapy and had completed additional training in trauma-specific treatment methods.

\begin{itemize}
    
\item \textit{The Therapist for PTSD related to childhood trauma \newline (childhood trauma therapist in the following)} \\
is a senior consultant at a university psychiatric hospital, specialising in psychiatry and psychotherapy, with additional certification in psychotraumatology. His clinical focus is on personality disorders, trauma-related disorders, and crisis intervention. With more than 20 years of professional practice, the therapist has combined clinical work with academic research, including contributions to improving trauma treatment in inpatient psychiatric care.

\item \textit{Therapist for military trauma (military therapist in the following)}\\ is the chief psychologist in a military hospital, with experience that began through international research projects with conflict survivors. Over the years, their expertise has grown around trauma therapy methods for armed forces personnel, including work on complex trauma and moral injury. The therapist’s clinical practice is complemented by scientific research on the relationship between combat-related trauma, guilt, and post-traumatic disorders.
\end{itemize}

\noindent\textbf{Developer.}
The developer is a Ph.D. researcher in human–computer interaction who focuses on the perception of virtual environments and how design choices shape the feeling of ``being there'' in VR. Since 2022, the developer has collaborated with trauma specialists to examine how individuals with complex trauma perceive and respond to virtual environments. Within this study, the developer created and iterated the VR exposure scenarios, supported therapists during selected sessions, and contributed written accounts of her experiences to the process analysis.\\

\noindent\textbf{Patients.}
We recruited eleven patients (Gender: 6 female, 5 male, Age: $M=37.27$, $SD=11.91$) with interpersonal and military trauma experiences to investigate possible differences in needs and challenges between the groups. Five patients had experienced military trauma, and six had experienced childhood trauma (see~Table \ref{tab:diagnostics}). The patients were diagnosed with PTSD based on multiple traumatic events according to the International Classification of Diseases 10 (ICD-10)~\cite{worldhealthorganizationICD10Version2019} and the Diagnostic and Statistical Manual of Mental Disorders 5 (DSM-5)~\cite{DiagnosticStatisticalManual2013}. However, given the lack of differentiation between PTSD and C-PTSD in ICD-10 and DSM-5, we assessed C-PTSD in line with the suggested ICD-11~\cite{worldhealthorganizationICD11} symptomatology during the clinical sessions. In line with the current discussion on this differentiation between simple and complex PTSD, the patients were all part of the latter group, suffering from the complex symptoms of PTSD and in addition disturbances of self-organization.
As this article is primarily aimed at the HCI community, we opted to use C-PTSD for clarity. 
All patients were inpatients at the start of the study. Participation was entirely voluntary and could be aborted at any point in the therapy. To protect anonymity while allowing readers to connect with the cases, we use pseudonyms instead of IDs when presenting patient stories. Alongside a summary figure of all eleven patients (Figure \ref{fig:participants}) and a demographics overview (Table \ref{tab:demographics}), we describe four patients in more detail to illustrate the kinds of trauma backgrounds, therapeutic goals, and VR exposure processes encountered.

\smallskip
\noindent\textit{Robert - Military Trauma}
\begin{figure*}{t}
    \begin{center}  
        \includegraphics[width=0.8\linewidth]{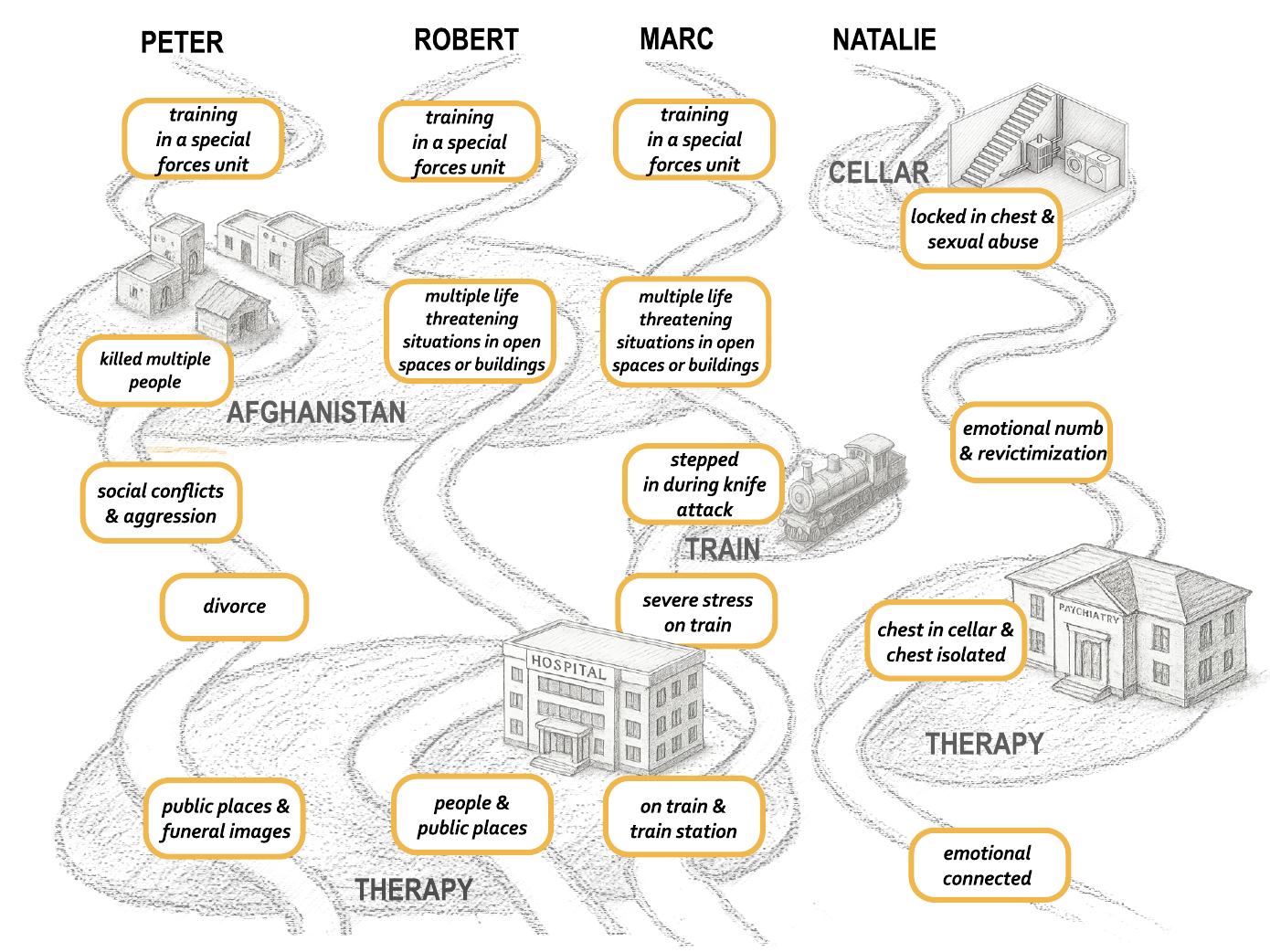}
        \caption{The journeys of Robert, Peter, Marc, and Natalie. Each journey shows key stages of the respective patient from trauma onset to life after therapy.}
    \end{center}
    \Description{Illustrates for different paths from the top to the bottom. Each path has texts that label a key stage. The paths of Robert, Peter, and Marc cross an area labeled with Afghanistan. The path of Marc further crosses an area representing his traumatic experience on a train. The fourth path of Natalie crosses an area representing the traumatic memories in the cellar. All four paths cross areas representing the inpatient stay at the psychiatry.}
    \label{fig:journeys}
\end{figure*}

\noindent Robert was admitted to inpatient treatment after decompensating from multiple deployment-related traumas and moral injuries, which had led to severe complex PTSD symptoms and strain on his marriage. After several trauma-focused inpatient intervals, his treatment turned to trauma-related agoraphobic symptoms, such as being stuck in crowds with a high level of threat, which he found most disruptive to daily life with his wife. Due to the high level of intrusions during prior exposure sessions, we opted against in vivo exposure and instead for a gradual VRET setting. For years, he had avoided family gatherings, walks, public places, and restaurants, as the presence of uncontrollable strangers triggered hyperarousal, anxiety, and avoidance. The therapeutic aim was to reduce his compulsive scanning of others, manage tension, and gradually re-engage with public situations.
He stated his goal clearly: to be able to sit in a café again with bearable stress levels. The VR exposures were structured toward this outcome. Starting in a neutral environment with only a few distant figures, scenarios were gradually intensified across five sessions. By the final session, he was able to sit in a virtual café and drink coffee. Physiological data mirrored this progress, with skin conductance levels rising to four times the baseline ($M_{Baseline}=1.01, MAX_{Exposure}=4.03$). After the study, Robert reported that regaining a sense of control encouraged him to try similar exposures in real life and he was indeed able to visit a café with manageable stress.\\

\noindent\textit{Peter - Military Trauma}\\
Peter is 42-year-old man was admitted as an inpatient after his partner left him due to intrusive experiences and frequent outbursts of anger. He was diagnosed with PTSD and profound moral injury, dominated by guilt. Clinically, this presented as a persistent sense of threat and a heightened risk of aggression in everyday situations, such as minor disagreements in a supermarket. He had also become socially withdrawn, driven by a harsh inner critic with punitive beliefs such as \textit{``you killed, that makes you a monster and unlovable''} and \textit{``you are a failure.''}
Early trauma-focused work addressed deployment experiences linked to fear of death and self-endangerment. He later agreed to join the feasibility study on individualized VR trigger stimuli. In sessions 1-3, Peter visited public places with people of different physical appearances to address typical feared situations in everyday life. In session $4$, he was shown pictures of people who reminded him of the dead people during his missions, each accompanied by a candle. This VR scenario targeted at retrieving avoided traumatic memories. Within this setting, he apologized to them for the harm he had caused, which enabled him to access emotions that had previously remained unreachable. This breakthrough was mirrored physiologically, with skin conductance levels rising by up to $111.79\%$.\\

\noindent\textit{Marc - Military Trauma}\\
\noindent Marc underwent inpatient trauma-focused treatment after experiencing numerous traumatic events and moral injuries during multiple deployments abroad. We clinically diagnosed PTSD. He perceived the treatment as positive for various reasons, but had to travel a long distance by public transport, which was an immense burden for him and was associated with a strong experience of tension and threat, a constant readiness to act in a potential emergency situation and subsequent exhaustion. For this reason, after several trauma-focused intervals on the military deployment experiences, the focus was now on dealing with trauma-related agoraphobic fears in public spaces, especially train journeys and train stations, in order to find an adequate way of dealing with the tension, irritability, maladaptive cognitive mechanisms, and avoidance tendencies, which he had experienced during similar traumatic events during deployment. We simulated different scenarios of a train ride to promote the retrieval of traumatic memories. Exposure was effective, which is illustrated by increases of skin conductance (SC) of up to $135\%$.\\

\noindent\textit{Natalie - Childhood trauma including sexual abuse}\\
\noindent Natalie is a woman in her early thirties with a history of repeated childhood sexual abuse who grew up in a violent household where conflict was punished by being locked in the cellar and sometimes inside a chest. Clinically, she presented with emotional numbing and difficulty accessing her feelings. In therapy, she agreed to confront triggers to reconnect with suppressed emotions.
Accordingly, all VR scenarios simulated aspects of traumatic memories in order to promote imaginary exposure and achieve emotional access to avoided traumatic memories. The first VR scene recreated a red-brick room, which she associated with a house where sexual assaults took place. This triggered only mild arousal, but allowed her to recall parental punishments and associate them with feelings of worthlessness that had made her vulnerable to abuse. A second scene was linked to these earlier memories and focused on a basement where she had been locked up as punishment, which she was able to describe in detail. Here she remembered that she often looked out of the window to escape in her mind. Because when she actually fled through this window, she was locked in a chest as punishment.
When the chest was isolated in a white VR environment, her emotional response intensified. She explained that in the full cellar scene she could distract herself by focusing on neutral elements or imagining escape, whereas the solitary chest forced her to confront its meaning directly. This proved to be the strongest trigger, allowing her to retrieve emotions and memories otherwise blocked. She described the experience as challenging but valued it as a way to face traumatic material she had long avoided through dysfunctional coping in daily life.

\subsection{Positionality}

Positionality was central to how this study evolved. Three authors, the developer (first author) and the two therapists (fifth and sixth), were directly involved in delivering the intervention and produced written accounts that became part of the data. Their dual roles as participants and authors shaped what was documented, how the study was designed, and how experiences were interpreted. These insider reflections form the core of the empirical material, offering rare access to the methodological and ethical challenges of VRET.
To balance this, three additional authors contributed with varying levels of distance. The second author, with a background in HCI and design research, brought a perspective focused on participatory and ethical approaches to immersive technologies and collaborated on the analysis without being involved in the intervention itself. The third and fourth authors, senior academics in HCI and XR, guided the framing of the study and situated the findings within broader research.

\noindent Qualitative traditions acknowledge that insider accounts can surface aspects of practice that might otherwise remain hidden, while also stressing the need for reflexivity to mitigate bias \cite{singh_exploring_2025}. Our approach reflects this balance: insider authors provided detailed, practice-based accounts, while external collaborators added critique, contextualization, and interpretive distance.
This composition shaped the paper in two ways. First, it required explicit reflection on roles, as therapists and the developer acted both as data sources and co-authors, while external collaborators interrogated and extended their interpretations. Second, it combined experiential depth with analytic distance, enabling us to surface the practical and ethical complexities of conducting VRET for C-PTSD.

\subsection{Data collection and analysis}

\subsubsection{Data collection}
Because patients were in inpatient care for C-PTSD, direct post-session interviews were not feasible. Engaging them in additional reflective conversations outside of therapy risked destabilizing progress and posed ethical concerns of retraumatization. For this reason, patients’ voices were captured indirectly through the clinical process itself: their behavior and comments during sessions, session documentation, psychometric measures, and therapist reflections. These were combined with first-hand accounts from the therapists and the developer, who were in the position to observe and experience the unfolding process.  We drew on multiple forms of data:

\smallskip
\noindent\textbf{Therapist and developer accounts.} After the sessions, both therapists and the developer produced structured written accounts of their experiences. These included what happened, how patients reacted, and their own emotional and practical experiences. Their accounts are a useful method in this context for three reasons: (1) they capture immediate impressions while events are fresh, (2) they record dimensions (such as professional uncertainty, communication breakdowns, or emotional load) that would not appear in standardized clinical notes, and (3) they provide insight into how professional roles and responsibilities were experienced in practice. These accounts form the central corpus for analysis.

\smallskip
\noindent\textbf{Session notes and study artifacts.} Therapists kept clinical notes of the sessions, and the developer archived design materials (scenario sketches, screenshots, and test applications). The artifacts allowed us to reconstruct how scenarios evolved and how decisions were made.

\smallskip
\noindent\textbf{Diagnostics.}
We carried out psychometric diagnostics before admission ($t_0$, pre-treatment), after each exposure session with regard to the VR experience ($t_1$), and after completion ($t_2$, end) of the treatment. Biofeedback was recorded and screened during exposure. More specifically, we collected the following data:

\begin{itemize}
    \item \textbf{Pre- and post treatment ($t_0$, $t_2$)}:
    \begin{itemize}
        \item \textbf{Borderline Symptomlist - 105}~\cite{bohusEntwicklungBorderlineSymptomListe2001}: Borderline symptoms assessed via 105 items that are answered on a 4-point Likert Scale from "not at all" to "very much". The score was determined as the average of all items as recommended by the authors. The score is the sum of items 1-95, ranging from 0 to 380.
        \item \textbf{Posttraumatic Symptom Scale - 10 items (PTSS-10)}~\cite{maercker2003posttraumatische}: PTSD diagnosis via 10 items that are answered on a 4-point Likert Scale from "not at all" = 0 to "often" = 3. PTSD is suspected from a score greater than $12.5$. 
        \item \textbf{Impact of Event Scale - revised (IES-R)}~\cite{maercker1998erfassung}: PTSD symptoms with subscales for intrusion, avoidance, and hyperarousal. Assessed via 22 items that are answered on a 4-point Likert scale from "not at all" = 0, to "often" = 5. The score was computed according to the instructions of the authors, ranging from $-5.06$ to $3.69$. A suspected diagnosis of PTSD is given for scores $>0$.  
        \item \textbf{Childhood Trauma Questionnaire (CTQ-SF)}~\cite{hagborgChildhoodTraumaQuestionnaire2022}: Childhood trauma. Includes 31 items with subscales for emotional abuse, physical abuse, sexual abuse, emotional neglect, and physical neglect. Answered on a 5-point Likert scale from "not at all" to "very often". The score is the sum of the five subscales, ranging from $25$ to $125$.
    \end{itemize}
    \item \textbf{During exposure} Heart rate ($BPM$) and skin conductance ($\si{\micro\siemens}$) were assessed and screened using the NeXus 10 MKII\footnote{\url{https://www.mindmedia.com/de/produkte/nexus-10-mkii/}, accessed: 05.09.25} by Mind media with the corresponding software BioTrace. Changes from session baseline were used as indicators of emotional arousal. Following \citet{posadaquintero2020eda}, skin conductance increases greater than $0.05\si{\micro\siemens}$ were classified as skin conductance responses and treated as clinically meaningful. A continuous rise during exposure signaled excessive demand, whereas a return to prior levels indicated habituation. Patients’ reactions to each scenario are reported in Table \ref{tab:demographics}.
    \item \textbf{Post exposure session ($t_1$)}
        \begin{itemize}
            \item \textbf{igroup Presence Questionnaire (IPQ)}~\cite{schubertExperiencePresenceFactor2001}: Perceived presence during virtual experience with subscales for Spatial Presence, Involvement, and Realness. Includes 14 items that are answered on a 7-point Likert Scale (-3,3) with individual labels. The score is the average of all item scores.
            \item \textbf{Single items on VR experience (Table \ref{tab:questionnaire_vrexperience}):} 9 custom items to explore VR experience and acceptance of VR in the context of exposure therapy. Items 1-7 were answered on a 7-point Likert Scale; items 8 and 9 were open questions.
        \end{itemize}
\end{itemize}

\noindent The measures were used descriptively to contextualize accounts and support the interpretation of incidents. For example, sudden spikes in skin conductance could indicate physiological evidence of triggering, which could then be cross-checked with therapists’ observations and patient comments. At the same time, we treated biometric data cautiously. For example, arousal does not map one-to-one to therapeutic progress or distress. A patient may show high arousal yet this might be due to feeling cold rather than being triggered, or vice versa. For this reason, diagnostic data were never treated as sole indicators but used as one layer within a broader triangulation.

\subsubsection{Data analysis}
The analytic process was conducted collaboratively by the first and second authors over six weeks. The first author, who was also the developer, contributed insider knowledge of how sessions unfolded, why particular VR design choices were made, and the emotional demands of direct participation. The second author, who had no role in the study setup or clinical delivery, provided distance and an external perspective. This combination enabled a balance between experiential depth and analytic detachment.

We followed the principles of reflexive thematic analysis~\cite{braun_one_2021, braun_reflecting_2019}, treating the insider and outsider perspectives as complementary. Rather than pursuing inter-coder reliability, the focus was on reflexive meaning-making. The analysis was further shaped by the critical incident technique~\cite{viergever_critical_2019, ppali_creating_2025}, which foregrounds events that significantly influence outcomes or reveal tensions not visible in routine processes. Incidents such as accidental triggers in default VR environments or uncertainty about responding to highly emotional disclosures guided theme development and were later expanded into vignettes that combined multiple perspectives, making it possible to examine both what occurred and how it was experienced by different stakeholders.

At the start of the process, the first author provided detailed context for the study, including technical and therapeutic notes. The second author then conducted inductive coding of therapist and developer accounts, generating codes that captured practical actions (e.g., ``adjusted VR exposure level''), emotional states (e.g., ``felt overwhelmed''), and methodological concerns (e.g., ``scenario description unclear''). Weekly discussions between the two authors allowed coded excerpts to be reviewed in detail. The first author supplied additional context about technical constraints, therapeutic protocols, and interpersonal dynamics, while the second author challenged interpretations and raised alternative readings. Each discussion was documented in analytic memos that recorded emerging patterns and disagreements.

Through this iterative process, codes were grouped into broader categories and refined into themes that captured recurring challenges such as role boundaries, unexpected triggers, and balancing safety with adaptability. Psychometric scores and biometric traces were used descriptively to situate qualitative accounts but were not treated as stand-alone indicators of progress, since arousal data can have multiple interpretations. The outcome was a set of seven themes that reflect the methodological and ethical complexities of conducting trauma-focused VR research, presented in the next section.

\section{{Findings}}
In the sections below we present our findings. We write in a descriptive way to give an understanding how certain processes unfolded, using also quotes from the entries of the therapists and developer.

\subsection{How do we find (and design) a trigger?}
\subsubsection{Finding the trigger.}

Identifying effective trauma triggers was one of the most complex aspects of the study. Unlike standardized VR scenarios often used for combat-related PTSD~\cite{rizzoVirtualRealityExposure2014}, triggers for C-PTSD were highly individual, fragmented, and often difficult for patients to articulate. To address this, we embedded methods from psychotherapy and software development into the therapeutic process, which unfolded through overlapping phases of memory identification, conversion, and iteration.

During memory identification, therapists worked with patients in the psycho-assessment phase to reconstruct formative experiences and detect scenarios that might evoke trauma recall. These initial narratives (e.g., as a cellar associated with childhood punishment or crowded public spaces avoided for years) were often fragmentary. Their therapeutic relevance could not be established until tested in VR. When memory processing stalled or remained incomplete, narratives were converted into design requirements for VR implementation. Most often, this followed a top-down approach, with therapists or patients providing summaries, sketches, or reference photos that were passed to the developer. This was favored for childhood trauma cases with lower resilience and for military patients with shorter inpatient stays. The conversion process acted as an additional confrontation step: translating narratives into design details forced patients to reflect on aspects that might not have surfaced in therapy conversations. At times, patients’ visual imagination proved more precise than therapist summaries. For example, Lydia and her therapist initially described a \textit{``room with red brick walls and blue carpeting''}. Early builds failed to trigger her, and only after multiple iterations did it become clear that the specific materials were crucial. A full month passed between the first description and a usable exposure.

\noindent In two cases, the developer collaborated with patients directly in remote sessions, using online images as prompts to refine details in real time. This approach surfaced misunderstandings and enabled the retrieval of traumatic memories that had not emerged earlier. During the reconstruction of Lydia’s childhood room, she suddenly recalled that her teddy bear had a blue bow tie. What began as a generic object became a focal point, prompting both the patient and the therapists to recognize its deeper significance. Collaborative conversion not only clarified design priorities but also encouraged patients to confront avoided aspects of their trauma.

    \begin{figure}
        \begin{center}
             \includegraphics[width=0.8\linewidth]{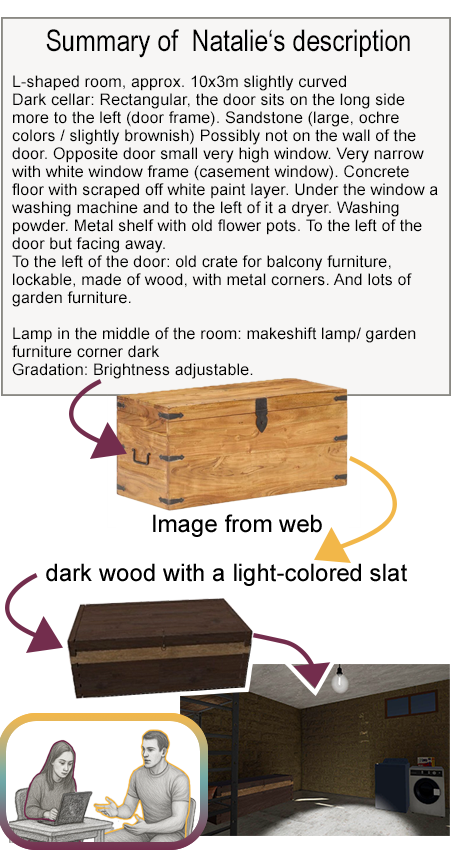}
             \caption{The different stages of the conversion of Natalie's scenario description to the final virtual reality scenarios. Stages include the textual description, a web image used for discussion, the resulting chest model, and the two VR scenarios including the chest.}
         \label{fig:require}
        \Description{The figure shows different stages of the design process of Natalie's scenario, starting with a textual description. Below is a web image that was used by the developer as an example. Further below is the 3D model of the chest that the developer designed. On the right and bottom of it are images of the two virtual reality scenarios with that chest. Once the neutral scene, once the cellar.}
         \end{center}
    \end{figure}

\noindent Outputs from this process ranged from textual notes and sketches to web images and personal photos (see~Figure \ref{fig:require}). Whether conversion was necessary in every case remained an open question. For complex scenarios, it clarified priorities and reduced wasted effort, but in simpler cases (e.g., where a single object acted as the trigger), it added little. Indeed, some of the most critical triggers only emerged during direct VR exposure. For instance, Peter was unaffected by scene complexity or crowd size but reacted strongly to the behavior of individual virtual characters, a pattern also observed with Lydia. Patients themselves were ambivalent, rating the helpfulness of participating in the design process neutrally ($M=3.38, SD=1.48$; see \hyperlink{VRQ7}{question 7 of the post-session questionnaire on VR experience}).

\noindent To reduce the risk of misimplementation, interim versions were shared with therapists as screenshots, 3D previews, or lightweight prototypes. These iterations often revealed unexpected priorities that shaped refinements. In the case of Lydia’s brick-wall scenario, successive iterations clarified that the bricks should be smaller, the carpet a specific azure blue, and the joints white for the scene to become therapeutically effective.

\subsubsection{Implementing the trigger.}
Once candidate triggers were identified, we prioritized fast and flexible iteration over polished builds. Most environments were developed in Unity with off-the-shelf assets, enabling runtime changes to textures, object placement, and crowd configurations (Table \ref{tab:scenarios}). This adaptability was essential for tailoring exposures. 
Scenarios were designed for both synchronous (in-session) and asynchronous (between-session) adjustments. Instead of compiling full applications, scenes were often simulated directly in Unity, which allowed rapid changes to objects and materials. Assets with runtime configuration options, such as the Citizens Pro 2024 package\footnote{\url{https://assetstore.unity.com/packages/3d/characters/citizens-pro-2024-143604}}, supported quick crowd modifications. Flexibility, however, came at a cost. When Peter asked for a denser marketplace, the developer expanded the spawn area but overlooked that his avatar was inside it, causing a virtual character to appear directly beside him and triggering unintended arousal.
Patient positioning proved to be as important as scene content. For Robert, therapeutic progress meant staying on the edge of a crowd rather than its center: \textit{``It’s enough for me to keep walking along the edge, where I can scan everything, but with a bearable sense of stress.''}. To reflect this, therapists began sessions by asking where patients would stand in real life, and the developer set their starting positions accordingly.

The required level of detail varied widely. Generic assets sometimes sufficed, as with Andrea, who reacted strongly to a standard dollhouse. In other cases, fidelity was crucial. Despite detailed descriptions, Lydia’s teddy bear only became effective after refinements. Public-space scenes showed similar contrasts: simple looped animations made Marc feel watched, while Peter dismissed the repetition as artificial and disengaged. Lightweight prototypes and early feedback helped avoid wasted effort, and imperfections were not always detrimental. As the developer reflected, \textit{``Predefined animation scripts caused unplanned triggers [...] some patients got really anxious about implausible behaviors. So my conclusion was that uncanny valley effects may even help by preventing unbearable levels of arousal.''}

Gradation strategies relied on straightforward manipulations such as avatar distance, object proximity, brightness, or crowd density and dynamics. Distance worked well in neutral scenes, while confined spaces required other tactics: Natalie progressed through brightness levels, Lydia added furniture piece by piece, and Robert varied crowd movement and density. Therapists also abstracted triggers into intermediate steps, as with Stephan, who first encountered an emergency helicopter model combined with the sound of military helicopters. Patients appreciated these strategies; as the military therapist noted, \textit{``The graded exposure seemed very helpful and allowed patients to finally confront themselves with triggers they had avoided before.''}.

Moreover, not all triggers were planned. Default VR lobbies sometimes provoked strong responses. The SteamVR mountain scene\footnote{\url{https://www.youtube.com/watch?v=2OSrcA_59Is}, accessed: 05.09.2025} reminded John of Afghanistan, while the Vive northern-lights lobby\footnote{\url{https://steamcommunity.com/sharedfiles/filedetails/?id=3014836195}, accessed: 05.09.2025} evoked both distress and positive memories depending on the patient. These reactions led us to replace defaults with neutral loading spaces. Escape routes also proved important. Even when unused, doors or alleys gave patients reassurance and control. As the developer explained, \textit{``It was not that the patients really used virtual escape opportunities, but some stated that they felt better when they were able to plan escape routes in VR.''}

The most striking example of C-PTSD’s complexity was Natalie. A detailed reconstruction of the cellar where she had been abused produced little effect, as neutral elements allowed distraction. By contrast, a single wooden chest, where she had been locked as a child, triggered strong emotional recall and surfaced repressed memories. As the childhood trauma therapist observed, \textit{``This is because she can distract herself better in the cellar with neutral elements.''}. Effort was therefore redirected to refining the chest, while other room details were scaled back. To manage spatial limits in the therapy room, her avatar was positioned directly in front of the cellar door, reducing the need for locomotion.

    \subsection{The struggle (and benefits) of not being an expert}

    Involving the developer directly in therapy sessions brought clear advantages but also revealed significant challenges. Feasibility checks and in-session adjustments shortened design–implementa\-tion cycles. For example, in Robert’s penultimate session, the developer suggested simulating a recently visited public space - an option neither patient nor therapist had considered technically possible. Dynamic tweaks also clarified triggers, such as recognizing that Peter reacted more strongly to suspicious behavior than to ethnic appearance. The developer could resolve technical problems (e.g., tracking) and temporarily handle grading or biofeedback, allowing the therapist to focus fully on therapy. Yet the absence of trauma-therapy expertise introduced risks that shaped both process and relationships.
    
Unlike therapists and patients, who had developed emotional distance through years of exposure, the developer was often deeply affected by patients’ narratives. During Peter’s baseline, she reported lingering second-hand arousal, \textit{``I never experienced a person radiating such intense levels of arousal without saying a word [...] even as I arrived back at my office, I still felt this undefinable arousal.''} Thoughts such as \textit{``the patient deserves the attention''} and concerns about appearing unprofessional discouraged her from requesting post-session debriefs. Communication also proved challenging. Without psychotherapy training, the developer was unsure about appropriate phrasing and the risk of unintentional triggers, compounded by patient-specific communication styles. In requirements interviews, therapists framed her role as an implementation expert, keeping discussions comparatively objective. But in military-trauma cases, where requirements were gathered during inpatient sessions, design talk became interwoven with emotional therapeutic discourse.

This unveiled uncertainty about sharing empathy. The developer often felt the need to acknowledge patients’ trust, \textit{``Their experiences were not only much worse than everything I’ve been told before [...] I felt the need to respond appropriately.''}. Some patients valued this, describing small gestures as \textit{``very validating'' [military therapist]}, while others preferred the developer to remain passive. Empathic engagement, however, increased patients' openness, further increasing the developer’s emotional load. Coping strategies such as talking to colleagues were constrained by confidentiality and the sensitivity of trauma narratives, \textit{``Trauma does not pay respect to ethic [...] I worried about causing prejudices through incomplete information [...] whether and how to communicate with outsiders.'' [developer]}.

The presence of an unfamiliar developer also altered patient-\newline therapist dynamics. Therapists observed that the intimacy of the therapeutic relationship was sometimes unsettled, particularly when guilt was involved: \textit{``the intimacy of the therapist-patient interaction was somewhat unsettled, particularly in cases where guilt played a role.'' [military therapist]}. Patients themselves appeared attentive to the developer’s reactions, \textit{``they knew I had a different background from the therapists. Some of them even tried to explain themselves to avoid confusion''}. In some cases, empathy from the developer improved trust and openness; in others, patients preferred a clearer separation. At the same time, the developer faced her own uncertainty about the patients. Initial impressions sometimes shaped her sense of safety, as when Marc’s stern expression felt intimidating, \textit{``I was immediately intimidated by his appearance [...] I avoided eye contact to prevent conflict.''}. Only later, when Marc explained that his demeanor stemmed from social expectations that he intervene in crises, did the developer realize the extent of her own prejudices, \textit{``as I understood that the reason [...] was just the feeling of being urged by the public [...] I felt really guilty about my prejudices.''}.

    \subsection{Co-designing the experience with patients}
    Therapists encouraged patients to take an active role in shaping the intervention. The goal was not only to achieve patient-centered outcomes but also to test whether patients valued collaboration in their therapy design. As the military therapist reflected: \textit{Being part of a new project, seeing their feedback implemented, and shaping their therapy was a positive aspect in the patient–therapist relationship.''}. The same therapist also described the satisfaction patients get in knowing their feedback might help others, \textit{``Another aspect of appreciation [...] can be found in the possibility for participating patients to give direct feedback on the study set-up and the linked feeling of helping others and shaping progress.''}.
    
Contributions varied widely. Some patients shared only minimal input, such as a short phone clip of their room, while others provided detailed written descriptions and reference images. During sessions, patients often suggested improvements on their own initiative. Lydia, for example, participated in a remote interview to reconstruct her childhood room live, offering extensive feedback and reporting that she valued the collaborative process ($M=4.6$). Similarly, Robert strongly appreciated the opportunity to co-design ($M=6.0$). His openness encouraged the developer to ask about design preferences, and the therapist allowed him to set his own exposure levels. By contrast, Marc also managed his gradation but misjudged its effects, requesting rapid increases that led to hyperarousal and days of fatigue. Control was therefore returned to the therapist. His feedback focused mainly on how much scenarios affected him rather than on design refinements, making it harder for the developer to adjust effectively. Peter took a similar stance, concentrating more on emotional responses than on design details, which led to a more trial-and-error process.

Patient involvement extended beyond the virtual design. Each VR session began by clarifying the day’s therapeutic goal, with scenarios adapted accordingly. Patients also influenced the physical setting - choosing whether to sit or stand, where to position themselves in the room, and whether to remain near an escape route. They decided when to start, pause, or end exposures, and how long to take breaks. While therapists guided these decisions, they often prioritized self-efficacy over strict exposure heuristics. As one explained: \textit{``While during classic exposure interventions, I would rate a high presence as more efficient, during the VR-sessions, I found self-efficacy to be the key for exposure efficacy.''}.

    \subsection{To VR or not to VR?}
    Across patients, VR proved to be an effective medium. Individuals such as Sabrina, who showed little response to static design previews, reacted strongly in VR - suggesting that immersive presentation is more impactful than 2D visualizations commonly used in therapy. None of the patients had used VR more than three times before, yet post-session questionnaires showed they valued it retrospectively ($M=5.34$; see~Table \ref{tab:presence}) and slightly preferred it over imaginative exposure. As one therapist summarized, \textit{``The new experience gave all of them a push in their self-efficacy and allowed the continuation on the exposed topics in the following sessions.''}.

   A key strength of VR was its adaptability. Patients often began with broad scenarios, and iterative adjustment isolated specific triggers. For Robert, repeated changes to crowd size and dynamics revealed which configurations provoked arousal - something impossible to achieve \textit{in situ}. For Lydia, an empty childhood room had little effect until furniture was added, instantly amplifying relevance. Such rapid modifications underscored VR’s value for titrated exposure. Yet flexibility also carried risks. Frequent adjustments raised the likelihood of technical glitches such as flickering textures or calibration errors, which disrupted immersion. Equipment had to be moved across therapy rooms, producing variability and occasional malfunctions. While therapists bridged delays with reassurance, interruptions tested patient trust. More concerning were accidental triggers. For instance, during a session of Robert, the developer once configured the dynamics of a virtual crowd incorrectly causing them to run around. This overstrained Robert, which provoked arousal. This and similar incidents risked undermining confidence in the therapy.
   
VR introduced other challenges, too. The headset itself posed barriers. Because patients could not see their surroundings, therapist presence and trust became crucial: \textit{``[...] since patients cannot see their surroundings while in the VR, it seems important that the therapist in the room is someone they know and trust.'' [military therapist]}. For therapists, goggles obstructed facial cues, making it harder to stay emotionally attuned: \textit{One difficulty that presented itself was given by the VR-goggles, which made it more difficult to assess their facial expressions and compared to other exposure techniques made it more difficult to stay emotionally in touch.'' [military therapist]}. In some cases, the hardware itself became a trigger. Marc and John associated the headset with military night-vision devices; for John, this reaction led to the study being aborted.

The setup itself further shaped experiences. Requirements analyses for Natalie and Lydia were conducted remotely, which provided useful input but felt misaligned to the developer: \textit{``the whole interview experience felt quite strange. It felt somehow like a test situation [...] misplaced as the patient shared information about a place where she experienced traumatic experiences over years. It was effective, but for me, this was not patient-centered design.''}. Exposure sessions were also constrained by the physical environment. Therapy rooms had to fit VR play area ($2m \times 1.5m$), biotracking equipment, and still leave escape routes clear. Spontaneous adjustments (e.g., moving the developer outside to reduce disclosure pressure) sometimes disrupted tracking or screening. These shifts highlighted how tightly therapeutic, logistical, and technical factors were intertwined.

    \subsection{Taking things one step at a time}

    \begin{figure}
        \begin{center}
        \includegraphics[width=0.9\linewidth]{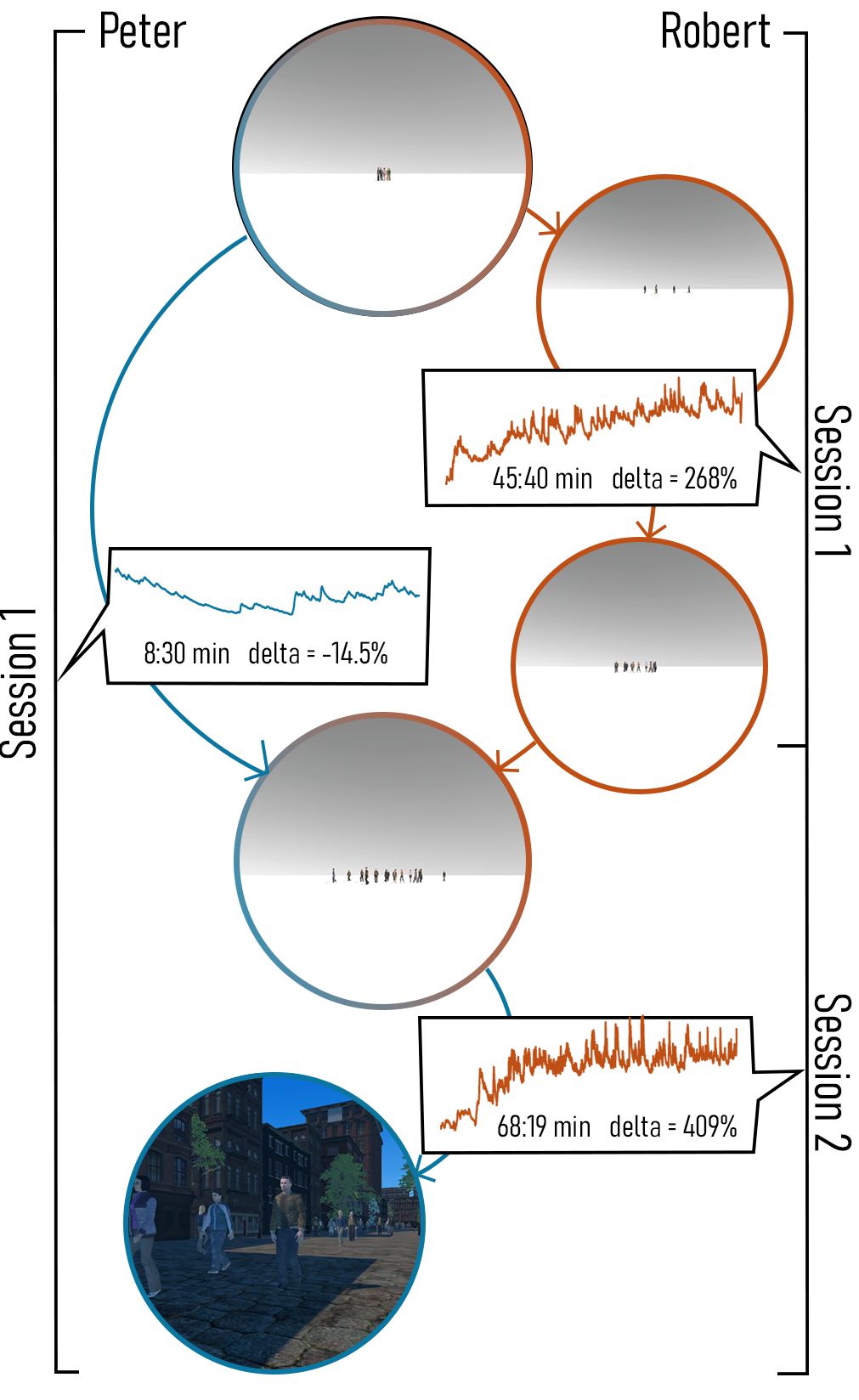}
        \caption{Comparison of Robert's and Peter's first gradation levels. Session durations are in minutes; the delta indicates the max. difference of the patients' skin conductance from the first minute in percent.}
        \label{fig:gradation_levels}
        \Description{The figure shows episodes of gradation of Robert and Peter. Images show different gradations. For each gradation image, the measured skin conductance is visualized.}
        \end{center}
    \end{figure}

    The complexity of scenarios and the use of gradation varied widely across patients. For Peter, the first session began with a crowded street where people walked past at a distance of 1-2 meters. This produced no measurable reaction ($\delta_{SC}\approx0$). In the next session, the scene was intensified with a busy marketplace, but this too was ineffective. By contrast, Robert started with just five people placed 20 meters away in a neutral scene. When the number was increased to fifteen, his response was strong ($\delta_{SC}\approx324\%$). Although he chose the steps himself, which gave him a sense of \textit{``self-efficacy'' [military therapist]}, the sudden adjustments were overwhelming. The developer therefore agreed to announce changes in advance — a lesson that came from realizing how differently Robert reacted compared to Peter. It became clear that responses could not be generalized across patients. For patients such as John, Stephan, and Charlotte, the risks of a participatory design process were considered too high. Because VR exposure for C-PTSD had no established precedent, therapists could not predict after-effects with certainty. To minimize the chance of unnoticed hyperarousal, they capped the number of scenario adaptations per session. Control over exposure levels also depended on each patient’s ability for self-evaluation. When appropriate, therapists allowed patients to decide when and how much to increase exposure, and to reverse changes if arousal became excessive. This flexibility reinforced self-efficacy but required careful monitoring. In sessions without the developer, scenarios had to be preset in advance. Sometimes this worked well: for John, no increases were needed, as the initial scenario was already intense enough to sustain all sessions. In other cases, such as Charlotte, habituation reduced effectiveness. After three sessions, her first scenario no longer provoked reactions, prompting the therapist to request a second scenario to maintain therapeutic value.

    \subsection{The value of visual triggers}
    Although there was initial uncertainty about whether VR would support memory work, both therapists and patients ultimately saw clear value in visual triggers. As the childhood trauma therapist reflected, \textit{``visual accompaniment makes it easier for seriously ill people in particular to access their emotional experiences and also to recount the associated memories.''}. Patients often expressed surprise at the effectiveness: Natalie noted, \textit{``I did not expect it to work this well!''}.
    Feedback suggested that VR helped counteract avoidance, a common barrier in trauma therapy. Andrea described that \textit{``suppression [was] no longer possible,''} while Lydia explained that she was \textit{``brought into old movement patterns''}, which enabled her and the therapist to address them directly. Therapists observed similar effects, noting that patients could retrieve memories long suppressed. As the childhood trauma therapist explained about Lydia, \textit{``These memories of the corner in her childhood room were no longer present before and conventional therapy would not have brought them to mind. It was only through the visualization that she remembered it again and was able to talk about what had happened. These bridges to her memory were very important and made conversations possible that would not have been possible or would not have happened before.''}.
    The therapeutic impact was not limited to the sessions. Some patients reported symptom improvements within days (Lisa after four days, Andrea after five days). Months later, Sabrina even sent a card reporting that she was coping with daily life again and had started a new occupation. Importantly, these outcomes were achieved with modest resources. Many of the most effective triggers relied on existing assets that were adapted with small changes to materials or shapes. For example, the dollhouse (Andrea), the bed with the red blanket (Lisa), and the chest (Natalie) in a neutral environment were all built from free online assets and could be implemented in 30 minutes to three hours. This shows that meaningful therapeutic exposure can be achieved without extensive technical complexity or cost.

\subsection{Successful triggering does not correlate with presence}

Contrary to our initial expectations, therapeutic effectiveness in VR exposure did not correlate with patients' sense of presence in the virtual environment. The developer initially hypothesized that successful triggering would increase presence levels, reasoning that emotionally demanding situations would consume attentional resources and reduce patients' ability to recognize the artificial nature of the VR environment. However, this assumption proved incorrect.

Patients consistently reported low presence scores  ($M=0.09$) and maintained awareness that they were experiencing a simulation throughout their exposure sessions (see~Table \ref{tab:presence}). Paradoxically, patients often reported lower presence scores during sessions that produced the strongest physiological reactions. For example, Robert experienced the most intense physical responses during sessions 1 and 3, with SC levels up to six times higher than baseline, yet remained cognitively aware of the virtual nature of the experience. Similarly, Peter and Marc repeatedly acknowledged they knew the scenarios were simulated, even while experiencing significant emotional responses.

This disconnect between presence and efficacy was perhaps most clearly demonstrated in Peter's case. Despite acknowledging that the VR setup did not appear immersive enough, he experienced his strongest triggers during sessions 3-4 when confronted with images of Afghan people in a neutral scenario. As the military therapist noted: \textit{``For Peter, the [...] set-up did not appear to be immersive enough. However, when confronted with images of people that resembled people during missions, he was instantly triggered. Step by step, he managed to look at these people and eventually even managed to apologize, which was integrated into an imagery rescripting session afterwards, that finally released the guilt, he had felt for many years''}. Notably, Peter actively used his awareness of the simulation to help regulate his emotional responses, treating the artificial nature as a coping resource rather than a therapeutic limitation.

The therapists viewed this maintained connection to reality as therapeutically beneficial. As the military therapist observed: \textit{``It was a very important experience that the patient was not completely absorbed by the situation in VR but maintained the connection to the outside world and the triggers in combination with the classic therapeutic work on memories worked very well''}. This accessibility allowed for real-time therapeutic intervention and dialogue during exposure, which would be compromised if patients became fully absorbed in the virtual environment. The military therapist further noted that for military trauma patients specifically, \textit{``the feeling of being in control seemed expedient''}, as trauma recovery often involves regaining personal agency and control.

The developer reflected on this unexpected finding, noting patients' ability to simultaneously acknowledge simulation while experiencing genuine emotional impact: \textit{``We heard statements like 'I know this is not real, and the person ahead of me probably looked at me because of an implementation flaw, but nevertheless, it really emotionally affects me, and I struggle to control my feelings right now.''}. This suggests that trauma triggers may operate at a more fundamental sensory or emotional level that bypasses cognitive recognition of artificiality, making high-fidelity immersion unnecessary for therapeutic effectiveness.

\section{{Discussion}}
Our feasibility study revealed that VRET for C-PTSD holds significant therapeutic promise but may require different implementation approaches than those used for single-incident trauma. While some patients achieved breakthroughs that surpassed years of prior conventional therapy, our findings also highlighted critical methodological and ethical challenges that the existing literature has not adequately addressed. The highly individualized nature of C-PTSD triggers, combined with the necessity of involving technical developers in therapeutic processes, created a complex ecosystem where traditional research protocols proved insufficient. In the sections that follow, we reflect on the key lessons we learned and discuss their implications for advancing VRET research with vulnerable populations.

\subsection{Who is in the lead?}
Leadership in collaborative VRET development was a negotiation rather than a fixed hierarchy, marked by tensions between clinical authority and technical expertise. Therapists typically led, drawing on their training to elicit unclear memories and safeguard patient welfare. Yet the technical demands of C-PTSD scenarios required shared authority, which pushed against traditional therapeutic boundaries.

The developer’s expertise often shifted dynamics by introducing technical options that therapists and patients had assumed impossible. The result was a blurring of clinical and technical discourse: patients justified design needs through emotional accounts, and the developer felt compelled to respond. Without clinical training, the developer was unprepared to manage such disclosures, leading to communication breakdowns that threatened both therapeutic integrity and researcher well-being. These challenges extend beyond trauma-focused research to HCI work in sensitive contexts. While HCI emphasizes role clarity and emotional preparation~\cite{houben_hci_2024}, and healthcare literature stresses interprofessional collaboration~\cite{huang2014samhsa, ball_integrating_2021}, existing frameworks assume all team members share a clinical background. That assumption collapses once developers enter therapeutic settings.

Patient control introduced further complexity. Military trauma patients often reported greater agency when directly involved in design, aligning with trauma recovery literature that identifies perceived self-efficacy as central to healing~\cite{Bandura1977-fc,herman_trauma_2015}. Yet empowerment carried risk: clinical judgment was still needed to prevent retraumatization, a tension not present in most co-design processes~\cite{donetto_experience-based_2015}. We also observed delayed effects, where little reaction in-session was followed by strong aftereffects days later. The temporal gap challenges participatory design practices that rely on immediate feedback to guide iteration.

Taken together, the findings suggest that effective collaboration in trauma-focused VR requires structured protocols that enable role shifting while maintaining therapeutic relationships. Pre-session planning can help anticipate emotional content, explicit communication strategies can clarify role transitions, and post-session debriefings should integrate both clinical and technical perspectives~\cite{clemensen_participatory_2007}. New assessment tools are also needed to track delayed trauma responses, ensuring that design decisions are informed by impacts beyond the immediate session. Developing such practices is essential for advancing collaborative VRET safely and effectively, while protecting patients, therapists, and technical staff alike.

\subsection{Less is more}

Our study revealed a counterintuitive finding that challenges common assumptions about VRET design. Unlike mono-trauma, where triggers often cluster around identifiable events, C-PTSD triggers were highly individualized. The question was not whether VR exposure works but how much individualization is required and whether it can be achieved in practice. Across design variants, including simplified scenarios for patients with lower resilience, single patient-specific objects often elicited stronger memory recall than elaborate environments. As childhood trauma therapist noted, \textit{``[...] we have learned that complex scenarios are not necessary, but that individualization is what counts.''}. The military therapist similarly observed, \textit{``[...] simple triggers, like a picture and a candle to trigger guilt, or a simple white room with only a handful of moving people to begin with, may be more beneficial in the treatment of C-PTSD.''}. Using free web assets, we produced prototypes within an hour and at no cost. In one striking case, a corset in a white room prompted more therapeutic progress than years of prior psychotherapy.

This aligns with findings by~\citet{Sherrill2019-io,Sherrill2020-mn,Sherrill2025-fo}. The authors recommend reducing complexity to a necessary minimum to prevent avoidance and enable focused confrontation. In HCI, simplicity is also a common design strategy to guide attention. Research on healthcare interfaces shows that reducing complexity lowers cognitive load and improves usability for vulnerable groups~\cite{alnanih_mapping_2016, edwards_user_2018, kushniruk_cognitive_2004}.

We also found that the therapeutic effect was decoupled from presence. Contrary to VR literature linking higher immersion to better outcomes~\cite{Sherrill2025-fo,slater_immersion_2018}. Although~\citet{Sherrill2025-fo} agree that it does not require complexity for effective exposure, they assume the necessity presence. Our findings revealed that stronger flashbacks did not align with higher presence scores. Ratings were generally low and often lowest in the most effective sessions, likely because active therapeutic dialogue kept patients engaged with the therapist. Patients remained reachable and responsive, which therapists identified as crucial for intervention. Some requested more immersive settings, often remarking \textit{``I always knew it was a simulation,''} yet high immersion was not necessary. In some cases, post-session hyperarousal followed muted in-session reactions, highlighting the temporal complexity of trauma responses.

The findings reflect deeper principles about trauma processing at the sensory rather than propositional level~\cite{md_body_2015}. Memory retrieval appeared to emerge from sensory activation rather than conscious recognition, echoing trauma literature on how memories are encoded. For design, the implication is to prioritize individualized and low-complexity stimuli that activate the sensory core of trauma, rather than investing in high-fidelity immersive worlds. In trauma-focused VR, less can be more: simple, targeted content may deliver greater therapeutic value than sophisticated but generalized environments.

\subsection{The design process is also therapy}
A key lesson from the study was that the act of designing exposure scenarios carried therapeutic value in its own right. Translating autobiographical memories into design requirements required patients to recall, reframe, and externalize their experiences. This created additional, graded confrontations with trauma material before entering VR. Several patients reported appreciating the opportunity to shape their own scenarios, and therapists noted that the use of exemplar images or simple sketches often helped with memory retrieval or clarifying vague accounts. Similar observations in participatory health design show that co-creation can itself support reflection and agency~\cite{donetto_experience-based_2015}. Our experience suggests that the same applies here. Design became more than preparation for therapy; it acted as a step within therapy, that offers a more seamless introduction into emotional processing and inhibitory learning~\cite{CRASKE201410,foaEmotionalProcessingFear1986}. This can be particularly valuable to overcome the high levels of avoidance that often characterize C-PTSD.

Because the study was exploratory, the process was intentionally loose, but it showed how design activities can be integrated into the therapeutic pathway. Scenario creation functioned as an intermediate stage between psychoeducation and VR exposure, offering patients a lower-stakes way of approaching traumatic memories while producing concrete artifacts for later use. The process echoed narrative approaches where constructing accounts of trauma supports integration~\cite{jongedijk_narrative_2014}, suggesting that design requirements may serve a similar purpose by translating memories into visual form. Looking ahead, future work could explore more structured ways of embedding design into therapy. Generative AI might provide rapid sketches or previews that help patients communicate vague or fragmented memories in situ \cite{ye_generative_2024, pataranutaporn_living_2023} The childhood trauma therapist reflected, \textit{``Even if it is enough that individualized triggers are visualized so that the patients can experience that they can deal with their negative life experiences, it would be desirable to find a framework that supports this even further. For example, an AI tool that visualizes an environment described by the patients in real time and a monitoring system that records the physiology of the stress experience in order to obtain feedback.''}. Such suggestions point to the potential of generative AI and monitoring systems to support real-time visualization, reduce communication gaps between patients, therapists, and developers, and tailor scenarios dynamically to patient resilience~\cite{degenhard2025rememberchallengesrisks}.

\subsection{Keeping stakeholders safe}
Another key lesson learned from our study is that safety in VRET for C-PTSD must extend beyond patients to all stakeholders. The exploratory nature of the work, coupled with the absence of established protocols, meant that unintentional triggers, prolonged hyperarousal, and emotional strain were experienced not only by patients but also by the developer. Safeguarding is therefore not just a clinical priority but also an HCI concern, requiring strategies that protect everyone involved.

For patients, therapist presence in every session proved essential. Unexpected flashbacks and boundary issues arose that a developer could not manage alone, and continuity with trusted therapists offered reassurance. Inpatient care was particularly important, enabling immediate support when physical or emotional reactions became overwhelming~\cite{herman1987recovery}. Clear communication also mattered: setting expectations that the study was exploratory, and that glitches or unanticipated effects might occur, helped patients accept deviations. Safety further depended on gradual exposure and physiological monitoring, since self-reports sometimes diverged from biofeedback. Contextual factors extended beyond the software. Some patients reacted to the default VR lobby, while others associated the headset with military night-vision gear. Discussing equipment and environments in advance, and retaining flexibility to adjust protocols, was critical. Together, these measures echo trauma therapy principles of controlled dosing~\cite{fordTraumaMemoryProcessing2018}, but adapted to the material and technological realities of VR.

Equally important was safeguarding the developer, whose well-being was initially underestimated. Research on secondary traumatization shows that indirect exposure to trauma can result in secondary traumatic stress (PTSD-like symptoms), vicarious traumatization (enduring shifts in beliefs about safety, trust, and control), and compassion fatigue (erosion of empathic capacity)\cite{figley1995compassion}. These risks are heightened when supervision and structured support are lacking \cite{jenkins2002secondary, bride2007prevalence, stamm2010concise}. They are not confined to clinicians: interpreters and other collaborators repeatedly exposed to trauma material also report stress and emotional strain~\cite{bontempo2012ounce}. In our study, patients sometimes shared emotional stories with the developer to explain design requirements. The developer reported that the processing of such stories was challenging. Support measures typically offered to novice clinicians, such as supervision and structured processing, could also be extended to developers and researchers.

The broader implication for HCI is that researchers are not detached observers but participants in sensitive contexts. While guidelines exist for ethnographers and qualitative researchers working with vulnerable groups~\cite{moncur_emotional_2013, feuston_researcher_2022, chancellor_sensitive_2019}, comparable guidance is lacking in technology-driven deployments where attention is on implementation and evaluation. This gap is concerning given the growing evidence of high stress, anxiety, and burnout among researchers, compounded by structural pressures in academia~\cite{torrisi_academic_2013, satinsky_systematic_2021, campbell_factors_2022}. Within HCI in particular, empathy and immersion are often central to research~\cite{wright_empathy_2008, altarriba_bertran_chasing_2019, balaam_emotion_2019}, yet the same care extended to participants is rarely directed toward researchers~\cite{wolters_emotional_2017}. Our findings suggest that safeguarding should be framed as a shared responsibility, with preparation, role clarity, and structured debriefing built into study protocols from the outset.

Although this project began as a feasibility trial, its most important insights came from the lived experiences of patients, therapists, and the developer. These revealed how easily role boundaries blur, how emotional burden can surface, and how unprepared researchers may be for the emotional intensity of therapeutic contexts. Future work should build on these lessons by treating researcher vulnerability as integral to ethical VRET research. Safety in VRET is relational and multi-directional. Protecting patients, therapists, developers, and researchers is essential for advancing the field.

\subsection{Keeping the balance}
Our study was among the first to explore the feasibility of VR exposure for individuals with C-PTSD. While it confirmed that VR can be effective and offered insight into how exposure scenarios can be designed, the study also showed how challenging it is to balance safety, flexibility, and therapeutic value. Therapists and the developer initially assumed that individualization would apply mainly to scenario content. In practice, it extended to the setting, communication, process, and design. No single study design emerged as ``most effective''. Instead, outcomes depended on continuous adjustment and balancing across competing demands. The work highlighted several balances that need to be negotiated rather than resolved:

\begin{itemize}
    \item \textbf{Ethics and safety vs. effectiveness.} Exposure therapy necessarily involves distressing recall, yet excessive caution may blunt therapeutic gains. In our study, some intense traumatic memories were followed by long-term improvement, echoing trauma research that emphasizes titrated exposure as central to efficacy. Case-by-case risk–benefit reviews, clear stop rules, and staged titration under therapist supervision remain essential.
    \item \textbf{Interactive vs. top-down design.} Involving patients directly in design helped refine scenarios and grading, aligning with participatory health design literature that values co-creation for accuracy and ownership~\cite{donetto_experience-based_2015}. However, for patients with severe trauma, participatory involvement risked accidental triggers and emotional intensity for the developer. In such contexts, therapist-led (top-down) design proved safer.
    \item \textbf{Implementation effort vs. benefit.} Starting simple and scaling effort only after identifying effective elements was both cost-efficient and therapeutically sound. High-fidelity detail or complexity did not consistently increase effect, aligning with HCI findings that simplicity can reduce cognitive load and support vulnerable users~\cite{alnanih_mapping_2016, edwards_user_2018, kushniruk_cognitive_2004}. Increasing interactivity selectively, only where scenario complexity demanded it, helped avoid wasted effort and unnecessary exposure.
    \item \textbf{Professionalism vs. empathy of the developer.} The developer’s presence required patients to disclose sensitive information to someone outside the trusted therapeutic dyad. Some benefited from the empathy shown, which fostered trust, but excessive openness risked oversharing and emotional strain. This tension echoes wider concerns in HCI about balancing rapport with role clarity in sensitive research~\cite{moncur_emotional_2013, feuston_researcher_2022}.
\end{itemize}

\noindent Taken together, these balances highlight that designing and deploying VRET for C-PTSD is not about prescribing a single ``best'' protocol but about navigating competing demands with flexibility, transparency, and safeguards.
Ethical research protocols provide important protection, but their rigidity can also limit opportunities to personalize processes and interventions. For VRET to be both safe and effective, ethics frameworks may need to evolve to accommodate flexible, case-sensitive adaptations while still maintaining accountability and oversight.

High levels of customization traditionally demand substantial resources, limiting the scalability of VRET. Our small-sample study made such customization feasible, but the results highlight constraints for broader deployment. Emerging generative AI systems, such as Stable Diffusion\footnote{\url{https://stablediffusionweb.com/de}
, accessed: 03.12.25}, can already produce diverse media and enable in situ creation of simple therapeutic scenarios similar to those used in our study. Over time, GAI could reduce reliance on developers and support collaborative exposure design across the therapy process.

\subsection{Implications for exposure design}
Despite the small sample, several refinements to conventional PTSD exposure design appear promising for VRET in C-PTSD.

\begin{itemize}
    \item \textbf{Individualization of scenario and process.} Depending on the patient's resilience and willingness to get engaged with the process, they should be involved in designing scenarios and processes. The design should consider the entire ecosystem of exposure (e.g. VR lobby design) in order to prevent unexpected exposure courses. Flexibility in both process and content - e.g. through dynamically adjustable assets - is likely more valuable than high-fidelity immersion.
    \item \textbf{The design process as part of the therapy.} The design process provides a gentle introduction to exposure, which can lower avoidance and promote self-efficacy. It should be considered as part of VRET design for C-PTSD. Exemplary visualizations (e.g. online media or prototypes) can lower communication barriers and promote effective therapeutic work. 
    \item \textbf{Less is more.} Even minimal stimuli can evoke memory and physiological responses. These should be combined with additional grading techniques such as distance or sensory variation. Grading should support a wide spectrum of complexity, beginning with minimal or even zero complexity when expectations alone prompt reactions.
    \item \textbf{Finding the right content.} Identifying trauma-relevant stimuli is often harder than designing them. To enable rapid iteration, implementation effort should stay low by using existing assets or GAI. Deep customization should be reserved for cases where simpler options fail.
\end{itemize}

\subsection{Reflecting on the methodology}
This study relied on an unusual methodological setup. Much of the empirical material came from those directly delivering the intervention (the developer and the two therapists) who acted at once as practitioners, participants, and authors. Their field notes and reflective reports offered rare access to the day-to-day realities of VRET, surfacing details that would be hard to capture otherwise. Yet this dual role inevitably shaped what was recorded and emphasized. To counterbalance this, a second author joined without prior involvement in therapy or technical delivery. This external perspective challenged insider assumptions and introduced interpretive distance. The analytic process was collaborative. The first author contributed contextual knowledge from direct involvement, while the second author conducted inductive coding and posed critical questions. Weekly discussions turned these differing readings into shared insights, documented through analytic memos. The result resembled but went beyond member-checking \cite{mckim2023-dy, birt_member_2016, mcdonald_reliability_2019}. The external author did not simply validate interpretations but actively reshaped them.

Two methodological choices were central. First, we followed reflexive thematic analysis \cite{braun_reflecting_2019,braun_one_2021}, which treats divergence between researchers as a resource rather than a threat to reliability. The friction between insider and outsider perspectives exposed blind spots and deepened interpretation. Second, we used the critical incident technique \cite{viergever_critical_2019,ppali_creating_2025} to anchor analysis in moments that disrupted routines, such as accidental triggers in default VR lobbies or uncertainty over how to respond when patients shared highly emotional disclosures. The incidents connected everyday challenges to broader methodological and ethical questions.

Time added another layer. During sessions, the developer’s role was to take notes of what happened; reflections on emotional impact often came later, sometimes reinterpreting earlier events. This shift shows how positionality is not fixed but moves between immediate observer, participant, and reflective narrator. Some experiences were captured most vividly in the moment, while others only surfaced after time had passed. As~\citet{singh_exploring_2025} argue, positionality in HCI is not a static declaration but a moving lens that shapes both data and analysis. Researchers are not only observers of sensitive encounters but may become entangled in them, experiencing emotional burden or shifts in their own self-understanding \cite{ppali_cite_2025}.
In this sense, journaling becomes more than documentation, it is a method for tracing positionality itself. Capturing impressions during sessions alongside delayed reflections makes visible how perspectives evolve and how knowledge is co-constructed. In our case, this revealed that positionality shifted not only between insider and outsider authors but also within the same individual over time.
We do not present this as a prescriptive model. Instead, we aim to share how combining insider accounts, external critique, and multi-layered reflection allowed us to engage with the methodological and ethical complexity of VRET. Making these tensions visible matters. It shows not only what happened in therapy (or any research context in fact), but also what it means to research it, both from within and alongside.

\section{{Limitations}}
Our work was exploratory based on a small sample of eleven inpatients and with no control group. The clinical picture of C-PTSD is highly diverse, and the course of therapy differed for every patient, which limits comparability and generalizability. Our sample was also restricted to two trauma types (childhood and military trauma) and even within these categories, the results varied. Cross-trauma implications should therefore be avoided. In addition, the two therapists applied different clinical approaches they had each found effective, meaning that findings may not transfer to other therapeutic methods.
Because patients were highly vulnerable, we did not conduct systematic post-session interviews. Their voices were captured indirectly through therapist observations, session notes, and developer reflections. The reliance on insider accounts brings potential bias, even though external collaborators provided critique and interpretive distance.
The absence of established protocols for VRET in C-PTSD also meant that many procedures were improvised, leading to ad hoc adaptations across cases. Finally, our focus was on feasibility and process rather than clinical outcomes, so no claims about efficacy can be made.
The motivation of this study is to share what was learned in practice rather than to provide guidelines. More structured, comparative work is needed to establish safe and effective frameworks for VRET with C-PTSD.

\section{{Conclusion}}

This feasibility study examined whether and how VRET for C-PTSD can be embedded into routine graded trauma exposure, while attending to the needs of patients, therapists, and developers. Building on guidelines for VRET for PTSD, we explored how scenario and process design should differ in terms of complexity, customizability, and adjustability to ensure a safe space that promotes engagement in the therapeutic process. Across eleven cases, simple object-based triggers often proved more effective than elaborate environments, presence was not necessary for therapeutic progress, and collaborative design itself became part of therapy. At the same time, the work exposed major risks: accidental triggers, delayed after-effects, and intense emotional impact of the developer. Involving a developer brought technical advantages but also demanded new role definitions, emotional safeguards, and structured debriefs.
Our findings point to three priorities for advancing this field: (1) designing simple and patient-specific scenarios rather than generalized high-fidelity ones, (2) formalizing roles and communication strategies when non-clinicians join therapy, and (3) building safety protocols that account for both patients and collaborators. Overall, the study supports a cautiously optimistic view: VRET for C-PTSD is feasible with simple but individualized scenarios. However, due to the complexity of this topic and the associated risks for patient and other stakeholders, extensive but ethically robust processes that protect all stakeholders while enabling individualized therapy are needed.

\begin{acks}
We thank all participants for taking part in our study and for their valuable reflections that substantially shaped this work. This work was also supported by the European Union’s Research and Innovation Programme under Grant Agreement No. 739578 and the Government of the Republic of Cyprus through the Deputy Ministry of Research, Innovation and Digital Policy. Views and opinions expressed are, however, those of the author(s) only and do not necessarily reflect those of the European Union. Neither the European Union nor the granting authority can be held responsible for them.
\end{acks}

\bibliographystyle{ACM-Reference-Format}

\begin{thebibliography}{133}


\ifx \showCODEN    \undefined \def \showCODEN     #1{\unskip}     \fi
\ifx \showISBNx    \undefined \def \showISBNx     #1{\unskip}     \fi
\ifx \showISBNxiii \undefined \def \showISBNxiii  #1{\unskip}     \fi
\ifx \showISSN     \undefined \def \showISSN      #1{\unskip}     \fi
\ifx \showLCCN     \undefined \def \showLCCN      #1{\unskip}     \fi
\ifx \shownote     \undefined \def \shownote      #1{#1}          \fi
\ifx \showarticletitle \undefined \def \showarticletitle #1{#1}   \fi
\ifx \showURL      \undefined \def \showURL       {\relax}        \fi
\providecommand\bibfield[2]{#2}
\providecommand\bibinfo[2]{#2}
\providecommand\natexlab[1]{#1}
\providecommand\showeprint[2][]{arXiv:#2}

\bibitem[ICH(1997)]%
        {ICHE6R21997}
 \bibinfo{year}{1997}\natexlab{}.
\newblock \bibinfo{title}{{{ICH E6}} ({{R2}}) {{Good}} Clinical Practice -
  {{Scientific}} Guideline {\textbar} {{European Medicines Agency}} ({{EMA}})}.
\newblock
  \bibinfo{howpublished}{https://www.ema.europa.eu/en/ich-e6-r2-good-clinical-practice-scientific-guideline}.
\newblock


\bibitem[Dia(2013)]%
        {DiagnosticStatisticalManual2013}
 \bibinfo{year}{2013}\natexlab{}.
\newblock \bibinfo{booktitle}{\emph{Diagnostic and Statistical Manual of Mental
  Disorders: {{DSM-5}}â„¢, 5th Ed.}}
\newblock \bibinfo{publisher}{American Psychiatric Publishing, Inc.},
  \bibinfo{address}{Arlington, VA, US}. xliv, 947 pages.
\newblock
\showISBNx{978-0-89042-554-1 (Hardcover); 978-0-89042-555-8 (Paperback)}
\href{https://doi.org/10.1176/appi.books.9780890425596}{doi:\nolinkurl{10.1176/appi.books.9780890425596}}


\bibitem[Alnanih and Ormandjieva(2016)]%
        {alnanih_mapping_2016}
\bibfield{author}{\bibinfo{person}{Reem Alnanih} {and} \bibinfo{person}{Olga
  Ormandjieva}.} \bibinfo{year}{2016}\natexlab{}.
\newblock \showarticletitle{Mapping {HCI} {Principles} to {Design} {Quality} of
  {Mobile} {User} {Interfaces} in {Healthcare} {Applications}}.
\newblock \bibinfo{journal}{\emph{Procedia Computer Science}}
  \bibinfo{volume}{94} (\bibinfo{date}{Jan.} \bibinfo{year}{2016}),
  \bibinfo{pages}{75--82}.
\newblock
\showISSN{1877-0509}
\href{https://doi.org/10.1016/j.procs.2016.08.014}{doi:\nolinkurl{10.1016/j.procs.2016.08.014}}


\bibitem[Altarriba~Bertran et~al\mbox{.}(2019)]%
        {altarriba_bertran_chasing_2019}
\bibfield{author}{\bibinfo{person}{Ferran Altarriba~Bertran},
  \bibinfo{person}{Elena MÃ¡rquez~Segura}, \bibinfo{person}{Jared Duval}, {and}
  \bibinfo{person}{Katherine Isbister}.} \bibinfo{year}{2019}\natexlab{}.
\newblock \showarticletitle{Chasing {Play} {Potentials}: {Towards} an
  {Increasingly} {Situated} and {Emergent} {Approach} to {Everyday} {Play}
  {Design}}. In \bibinfo{booktitle}{\emph{Proceedings of the 2019 on
  {Designing} {Interactive} {Systems} {Conference}}}. \bibinfo{publisher}{ACM},
  \bibinfo{address}{San Diego CA USA}, \bibinfo{pages}{1265--1277}.
\newblock
\showISBNx{978-1-4503-5850-7}
\href{https://doi.org/10.1145/3322276.3322325}{doi:\nolinkurl{10.1145/3322276.3322325}}


\bibitem[Anderson et~al\mbox{.}(2003)]%
        {andersonVirtualRealityExposure2003}
\bibfield{author}{\bibinfo{person}{Page Anderson}, \bibinfo{person}{Barbara~O.
  Rothbaum}, {and} \bibinfo{person}{Larry~F. Hodges}.}
  \bibinfo{year}{2003}\natexlab{}.
\newblock \showarticletitle{VR Exposure in the Treatment of Social
  Anxiety}.
\newblock \bibinfo{journal}{\emph{Cognitive and Behavioral Practice}}
  \bibinfo{volume}{10}, \bibinfo{number}{3} (\bibinfo{date}{June}
  \bibinfo{year}{2003}), \bibinfo{pages}{240--247}.
\newblock
\showISSN{1077-7229}
\href{https://doi.org/10.1016/S1077-7229(03)80036-6}{doi:\nolinkurl{10.1016/S1077-7229(03)80036-6}}


\bibitem[Balaam et~al\mbox{.}(2019)]%
        {balaam_emotion_2019}
\bibfield{author}{\bibinfo{person}{Madeline Balaam}, \bibinfo{person}{Rob
  Comber}, \bibinfo{person}{Rachel~E. Clarke}, \bibinfo{person}{Charles
  Windlin}, \bibinfo{person}{Anna StÃ¥hl}, \bibinfo{person}{Kristina HÃ¶Ã¶k},
  {and} \bibinfo{person}{Geraldine Fitzpatrick}.}
  \bibinfo{year}{2019}\natexlab{}.
\newblock \showarticletitle{Emotion {Work} in {Experience}-{Centered}
  {Design}}. In \bibinfo{booktitle}{\emph{Proceedings of the 2019 {CHI}
  {Conference} on {Human} {Factors} in {Computing} {Systems}}}
  \emph{(\bibinfo{series}{{CHI} '19})}. \bibinfo{publisher}{Association for
  Computing Machinery}, \bibinfo{address}{New York, NY, USA},
  \bibinfo{pages}{1--12}.
\newblock
\showISBNx{978-1-4503-5970-2}
\href{https://doi.org/10.1145/3290605.3300832}{doi:\nolinkurl{10.1145/3290605.3300832}}


\bibitem[Ball et~al\mbox{.}(2021)]%
        {ball_integrating_2021}
\bibfield{author}{\bibinfo{person}{Annahita Ball},
  \bibinfo{person}{Elizabeth~A. Bowen}, {and}
  \bibinfo{person}{Annette~Semanchin Jones}.} \bibinfo{year}{2021}\natexlab{}.
\newblock \showarticletitle{Integrating {Trauma}-{Informed} {Care} and
  {Collective} {Impact}: {Perspectives} of {Service} {Providers} {Working} with
  {Cross}-{System} {Youth}}.
\newblock \bibinfo{journal}{\emph{Journal of the Society for Social Work and
  Research}} \bibinfo{volume}{12}, \bibinfo{number}{1} (\bibinfo{date}{March}
  \bibinfo{year}{2021}), \bibinfo{pages}{59--81}.
\newblock
\showISSN{2334-2315}
\href{https://doi.org/10.1086/712960}{doi:\nolinkurl{10.1086/712960}}
\newblock
\shownote{Publisher: The University of Chicago Press}.


\bibitem[Bandura(1977)]%
        {Bandura1977-fc}
\bibfield{author}{\bibinfo{person}{Albert Bandura}.}
  \bibinfo{year}{1977}\natexlab{}.
\newblock \showarticletitle{Self-efficacy: Toward a unifying theory of
  behavioral change}.
\newblock \bibinfo{journal}{\emph{Psychol. Rev.}} \bibinfo{volume}{84},
  \bibinfo{number}{2} (\bibinfo{year}{1977}), \bibinfo{pages}{191--215}.
\newblock


\bibitem[Bandura(1989)]%
        {bandura1989regulation}
\bibfield{author}{\bibinfo{person}{Albert Bandura}.}
  \bibinfo{year}{1989}\natexlab{}.
\newblock \showarticletitle{Regulation of cognitive processes through perceived
  self-efficacy.}
\newblock \bibinfo{journal}{\emph{Developmental psychology}}
  \bibinfo{volume}{25}, \bibinfo{number}{5} (\bibinfo{year}{1989}),
  \bibinfo{pages}{729}.
\newblock


\bibitem[Ba{\~n}os et~al\mbox{.}(1999)]%
        {banosPsychologicalVariablesReality1999}
\bibfield{author}{\bibinfo{person}{Rosa Ba{\~n}os}, \bibinfo{person}{Cristina
  Botella}, \bibinfo{person}{Azucena {Garcia-Palacios}},
  \bibinfo{person}{Helena Villa}, \bibinfo{person}{Conxa Perpi{\~n}{\'a}},
  {and} \bibinfo{person}{M Gallardo}.} \bibinfo{year}{1999}\natexlab{}.
\newblock \showarticletitle{Psychological {{Variables}} and {{Reality
  Judgment}} in {{Virtual Environments}}: {{The Roles}} of {{Absorption}} and
  {{Dissociation}}}.
\newblock \bibinfo{journal}{\emph{Cyberpsychology \& behavior : the impact of
  the Internet, multimedia and virtual reality on behavior and society}}
  \bibinfo{volume}{2} (\bibinfo{date}{Feb.} \bibinfo{year}{1999}),
  \bibinfo{pages}{143--8}.
\newblock
\href{https://doi.org/10.1089/cpb.1999.2.143}{doi:\nolinkurl{10.1089/cpb.1999.2.143}}


\bibitem[Beidel et~al\mbox{.}(2017)]%
        {Beidel2017Trauma}
\bibfield{author}{\bibinfo{person}{D. Beidel}, \bibinfo{person}{B. Frueh},
  \bibinfo{person}{Sandra~M. Neer}, \bibinfo{person}{Clint~A. Bowers},
  \bibinfo{person}{Benjamin~J Trachik}, \bibinfo{person}{T.~W. Uhde}, {and}
  \bibinfo{person}{Anouk~L. Grubaugh}.} \bibinfo{year}{2017}\natexlab{}.
\newblock \showarticletitle{Trauma management therapy with virtual-reality
  augmented exposure therapy for combat-related PTSD: A randomized controlled
  trial.}
\newblock \bibinfo{journal}{\emph{Journal of anxiety disorders}}
  \bibinfo{volume}{61} (\bibinfo{year}{2017}), \bibinfo{pages}{64--74}.
\newblock
\href{https://doi.org/10.1016/j.janxdis.2017.08.005}{doi:\nolinkurl{10.1016/j.janxdis.2017.08.005}}


\bibitem[Benbow and Anderson(2019)]%
        {benbowMetaanalyticExaminationAttrition2019}
\bibfield{author}{\bibinfo{person}{Amanda~A. Benbow} {and}
  \bibinfo{person}{Page~L. Anderson}.} \bibinfo{year}{2019}\natexlab{}.
\newblock \showarticletitle{A Meta-Analytic Examination of Attrition in Virtual
  Reality Exposure Therapy for Anxiety Disorders}.
\newblock \bibinfo{journal}{\emph{Journal of Anxiety Disorders}}
  \bibinfo{volume}{61} (\bibinfo{date}{Jan.} \bibinfo{year}{2019}),
  \bibinfo{pages}{18--26}.
\newblock
\showISSN{0887-6185}
\href{https://doi.org/10.1016/j.janxdis.2018.06.006}{doi:\nolinkurl{10.1016/j.janxdis.2018.06.006}}


\bibitem[Benight and Bandura(2004)]%
        {BENIGHT20041129}
\bibfield{author}{\bibinfo{person}{Charles~C. Benight} {and}
  \bibinfo{person}{Albert Bandura}.} \bibinfo{year}{2004}\natexlab{}.
\newblock \showarticletitle{Social Cognitive Theory of Posttraumatic Recovery:
  The Role of Perceived Self-Efficacy}.
\newblock \bibinfo{journal}{\emph{Behaviour Research and Therapy}}
  \bibinfo{volume}{42}, \bibinfo{number}{10} (\bibinfo{year}{2004}),
  \bibinfo{pages}{1129--1148}.
\newblock
\showISSN{0005-7967}
\href{https://doi.org/10.1016/j.brat.2003.08.008}{doi:\nolinkurl{10.1016/j.brat.2003.08.008}}


\bibitem[Berntsen and Rubin(2007)]%
        {berntsen2007trauma}
\bibfield{author}{\bibinfo{person}{Dorthe Berntsen} {and}
  \bibinfo{person}{David~C Rubin}.} \bibinfo{year}{2007}\natexlab{}.
\newblock \showarticletitle{When a trauma becomes a key to identity: Enhanced
  integration of trauma memories predicts posttraumatic stress disorder
  symptoms}.
\newblock \bibinfo{journal}{\emph{Applied cognitive psychology}}
  \bibinfo{volume}{21}, \bibinfo{number}{4} (\bibinfo{year}{2007}),
  \bibinfo{pages}{417--431}.
\newblock


\bibitem[Billings and Nicholls(2025)]%
        {Billings.2025}
\bibfield{author}{\bibinfo{person}{Jo Billings} {and} \bibinfo{person}{Helen
  Nicholls}.} \bibinfo{year}{2025}\natexlab{}.
\newblock \showarticletitle{PTSD and complex PTSD, current treatments and
  debates: a review of reviews}.
\newblock \bibinfo{journal}{\emph{British Medical Bulletin}}
  \bibinfo{volume}{156}, \bibinfo{number}{1} (\bibinfo{year}{2025}),
  \bibinfo{pages}{ldaf015}.
\newblock
\showISSN{0007-1420}
\href{https://doi.org/10.1093/bmb/ldaf015}{doi:\nolinkurl{10.1093/bmb/ldaf015}}


\bibitem[Birt et~al\mbox{.}(2016)]%
        {birt_member_2016}
\bibfield{author}{\bibinfo{person}{Linda Birt}, \bibinfo{person}{Suzanne
  Scott}, \bibinfo{person}{Debbie Cavers}, \bibinfo{person}{Christine
  Campbell}, {and} \bibinfo{person}{Fiona Walter}.}
  \bibinfo{year}{2016}\natexlab{}.
\newblock \showarticletitle{Member {Checking}: {A} {Tool} to {Enhance}
  {Trustworthiness} or {Merely} a {Nod} to {Validation}?}
\newblock \bibinfo{journal}{\emph{Qualitative Health Research}}
  \bibinfo{volume}{26}, \bibinfo{number}{13} (\bibinfo{date}{Nov.}
  \bibinfo{year}{2016}), \bibinfo{pages}{1802--1811}.
\newblock
\showISSN{1049-7323}
\href{https://doi.org/10.1177/1049732316654870}{doi:\nolinkurl{10.1177/1049732316654870}}


\bibitem[Bohus et~al\mbox{.}(2001)]%
        {bohusEntwicklungBorderlineSymptomListe2001}
\bibfield{author}{\bibinfo{person}{Martin Bohus}, \bibinfo{person}{Matthias~F.
  Limberger}, \bibinfo{person}{Ingrid Frank, Ulrike;~Sender},
  \bibinfo{person}{Tanja Gratwohl}, {and} \bibinfo{person}{Rolf-Dieter
  Stieglitz}.} \bibinfo{year}{2001}\natexlab{}.
\newblock \showarticletitle{{Entwicklung der Borderline- Symptom-Liste}}.
\newblock \bibinfo{journal}{\emph{Psychother Psychosom Med Psychol}}
  \bibinfo{volume}{51}, \bibinfo{number}{05} (\bibinfo{date}{Dec.}
  \bibinfo{year}{2001}), \bibinfo{pages}{201--211}.
\newblock
\showISSN{0937-2032}
\href{https://doi.org/10.1055/s-2001-13281}{doi:\nolinkurl{10.1055/s-2001-13281}}


\bibitem[Bohus and Vonderlin(2024)]%
        {bohusDialektischBehavioraleTherapie2024}
\bibfield{author}{\bibinfo{person}{Martin Bohus} {and} \bibinfo{person}{Ruben
  Vonderlin}.} \bibinfo{year}{2024}\natexlab{}.
\newblock \showarticletitle{Dialektisch Behaviorale {{Therapie}} F{\"u}r
  Komplexe Posttraumatische {{Belastungsst{\"o}rung}} ({{DBT-PTBS}}):
  Ein~Evidenzbasiertes St{\"o}rungsspezfisches {{Behandlungsprogramm}}}.
\newblock \bibinfo{journal}{\emph{Der Nervenarzt}} \bibinfo{volume}{95},
  \bibinfo{number}{7} (\bibinfo{date}{July} \bibinfo{year}{2024}),
  \bibinfo{pages}{630--638}.
\newblock
\showISSN{1433-0407}
\href{https://doi.org/10.1007/s00115-024-01680-y}{doi:\nolinkurl{10.1007/s00115-024-01680-y}}


\bibitem[Bontempo and Malcolm(2012)]%
        {bontempo2012ounce}
\bibfield{author}{\bibinfo{person}{Karen Bontempo} {and} \bibinfo{person}{Karen
  Malcolm}.} \bibinfo{year}{2012}\natexlab{}.
\newblock \showarticletitle{An ounce of prevention is worth a pound of cure}.
\newblock \bibinfo{journal}{\emph{In Our Hands. Educating Healthcare
  Interpreters}} (\bibinfo{year}{2012}), \bibinfo{pages}{105--130}.
\newblock


\bibitem[Botella et~al\mbox{.}(1998)]%
        {botellaVirtualRealityTreatment1998}
\bibfield{author}{\bibinfo{person}{Cristina Botella},
  \bibinfo{person}{Rosa~Mar{\'i}a Ba{\~n}os}, \bibinfo{person}{Conxa
  Perpi{\~n}{\'a}}, \bibinfo{person}{Helena Villa}, \bibinfo{person}{Mariano
  Alca{\~n}iz}, {and} \bibinfo{person}{A. Rey}.}
  \bibinfo{year}{1998}\natexlab{}.
\newblock \showarticletitle{Virtual Reality Treatment of Claustrophobia: A Case
  Report.}
\newblock \bibinfo{journal}{\emph{Behaviour research and therapy}}
  \bibinfo{volume}{36 2} (\bibinfo{year}{1998}), \bibinfo{pages}{239--46}.
\newblock
\href{https://doi.org/10.1016/S0005-7967(97)10006-7}{doi:\nolinkurl{10.1016/S0005-7967(97)10006-7}}


\bibitem[Botella et~al\mbox{.}(2015)]%
        {Botella2015Virtual}
\bibfield{author}{\bibinfo{person}{Cristina Botella}, \bibinfo{person}{Berenice
  Serrano}, \bibinfo{person}{Rosa BaÃ±os}, {and} \bibinfo{person}{Azucena
  GarcÃ­a-Palacios}.} \bibinfo{year}{2015}\natexlab{}.
\newblock \showarticletitle{Virtual reality exposure-based therapy for the
  treatment of post-traumatic stress disorder: a review of its efficacy, the
  adequacy of the treatment protocol, and its acceptability}.
\newblock \bibinfo{journal}{\emph{Neuropsychiatric Disease and Treatment}}
  \bibinfo{volume}{11} (\bibinfo{year}{2015}), \bibinfo{pages}{2533 -- 2545}.
\newblock
\href{https://doi.org/10.2147/NDT.S89542}{doi:\nolinkurl{10.2147/NDT.S89542}}


\bibitem[Bowen et~al\mbox{.}(2009)]%
        {bowen2009we}
\bibfield{author}{\bibinfo{person}{Deborah~J Bowen}, \bibinfo{person}{Matthew
  Kreuter}, \bibinfo{person}{Bonnie Spring}, \bibinfo{person}{Ludmila
  {Cofta-Woerpel}}, \bibinfo{person}{Laura Linnan}, \bibinfo{person}{Diane
  Weiner}, \bibinfo{person}{Suzanne Bakken}, \bibinfo{person}{Cecilia~Patrick
  Kaplan}, \bibinfo{person}{Linda Squiers}, \bibinfo{person}{Cecilia Fabrizio},
  {et~al\mbox{.}}} \bibinfo{year}{2009}\natexlab{}.
\newblock \showarticletitle{How We Design Feasibility Studies}.
\newblock \bibinfo{journal}{\emph{American journal of preventive medicine}}
  \bibinfo{volume}{36}, \bibinfo{number}{5} (\bibinfo{year}{2009}),
  \bibinfo{pages}{452--457}.
\newblock
\href{https://doi.org/10.1016/j.amepre.2009.02.002}{doi:\nolinkurl{10.1016/j.amepre.2009.02.002}}


\bibitem[Braun and Clarke(2019)]%
        {braun_reflecting_2019}
\bibfield{author}{\bibinfo{person}{Virginia Braun} {and}
  \bibinfo{person}{Victoria Clarke}.} \bibinfo{year}{2019}\natexlab{}.
\newblock \showarticletitle{Reflecting on reflexive thematic analysis}.
\newblock \bibinfo{journal}{\emph{Qualitative Research in Sport, Exercise and
  Health}} \bibinfo{volume}{11}, \bibinfo{number}{4} (\bibinfo{date}{Aug.}
  \bibinfo{year}{2019}), \bibinfo{pages}{589--597}.
\newblock
\showISSN{2159-676X}
\href{https://doi.org/10.1080/2159676X.2019.1628806}{doi:\nolinkurl{10.1080/2159676X.2019.1628806}}
\newblock
\shownote{Publisher: Routledge \_eprint:
  https://doi.org/10.1080/2159676X.2019.1628806}.


\bibitem[Braun and Clarke(2021)]%
        {braun_one_2021}
\bibfield{author}{\bibinfo{person}{Virginia Braun} {and}
  \bibinfo{person}{Victoria Clarke}.} \bibinfo{year}{2021}\natexlab{}.
\newblock \showarticletitle{One size fits all? {What} counts as quality
  practice in (reflexive) thematic analysis?}
\newblock \bibinfo{journal}{\emph{Qualitative Research in Psychology}}
  \bibinfo{volume}{18}, \bibinfo{number}{3} (\bibinfo{year}{2021}),
  \bibinfo{pages}{328--352}.
\newblock
\showISSN{1478-0895}
\href{https://doi.org/10.1080/14780887.2020.1769238}{doi:\nolinkurl{10.1080/14780887.2020.1769238}}
\newblock
\shownote{Place: United Kingdom Publisher: Taylor \& Francis}.


\bibitem[Brewin et~al\mbox{.}(2025)]%
        {brewin2025post}
\bibfield{author}{\bibinfo{person}{Chris~R Brewin}, \bibinfo{person}{Lukoye
  Atwoli}, \bibinfo{person}{Jonathan~I Bisson}, \bibinfo{person}{Sandro Galea},
  \bibinfo{person}{Karestan Koenen}, {and} \bibinfo{person}{Roberto
  {Lewis-Fern{\'a}ndez}}.} \bibinfo{year}{2025}\natexlab{}.
\newblock \showarticletitle{Post-Traumatic Stress Disorder: Evolving
  Conceptualization and Evidence, and Future Research Directions}.
\newblock \bibinfo{journal}{\emph{World Psychiatry}} \bibinfo{volume}{24},
  \bibinfo{number}{1} (\bibinfo{year}{2025}), \bibinfo{pages}{52--80}.
\newblock
\href{https://doi.org/10.1002/wps.21269}{doi:\nolinkurl{10.1002/wps.21269}}


\bibitem[Bride(2007)]%
        {bride2007prevalence}
\bibfield{author}{\bibinfo{person}{Brian~E Bride}.}
  \bibinfo{year}{2007}\natexlab{}.
\newblock \showarticletitle{Prevalence of secondary traumatic stress among
  social workers}.
\newblock \bibinfo{journal}{\emph{Social work}} \bibinfo{volume}{52},
  \bibinfo{number}{1} (\bibinfo{year}{2007}), \bibinfo{pages}{63--70}.
\newblock


\bibitem[Campbell et~al\mbox{.}(2022)]%
        {campbell_factors_2022}
\bibfield{author}{\bibinfo{person}{Fiona Campbell}, \bibinfo{person}{Lindsay
  Blank}, \bibinfo{person}{Anna Cantrell}, \bibinfo{person}{Susan Baxter},
  \bibinfo{person}{Christopher Blackmore}, \bibinfo{person}{Jan Dixon}, {and}
  \bibinfo{person}{Elizabeth Goyder}.} \bibinfo{year}{2022}\natexlab{}.
\newblock \showarticletitle{Factors that influence mental health of university
  and college students in the {UK}: a systematic review}.
\newblock \bibinfo{journal}{\emph{BMC Public Health}} \bibinfo{volume}{22},
  \bibinfo{number}{1} (\bibinfo{year}{2022}), \bibinfo{pages}{1778}.
\newblock


\bibitem[Carl et~al\mbox{.}(2019)]%
        {Carl2019-jk}
\bibfield{author}{\bibinfo{person}{Emily Carl}, \bibinfo{person}{Aliza~T
  Stein}, \bibinfo{person}{Andrew Levihn-Coon}, \bibinfo{person}{Jamie~R
  Pogue}, \bibinfo{person}{Barbara Rothbaum}, \bibinfo{person}{Paul Emmelkamp},
  \bibinfo{person}{Gordon J~G Asmundson}, \bibinfo{person}{Per Carlbring},
  {and} \bibinfo{person}{Mark~B Powers}.} \bibinfo{year}{2019}\natexlab{}.
\newblock \showarticletitle{Virtual reality exposure therapy for anxiety and
  related disorders: A meta-analysis of randomized controlled trials}.
\newblock \bibinfo{journal}{\emph{J. Anxiety Disord.}}  \bibinfo{volume}{61}
  (\bibinfo{date}{Jan.} \bibinfo{year}{2019}), \bibinfo{pages}{27--36}.
\newblock
\href{https://doi.org/10.1016/j.janxdis.2018.08.003}{doi:\nolinkurl{10.1016/j.janxdis.2018.08.003}}


\bibitem[Carlin et~al\mbox{.}(1997)]%
        {carlinVirtualRealityTactile1997}
\bibfield{author}{\bibinfo{person}{Albert Carlin}, \bibinfo{person}{Hunter~G.
  Hoffman}, {and} \bibinfo{person}{Suzanne~J. Weghorst}.}
  \bibinfo{year}{1997}\natexlab{}.
\newblock \showarticletitle{Virtual Reality and Tactile Augmentation in the
  Treatment of Spider Phobia: A Case Report.}
\newblock \bibinfo{journal}{\emph{Behaviour research and therapy}}
  \bibinfo{volume}{35 2} (\bibinfo{year}{1997}), \bibinfo{pages}{153--8}.
\newblock
\href{https://doi.org/10.1016/S0005-7967(96)00085-X}{doi:\nolinkurl{10.1016/S0005-7967(96)00085-X}}


\bibitem[Cesari et~al\mbox{.}(2023)]%
        {CESARI2023175}
\bibfield{author}{\bibinfo{person}{Valentina Cesari}, \bibinfo{person}{Sergio
  Frumento}, \bibinfo{person}{Andrea Leo}, \bibinfo{person}{Marina Baroni},
  \bibinfo{person}{Grazia Rutigliano}, \bibinfo{person}{Angelo Gemignani},
  {and} \bibinfo{person}{Danilo Menicucci}.} \bibinfo{year}{2023}\natexlab{}.
\newblock \showarticletitle{Functional Correlates of Subliminal Stimulation in
  {{Posttraumatic Stress Disorder}}: {{Systematic}} Review and Meta-Analysis}.
\newblock \bibinfo{journal}{\emph{Journal of Affective Disorders}}
  \bibinfo{volume}{337} (\bibinfo{year}{2023}), \bibinfo{pages}{175--185}.
\newblock
\showISSN{0165-0327}
\href{https://doi.org/10.1016/j.jad.2023.05.047}{doi:\nolinkurl{10.1016/j.jad.2023.05.047}}


\bibitem[Chancellor et~al\mbox{.}(2019)]%
        {chancellor_sensitive_2019}
\bibfield{author}{\bibinfo{person}{Stevie Chancellor}, \bibinfo{person}{Nazanin
  Andalibi}, \bibinfo{person}{Lindsay Blackwell}, \bibinfo{person}{David
  Nemer}, {and} \bibinfo{person}{Wendy Moncur}.}
  \bibinfo{year}{2019}\natexlab{}.
\newblock \showarticletitle{Sensitive {Research}, {Practice} and {Design} in
  {HCI}}. In \bibinfo{booktitle}{\emph{Extended {Abstracts} of the 2019 {CHI}
  {Conference} on {Human} {Factors} in {Computing} {Systems}}}
  \emph{(\bibinfo{series}{{CHI} {EA} '19})}. \bibinfo{publisher}{Association
  for Computing Machinery}, \bibinfo{address}{New York, NY, USA},
  \bibinfo{pages}{1--8}.
\newblock
\showISBNx{978-1-4503-5971-9}
\href{https://doi.org/10.1145/3290607.3299003}{doi:\nolinkurl{10.1145/3290607.3299003}}


\bibitem[Chang(2016)]%
        {chang2016autoethnography}
\bibfield{author}{\bibinfo{person}{Heewon Chang}.}
  \bibinfo{year}{2016}\natexlab{}.
\newblock \bibinfo{booktitle}{\emph{Autoethnography as Method}
  (\bibinfo{edition}{1} ed.)}.
\newblock \bibinfo{publisher}{Routledge}, \bibinfo{address}{New York}.
\newblock
\showISBNx{978-1-315-43337-0}


\bibitem[Charlton and Thompson(1996)]%
        {charltonWaysCopingPsychological1996}
\bibfield{author}{\bibinfo{person}{P.~F.~C. Charlton} {and}
  \bibinfo{person}{J.~A. Thompson}.} \bibinfo{year}{1996}\natexlab{}.
\newblock \showarticletitle{Ways of Coping with Psychological Distress after
  Trauma}.
\newblock \bibinfo{journal}{\emph{British Journal of Clinical Psychology}}
  \bibinfo{volume}{35}, \bibinfo{number}{4} (\bibinfo{date}{Nov.}
  \bibinfo{year}{1996}), \bibinfo{pages}{517--530}.
\newblock
\showISSN{0144-6657}
\href{https://doi.org/10.1111/j.2044-8260.1996.tb01208.x}{doi:\nolinkurl{10.1111/j.2044-8260.1996.tb01208.x}}


\bibitem[Chu and Dill(1990)]%
        {chuDissociativeSymptomsRelation1990}
\bibfield{author}{\bibinfo{person}{James Chu} {and} \bibinfo{person}{Diana
  Dill}.} \bibinfo{year}{1990}\natexlab{}.
\newblock \showarticletitle{Dissociative {{Symptoms}} in {{Relation}} to
  {{Childhood Physical}} and {{Sexual Abuse}}}.
\newblock \bibinfo{journal}{\emph{The American journal of psychiatry}}
  \bibinfo{volume}{147} (\bibinfo{date}{Aug.} \bibinfo{year}{1990}),
  \bibinfo{pages}{887--92}.
\newblock
\href{https://doi.org/10.1176/ajp.147.7.887}{doi:\nolinkurl{10.1176/ajp.147.7.887}}


\bibitem[Clemensen et~al\mbox{.}(2007)]%
        {clemensen_participatory_2007}
\bibfield{author}{\bibinfo{person}{Jane Clemensen}, \bibinfo{person}{Simon~B.
  Larsen}, \bibinfo{person}{Morten Kyng}, {and} \bibinfo{person}{Marit
  Kirkevold}.} \bibinfo{year}{2007}\natexlab{}.
\newblock \showarticletitle{Participatory design in health sciences: {Using}
  cooperative experimental methods in developing health services and computer
  technology}.
\newblock \bibinfo{journal}{\emph{Qualitative Health Research}}
  \bibinfo{volume}{17}, \bibinfo{number}{1} (\bibinfo{date}{Jan.}
  \bibinfo{year}{2007}), \bibinfo{pages}{122--130}.
\newblock
\showISSN{1049-7323}
\href{https://doi.org/10.1177/1049732306293664}{doi:\nolinkurl{10.1177/1049732306293664}}


\bibitem[Cloitre et~al\mbox{.}(2011)]%
        {cloitre2011treatment}
\bibfield{author}{\bibinfo{person}{Marylene Cloitre},
  \bibinfo{person}{Christine~A Courtois}, \bibinfo{person}{Anthony
  Charuvastra}, \bibinfo{person}{Richard Carapezza}, \bibinfo{person}{Bradley~C
  Stolbach}, {and} \bibinfo{person}{Bonnie~L Green}.}
  \bibinfo{year}{2011}\natexlab{}.
\newblock \showarticletitle{Treatment of complex PTSD: Results of the ISTSS
  expert clinician survey on best practices}.
\newblock \bibinfo{journal}{\emph{Journal of traumatic stress}}
  \bibinfo{volume}{24}, \bibinfo{number}{6} (\bibinfo{year}{2011}),
  \bibinfo{pages}{615--627}.
\newblock


\bibitem[Cloitre et~al\mbox{.}(2014)]%
        {cloitreDistinguishingPTSDComplex2014}
\bibfield{author}{\bibinfo{person}{Maryl{\'e}ne Cloitre},
  \bibinfo{person}{Donn~W. Garvert}, \bibinfo{person}{Brandon Weiss},
  \bibinfo{person}{Eve~B. Carlson}, {and} \bibinfo{person}{Richard~A. Bryant}.}
  \bibinfo{year}{2014}\natexlab{}.
\newblock \showarticletitle{Distinguishing {{PTSD}}, Complex {{PTSD}}, and
  Borderline Personality Disorder: {{A}} Latent Class Analysis.}
\newblock \bibinfo{journal}{\emph{European Journal of Psychotraumatology}}
  \bibinfo{volume}{5} (\bibinfo{year}{2014}).
\newblock
\showISSN{2000-8066(Electronic)}


\bibitem[Craske et~al\mbox{.}(2014)]%
        {CRASKE201410}
\bibfield{author}{\bibinfo{person}{Michelle~G. Craske},
  \bibinfo{person}{Michael Treanor}, \bibinfo{person}{Christopher~C. Conway},
  \bibinfo{person}{Tomislav Zbozinek}, {and} \bibinfo{person}{Bram Vervliet}.}
  \bibinfo{year}{2014}\natexlab{}.
\newblock \showarticletitle{Maximizing Exposure Therapy: {{An}} Inhibitory
  Learning Approach}.
\newblock \bibinfo{journal}{\emph{Behaviour Research and Therapy}}
  \bibinfo{volume}{58} (\bibinfo{year}{2014}), \bibinfo{pages}{10--23}.
\newblock
\showISSN{0005-7967}
\href{https://doi.org/10.1016/j.brat.2014.04.006}{doi:\nolinkurl{10.1016/j.brat.2014.04.006}}


\bibitem[Cushing et~al\mbox{.}(2024)]%
        {cushing2024Metacognition}
\bibfield{author}{\bibinfo{person}{Cody Cushing}, \bibinfo{person}{Hakwan Lau},
  \bibinfo{person}{Stefan Hofmann}, \bibinfo{person}{Joseph Ledoux}, {and}
  \bibinfo{person}{Vincent Taschereau-Dumouchel}.}
  \bibinfo{year}{2024}\natexlab{}.
\newblock \showarticletitle{Metacognition as a window into subjective affective
  experience}.
\newblock \bibinfo{journal}{\emph{Psychiatry and clinical neurosciences}}
  \bibinfo{volume}{78} (\bibinfo{date}{06} \bibinfo{year}{2024}).
\newblock
\href{https://doi.org/10.1111/pcn.13683}{doi:\nolinkurl{10.1111/pcn.13683}}


\bibitem[Degenhard et~al\mbox{.}(2025)]%
        {degenhard2025rememberchallengesrisks}
\bibfield{author}{\bibinfo{person}{Annalisa Degenhard}, \bibinfo{person}{Stefan
  TschÃ¶ke}, \bibinfo{person}{Michael Rietzler}, {and} \bibinfo{person}{Enrico
  Rukzio}.} \bibinfo{year}{2025}\natexlab{}.
\newblock \bibinfo{title}{Describe Me Something You Do Not Remember -
  Challenges and Risks of Exposure Design Using Generative Artificial
  Intelligence for Therapy of Complex Post-traumatic Disorder}.
\newblock
\showeprint[arxiv]{2505.20796}~[cs.HC]
\urldef\tempurl%
\url{https://arxiv.org/abs/2505.20796}
\showURL{%
\tempurl}


\bibitem[Difede et~al\mbox{.}(2007)]%
        {difedeVirtualRealityExposure2007}
\bibfield{author}{\bibinfo{person}{Joann Difede}, \bibinfo{person}{Judith
  Cukor}, \bibinfo{person}{Nimali Jayasinghe}, \bibinfo{person}{Ivy Patt},
  \bibinfo{person}{Sharon Jedel}, \bibinfo{person}{Lisa Spielman},
  \bibinfo{person}{Cezar Giosan}, {and} \bibinfo{person}{Hunter Hoffman}.}
  \bibinfo{year}{2007}\natexlab{}.
\newblock \showarticletitle{Virtual {{Reality Exposure Therapy}} for the
  {{Treatment}} of {{Posttraumatic Stress Disorder Following September}} 11,
  2001}.
\newblock \bibinfo{journal}{\emph{The Journal of clinical psychiatry}}
  \bibinfo{volume}{68} (\bibinfo{date}{Nov.} \bibinfo{year}{2007}),
  \bibinfo{pages}{1639--47}.
\newblock
\href{https://doi.org/10.4088/JCP.v68n1102}{doi:\nolinkurl{10.4088/JCP.v68n1102}}


\bibitem[Difede et~al\mbox{.}(2002)]%
        {difedeInnovativeUseVirtual2002}
\bibfield{author}{\bibinfo{person}{JoAnn Difede}, \bibinfo{person}{Hunter~G.
  Hoffman}, {and} \bibinfo{person}{Nimale Jaysinghe}.}
  \bibinfo{year}{2002}\natexlab{}.
\newblock \showarticletitle{Innovative Use of Virtual Reality Technology in the
  Treatment of {{PTSD}} in the Aftermath of {{September}} 11.}
\newblock \bibinfo{journal}{\emph{Psychiatric services}}  \bibinfo{volume}{53
  9} (\bibinfo{year}{2002}), \bibinfo{pages}{1083--5}.
\newblock
\href{https://doi.org/10.1176/APPI.PS.53.9.1083}{doi:\nolinkurl{10.1176/APPI.PS.53.9.1083}}


\bibitem[Donetto et~al\mbox{.}(2015)]%
        {donetto_experience-based_2015}
\bibfield{author}{\bibinfo{person}{Sara Donetto}, \bibinfo{person}{Paola
  Pierri}, \bibinfo{person}{Vicki Tsianakas}, {and} \bibinfo{person}{Glenn
  Robert}.} \bibinfo{year}{2015}\natexlab{}.
\newblock \showarticletitle{Experience-based {Co}-design and {Healthcare}
  {Improvement}: {Realizing} {Participatory} {Design} in the {Public}
  {Sector}}.
\newblock \bibinfo{journal}{\emph{The Design Journal}} \bibinfo{volume}{18},
  \bibinfo{number}{2} (\bibinfo{date}{June} \bibinfo{year}{2015}),
  \bibinfo{pages}{227--248}.
\newblock
\showISSN{1460-6925}
\href{https://doi.org/10.2752/175630615X14212498964312}{doi:\nolinkurl{10.2752/175630615X14212498964312}}
\newblock
\shownote{Publisher: Routledge \_eprint:
  https://doi.org/10.2752/175630615X14212498964312}.


\bibitem[Dunsmoor et~al\mbox{.}(2022)]%
        {dunsmoorLaboratoryModelsPosttraumatic2022}
\bibfield{author}{\bibinfo{person}{Joseph~E. Dunsmoor},
  \bibinfo{person}{Josh~M. Cisler}, \bibinfo{person}{Gregory~A. Fonzo},
  \bibinfo{person}{Suzannah~K. Creech}, {and} \bibinfo{person}{Charles~B.
  Nemeroff}.} \bibinfo{year}{2022}\natexlab{}.
\newblock \showarticletitle{Laboratory Models of Post-Traumatic Stress
  Disorder: {{The}} Elusive Bridge to Translation}.
\newblock \bibinfo{journal}{\emph{Neuron}} \bibinfo{volume}{110},
  \bibinfo{number}{11} (\bibinfo{date}{June} \bibinfo{year}{2022}),
  \bibinfo{pages}{1754--1776}.
\newblock
\showISSN{0896-6273}
\href{https://doi.org/10.1016/j.neuron.2022.03.001}{doi:\nolinkurl{10.1016/j.neuron.2022.03.001}}


\bibitem[Edwards(2018)]%
        {edwards_user_2018}
\bibfield{author}{\bibinfo{person}{Sydney Edwards}.}
  \bibinfo{year}{2018}\natexlab{}.
\newblock \bibinfo{title}{User {Centered} {Simplicity}: {A} {Design}
  {Manifesto}}.
\newblock
\urldef\tempurl%
\url{https://medium.com/@edwar3se/user-centered-simplicity-a-design-manifesto-21f92402894d}
\showURL{%
\tempurl}


\bibitem[Emmelkamp et~al\mbox{.}(2001)]%
        {emmelkampVirtualRealityTreatment2001}
\bibfield{author}{\bibinfo{person}{Paul Emmelkamp}, \bibinfo{person}{Mary
  Bruynzeel}, \bibinfo{person}{Leonie Drost}, {and} \bibinfo{person}{C Mast}.}
  \bibinfo{year}{2001}\natexlab{}.
\newblock \showarticletitle{Virtual {{Reality Treatment}} in {{Acrophobia}}:
  {{A Comparison}} with {{Exposure}} in {{Vivo}}}.
\newblock \bibinfo{journal}{\emph{Cyberpsychology \& behavior : the impact of
  the Internet, multimedia and virtual reality on behavior and society}}
  \bibinfo{volume}{4} (\bibinfo{date}{July} \bibinfo{year}{2001}),
  \bibinfo{pages}{335--9}.
\newblock
\href{https://doi.org/10.1089/109493101300210222}{doi:\nolinkurl{10.1089/109493101300210222}}


\bibitem[Feuston et~al\mbox{.}(2022)]%
        {feuston_researcher_2022}
\bibfield{author}{\bibinfo{person}{Jessica~L. Feuston}, \bibinfo{person}{Arpita
  Bhattacharya}, \bibinfo{person}{Nazanin Andalibi},
  \bibinfo{person}{Elizabeth~A. Ankrah}, \bibinfo{person}{Sheena Erete},
  \bibinfo{person}{Mark Handel}, \bibinfo{person}{Wendy Moncur},
  \bibinfo{person}{Sarah Vieweg}, {and} \bibinfo{person}{Jed~R. Brubaker}.}
  \bibinfo{year}{2022}\natexlab{}.
\newblock \showarticletitle{Researcher {Wellbeing} and {Best} {Practices} in
  {Emotionally} {Demanding} {Research}}. In \bibinfo{booktitle}{\emph{{CHI}
  {Conference} on {Human} {Factors} in {Computing} {Systems} {Extended}
  {Abstracts}}}. \bibinfo{publisher}{ACM}, \bibinfo{address}{New Orleans LA
  USA}, \bibinfo{pages}{1--6}.
\newblock
\showISBNx{978-1-4503-9156-6}
\href{https://doi.org/10.1145/3491101.3503742}{doi:\nolinkurl{10.1145/3491101.3503742}}


\bibitem[Figley(1995)]%
        {figley1995compassion}
\bibfield{author}{\bibinfo{person}{Charles~R Figley}.}
  \bibinfo{year}{1995}\natexlab{}.
\newblock \bibinfo{booktitle}{\emph{Compassion Fatigue: Coping with Secondary
  Traumatic Stress Disorder in Those who Treat the Traumatized}}.
\newblock Number~23. \bibinfo{publisher}{Psychology Press}.
\newblock


\bibitem[Foa and Kozak(1986)]%
        {foaEmotionalProcessingFear1986}
\bibfield{author}{\bibinfo{person}{Edna Foa} {and} \bibinfo{person}{M.J.
  Kozak}.} \bibinfo{year}{1986}\natexlab{}.
\newblock \showarticletitle{Emotional {{Processing}} of {{Fear}}. {{Exposure}}
  to {{Corrective Information}}}.
\newblock \bibinfo{journal}{\emph{Psychological bulletin}}
  \bibinfo{volume}{99} (\bibinfo{date}{Jan.} \bibinfo{year}{1986}),
  \bibinfo{pages}{20}.
\newblock
\href{https://doi.org/10.1037//0033-2909.99.1.20}{doi:\nolinkurl{10.1037//0033-2909.99.1.20}}


\bibitem[Ford(2018)]%
        {fordTraumaMemoryProcessing2018}
\bibfield{author}{\bibinfo{person}{Julian~D. Ford}.}
  \bibinfo{year}{2018}\natexlab{}.
\newblock \showarticletitle{Trauma {{Memory Processing}} in {{Posttraumatic
  Stress Disorder Psychotherapy}}: {{A Unifying Framework}}}.
\newblock \bibinfo{journal}{\emph{Journal of Traumatic Stress}}
  \bibinfo{volume}{31}, \bibinfo{number}{6} (\bibinfo{date}{Dec.}
  \bibinfo{year}{2018}), \bibinfo{pages}{933--942}.
\newblock
\showISSN{0894-9867}
\href{https://doi.org/10.1002/jts.22344}{doi:\nolinkurl{10.1002/jts.22344}}


\bibitem[Glicksohn and Avnon(1997)]%
        {glicksohnExplorationsVirtualReality1997}
\bibfield{author}{\bibinfo{person}{Joseph Glicksohn} {and}
  \bibinfo{person}{Michal Avnon}.} \bibinfo{year}{1997}\natexlab{}.
\newblock \showarticletitle{Explorations in {{Virtual Reality}}:
  {{Absorption}}, {{Cognition}} and {{Altered State}} of {{Consciousness}}}.
\newblock \bibinfo{journal}{\emph{Imagination, Cognition and Personality}}
  \bibinfo{volume}{17}, \bibinfo{number}{2} (\bibinfo{date}{Oct.}
  \bibinfo{year}{1997}), \bibinfo{pages}{141--151}.
\newblock
\showISSN{0276-2366}
\href{https://doi.org/10.2190/FTUU-GLC5-GBT8-9RUW}{doi:\nolinkurl{10.2190/FTUU-GLC5-GBT8-9RUW}}


\bibitem[GonÃ§alves et~al\mbox{.}(2012)]%
        {Gonalves2012Efficacy}
\bibfield{author}{\bibinfo{person}{Ana~LÃºcia GonÃ§alves, Raquel~andPedrozo},
  \bibinfo{person}{Evandro S.~F. Coutinho}, \bibinfo{person}{Ivan Figueira},
  {and} \bibinfo{person}{Paula Ventura}.} \bibinfo{year}{2012}\natexlab{}.
\newblock \showarticletitle{Efficacy of Virtual Reality Exposure Therapy in the
  Treatment of PTSD: A Systematic Review}.
\newblock \bibinfo{journal}{\emph{PLoS ONE}}  \bibinfo{volume}{7}
  (\bibinfo{year}{2012}).
\newblock
\href{https://doi.org/10.1371/journal.pone.0048469}{doi:\nolinkurl{10.1371/journal.pone.0048469}}


\bibitem[Haft and Sherrill(2025)]%
        {Haft2025-ko}
\bibfield{author}{\bibinfo{person}{S~M Haft} {and} \bibinfo{person}{A~M
  Sherrill}.} \bibinfo{year}{2025}\natexlab{}.
\newblock \showarticletitle{Virtual reality exposure}.
\newblock \bibinfo{journal}{\emph{The Sage Encyclopedia of Mood and Anxiety
  Disorders}}  \bibinfo{volume}{3} (\bibinfo{year}{2025}),
  \bibinfo{pages}{1401--1403}.
\newblock


\bibitem[Hagborg et~al\mbox{.}(2022)]%
        {hagborgChildhoodTraumaQuestionnaire2022}
\bibfield{author}{\bibinfo{person}{Johan~Melander Hagborg},
  \bibinfo{person}{Torbj{\"o}rn Kalin}, {and} \bibinfo{person}{Arne Gerdner}.}
  \bibinfo{year}{2022}\natexlab{}.
\newblock \showarticletitle{The {{Childhood Trauma Questionnaire}}---{{Short
  Form}} ({{CTQ-SF}}) Used with Adolescents -- Methodological Report from
  Clinical and Community Samples}.
\newblock \bibinfo{journal}{\emph{Journal of Child \& Adolescent Trauma}}
  \bibinfo{volume}{15}, \bibinfo{number}{4} (\bibinfo{date}{Dec.}
  \bibinfo{year}{2022}), \bibinfo{pages}{1199--1213}.
\newblock
\showISSN{1936-153X}
\href{https://doi.org/10.1007/s40653-022-00443-8}{doi:\nolinkurl{10.1007/s40653-022-00443-8}}


\bibitem[Harrington et~al\mbox{.}(2019)]%
        {harrington_deconstructing_2019}
\bibfield{author}{\bibinfo{person}{Christina Harrington},
  \bibinfo{person}{Sheena Erete}, {and} \bibinfo{person}{Anne~Marie Piper}.}
  \bibinfo{year}{2019}\natexlab{}.
\newblock \showarticletitle{Deconstructing {Community}-{Based} {Collaborative}
  {Design}: {Towards} {More} {Equitable} {Participatory} {Design}
  {Engagements}}.
\newblock \bibinfo{journal}{\emph{Proc. ACM Hum.-Comput. Interact.}}
  \bibinfo{volume}{3}, \bibinfo{number}{CSCW} (\bibinfo{date}{Nov.}
  \bibinfo{year}{2019}), \bibinfo{pages}{216:1--216:25}.
\newblock
\href{https://doi.org/10.1145/3359318}{doi:\nolinkurl{10.1145/3359318}}


\bibitem[Harris et~al\mbox{.}(2003)]%
        {harrisBriefVirtualReality2003}
\bibfield{author}{\bibinfo{person}{Sandra Harris}, \bibinfo{person}{Robert
  Kemmerling}, {and} \bibinfo{person}{Max North}.}
  \bibinfo{year}{2003}\natexlab{}.
\newblock \showarticletitle{Brief {{Virtual Reality Therapy}} for {{Public
  Speaking Anxiety}}}.
\newblock \bibinfo{journal}{\emph{Cyberpsychology \& behavior : the impact of
  the Internet, multimedia and virtual reality on behavior and society}}
  \bibinfo{volume}{5} (\bibinfo{date}{Jan.} \bibinfo{year}{2003}),
  \bibinfo{pages}{543--50}.
\newblock
\href{https://doi.org/10.1089/109493102321018187}{doi:\nolinkurl{10.1089/109493102321018187}}


\bibitem[Herman(2015)]%
        {herman_trauma_2015}
\bibfield{author}{\bibinfo{person}{Judith Herman}.}
  \bibinfo{year}{2015}\natexlab{}.
\newblock \bibinfo{booktitle}{\emph{Trauma and recovery: {The} aftermath of
  violenceâ€”from domestic abuse to political terror}}.
\newblock \bibinfo{publisher}{Basic Books/Hachette Book Group},
  \bibinfo{address}{New York, NY, US}.
\newblock
\showISBNx{978-0-465-06171-6 978-0-465-09873-6}
\newblock
\shownote{Pages: ix, 326}.


\bibitem[Herman(1992)]%
        {Herman.1992}
\bibfield{author}{\bibinfo{person}{Judith~Lewis Herman}.}
  \bibinfo{year}{1992}\natexlab{}.
\newblock \showarticletitle{Complex PTSD: A syndrome in survivors of prolonged
  and repeated trauma}.
\newblock \bibinfo{journal}{\emph{Journal of Traumatic Stress}}
  \bibinfo{volume}{5}, \bibinfo{number}{3} (\bibinfo{year}{1992}),
  \bibinfo{pages}{377--391}.
\newblock
\href{https://doi.org/10.1002/jts.2490050305}{doi:\nolinkurl{10.1002/jts.2490050305}}


\bibitem[Herman and Schatzow(1987)]%
        {herman1987recovery}
\bibfield{author}{\bibinfo{person}{J.~L. Herman} {and} \bibinfo{person}{E.
  Schatzow}.} \bibinfo{year}{1987}\natexlab{}.
\newblock \showarticletitle{Recovery and Verification of Memories of Childhood
  Sexual Trauma}.
\newblock \bibinfo{journal}{\emph{Psychoanalytic Psychology}}
  \bibinfo{volume}{4}, \bibinfo{number}{1} (\bibinfo{year}{1987}).
\newblock


\bibitem[Hodges et~al\mbox{.}(1995)]%
        {hodgesVirtualEnvironmentsTreating1995}
\bibfield{author}{\bibinfo{person}{L.F. Hodges}, \bibinfo{person}{R. Kooper},
  \bibinfo{person}{T.C. Meyer}, \bibinfo{person}{B.O. Rothbaum},
  \bibinfo{person}{D. Opdyke}, \bibinfo{person}{J.J. {de Graaff}},
  \bibinfo{person}{J.S. Williford}, {and} \bibinfo{person}{M.M. North}.}
  \bibinfo{year}{1995}\natexlab{}.
\newblock \showarticletitle{Virtual Environments for Treating the Fear of
  Heights}.
\newblock \bibinfo{journal}{\emph{Computer}} \bibinfo{volume}{28},
  \bibinfo{number}{7} (\bibinfo{date}{July} \bibinfo{year}{1995}),
  \bibinfo{pages}{27--34}.
\newblock
\showISSN{1558-0814}
\href{https://doi.org/10.1109/2.391038}{doi:\nolinkurl{10.1109/2.391038}}


\bibitem[Hoppen et~al\mbox{.}(2024)]%
        {HOPPEN2024112}
\bibfield{author}{\bibinfo{person}{Thole~H Hoppen}, \bibinfo{person}{Richard
  {Meiser-Stedman}}, \bibinfo{person}{Ahlke Kip},
  \bibinfo{person}{Marianne~Skogbrott Birkeland}, {and}
  \bibinfo{person}{Nexhmedin Morina}.} \bibinfo{year}{2024}\natexlab{}.
\newblock \showarticletitle{The Efficacy of Psychological Interventions for
  Adult Post-Traumatic Stress Disorder Following Exposure to Single versus
  Multiple Traumatic Events: A Meta-Analysis of Randomised Controlled Trials}.
\newblock \bibinfo{journal}{\emph{The Lancet Psychiatry}} \bibinfo{volume}{11},
  \bibinfo{number}{2} (\bibinfo{year}{2024}), \bibinfo{pages}{112--122}.
\newblock
\showISSN{2215-0366}
\href{https://doi.org/10.1016/S2215-0366(23)00373-5}{doi:\nolinkurl{10.1016/S2215-0366(23)00373-5}}


\bibitem[Houben et~al\mbox{.}(2024)]%
        {houben_hci_2024}
\bibfield{author}{\bibinfo{person}{Maarten Houben}, \bibinfo{person}{Minha
  Lee}, \bibinfo{person}{Sarah Foley}, \bibinfo{person}{Kellie Morrissey},
  {and} \bibinfo{person}{Rens Brankaert}.} \bibinfo{year}{2024}\natexlab{}.
\newblock \showarticletitle{{HCI} {Research} in {Sensitive} {Settings}:
  {Learning} {Researcher} {Reflexivity}, {Ethical} {Conduct} and {Empathy} in
  {Participatory} {Design} {Approaches}}. In \bibinfo{booktitle}{\emph{Extended
  {Abstracts} of the {CHI} {Conference} on {Human} {Factors} in {Computing}
  {Systems}}} \emph{(\bibinfo{series}{{CHI} {EA} '24})}.
  \bibinfo{publisher}{Association for Computing Machinery},
  \bibinfo{address}{New York, NY, USA}, \bibinfo{pages}{1--4}.
\newblock
\showISBNx{9798400703317}
\href{https://doi.org/10.1145/3613905.3636280}{doi:\nolinkurl{10.1145/3613905.3636280}}


\bibitem[Huang et~al\mbox{.}(2014)]%
        {huang2014samhsa}
\bibfield{author}{\bibinfo{person}{Larke~N Huang}, \bibinfo{person}{Rebecca
  Flatow}, \bibinfo{person}{Tenly Biggs}, \bibinfo{person}{Sara Afayee},
  \bibinfo{person}{Kelley Smith}, \bibinfo{person}{Thomas Clark}, {and}
  \bibinfo{person}{Mary Blake}.} \bibinfo{year}{2014}\natexlab{}.
\newblock \showarticletitle{{{SAMHSA}}'s Concept of Truama and Guidance for a
  Trauma-Informed Approach}.
\newblock  (\bibinfo{year}{2014}).
\newblock


\bibitem[Hyland et~al\mbox{.}(2020)]%
        {hyland2020relationship}
\bibfield{author}{\bibinfo{person}{Philip Hyland}, \bibinfo{person}{Mark
  Shevlin}, \bibinfo{person}{Claire Fyvie}, \bibinfo{person}{Maryl{\`e}ne
  Cloitre}, {and} \bibinfo{person}{Thanos Karatzias}.}
  \bibinfo{year}{2020}\natexlab{}.
\newblock \showarticletitle{The relationship between ICD-11 PTSD, complex PTSD
  and dissociative experiences}.
\newblock \bibinfo{journal}{\emph{Journal of Trauma \& Dissociation}}
  \bibinfo{volume}{21}, \bibinfo{number}{1} (\bibinfo{year}{2020}),
  \bibinfo{pages}{62--72}.
\newblock


\bibitem[Jenkins and Baird(2002)]%
        {jenkins2002secondary}
\bibfield{author}{\bibinfo{person}{Sharon~Rae Jenkins} {and}
  \bibinfo{person}{Stephanie Baird}.} \bibinfo{year}{2002}\natexlab{}.
\newblock \showarticletitle{Secondary traumatic stress and vicarious trauma: A
  validational study}.
\newblock \bibinfo{journal}{\emph{Journal of Traumatic Stress: Official
  Publication of The International Society for Traumatic Stress Studies}}
  \bibinfo{volume}{15}, \bibinfo{number}{5} (\bibinfo{year}{2002}),
  \bibinfo{pages}{423--432}.
\newblock


\bibitem[Jongedijk(2014)]%
        {jongedijk_narrative_2014}
\bibfield{author}{\bibinfo{person}{Ruud~A. Jongedijk}.}
  \bibinfo{year}{2014}\natexlab{}.
\newblock \showarticletitle{Narrative exposure therapy: an evidence-based
  treatment for multiple and complex trauma}.
\newblock \bibinfo{journal}{\emph{European Journal of Psychotraumatology}}
  \bibinfo{volume}{5} (\bibinfo{date}{Dec.} \bibinfo{year}{2014}),
  \bibinfo{pages}{10.3402/ejpt.v5.26522}.
\newblock
\showISSN{2000-8066}
\href{https://doi.org/10.3402/ejpt.v5.26522}{doi:\nolinkurl{10.3402/ejpt.v5.26522}}


\bibitem[Josman et~al\mbox{.}(2008)]%
        {josmanBusWorldAnalogPilot2008}
\bibfield{author}{\bibinfo{person}{Naomi Josman}, \bibinfo{person}{Ayelet
  Reisberg}, \bibinfo{person}{Patrice Weiss}, \bibinfo{person}{Azucena
  {Garcia-Palacios}}, {and} \bibinfo{person}{Hunter Hoffman}.}
  \bibinfo{year}{2008}\natexlab{}.
\newblock \showarticletitle{{{BusWorld}}: {{An Analog Pilot Test}} of a
  {{Virtual Environment Designed}} to {{Treat Posttraumatic Stress Disorder
  Originating}} from a {{Terrorist Suicide Bomb Attack}}}.
\newblock \bibinfo{journal}{\emph{Cyberpsychology \& behavior : the impact of
  the Internet, multimedia and virtual reality on behavior and society}}
  \bibinfo{volume}{11} (\bibinfo{date}{Dec.} \bibinfo{year}{2008}),
  \bibinfo{pages}{775--7}.
\newblock
\href{https://doi.org/10.1089/cpb.2008.0048}{doi:\nolinkurl{10.1089/cpb.2008.0048}}


\bibitem[Josman et~al\mbox{.}(2006)]%
        {josmanBusWorldDesigningVirtual2006}
\bibfield{author}{\bibinfo{person}{Naomi Josman}, \bibinfo{person}{Eli Somer},
  \bibinfo{person}{Ayelet Reisberg}, \bibinfo{person}{Patrice Weiss},
  \bibinfo{person}{Azucena {Garcia-Palacios}}, {and} \bibinfo{person}{Hunter
  Hoffman}.} \bibinfo{year}{2006}\natexlab{}.
\newblock \showarticletitle{{{BusWorld}}: {{Designing}} a {{Virtual
  Environment}} for {{Post-Traumatic Stress Disorder}} in {{Israel}}: {{A
  Protocol}}}.
\newblock \bibinfo{journal}{\emph{Cyberpsychology \& behavior : the impact of
  the Internet, multimedia and virtual reality on behavior and society}}
  \bibinfo{volume}{9} (\bibinfo{date}{May} \bibinfo{year}{2006}),
  \bibinfo{pages}{241--4}.
\newblock
\href{https://doi.org/10.1089/cpb.2006.9.241}{doi:\nolinkurl{10.1089/cpb.2006.9.241}}


\bibitem[Karatzias et~al\mbox{.}(2017)]%
        {karatzias2017ptsd}
\bibfield{author}{\bibinfo{person}{Thanos Karatzias}, \bibinfo{person}{Marylene
  Cloitre}, \bibinfo{person}{Andreas Maercker}, \bibinfo{person}{Evaldas
  Kazlauskas}, \bibinfo{person}{Mark Shevlin}, \bibinfo{person}{Philip Hyland},
  \bibinfo{person}{Jonathan~I Bisson}, \bibinfo{person}{Neil~P Roberts}, {and}
  \bibinfo{person}{Chris~R Brewin}.} \bibinfo{year}{2017}\natexlab{}.
\newblock \showarticletitle{PTSD and Complex PTSD: ICD-11 updates on concept
  and measurement in the UK, USA, Germany and Lithuania}.
\newblock \bibinfo{journal}{\emph{European journal of psychotraumatology}}
  \bibinfo{volume}{8}, \bibinfo{number}{sup7} (\bibinfo{year}{2017}),
  \bibinfo{pages}{1418103}.
\newblock


\bibitem[Kim and Park(2012)]%
        {kimDevelopmentHealthInformation2012}
\bibfield{author}{\bibinfo{person}{Jeongeun Kim} {and}
  \bibinfo{person}{Hyeoun-Ae Park}.} \bibinfo{year}{2012}\natexlab{}.
\newblock \showarticletitle{Development of a {{Health Information Technology
  Acceptance Model Using Consumers}}' {{Health Behavior Intention}}}.
\newblock \bibinfo{journal}{\emph{J Med Internet Res}} \bibinfo{volume}{14},
  \bibinfo{number}{5} (\bibinfo{date}{Oct.} \bibinfo{year}{2012}),
  \bibinfo{pages}{e133}.
\newblock
\showISSN{1438-8871}
\showeprint[pubmed]{23026508}
\href{https://doi.org/10.2196/jmir.2143}{doi:\nolinkurl{10.2196/jmir.2143}}


\bibitem[Knaust et~al\mbox{.}(2020)]%
        {knaust_virtual_2020}
\bibfield{author}{\bibinfo{person}{Thiemo Knaust}, \bibinfo{person}{Anna
  Felnhofer}, \bibinfo{person}{Oswald~D. Kothgassner}, \bibinfo{person}{Helge
  HÃ¶llmer}, \bibinfo{person}{Robert-Jacek Gorzka}, {and}
  \bibinfo{person}{Holger Schulz}.} \bibinfo{year}{2020}\natexlab{}.
\newblock \showarticletitle{Virtual {Trauma} {Interventions} for the
  {Treatment} of {Post}-traumatic {Stress} {Disorders}: {A} {Scoping}
  {Review}}.
\newblock \bibinfo{journal}{\emph{Frontiers in Psychology}}
  \bibinfo{volume}{11} (\bibinfo{date}{Nov.} \bibinfo{year}{2020}),
  \bibinfo{pages}{562506}.
\newblock
\showISSN{1664-1078}
\href{https://doi.org/10.3389/fpsyg.2020.562506}{doi:\nolinkurl{10.3389/fpsyg.2020.562506}}


\bibitem[Kushniruk and Patel(2004)]%
        {kushniruk_cognitive_2004}
\bibfield{author}{\bibinfo{person}{Andre~W. Kushniruk} {and}
  \bibinfo{person}{Vimla~L. Patel}.} \bibinfo{year}{2004}\natexlab{}.
\newblock \showarticletitle{Cognitive and usability engineering methods for the
  evaluation of clinical information systems}.
\newblock \bibinfo{journal}{\emph{Journal of Biomedical Informatics}}
  \bibinfo{volume}{37}, \bibinfo{number}{1} (\bibinfo{date}{Feb.}
  \bibinfo{year}{2004}), \bibinfo{pages}{56--76}.
\newblock
\showISSN{1532-0464}
\href{https://doi.org/10.1016/j.jbi.2004.01.003}{doi:\nolinkurl{10.1016/j.jbi.2004.01.003}}


\bibitem[Lanius et~al\mbox{.}(2017)]%
        {LANIUS2017109}
\bibfield{author}{\bibinfo{person}{Ruth~A Lanius}, \bibinfo{person}{Daniela
  Rabellino}, \bibinfo{person}{Jenna~E Boyd}, \bibinfo{person}{Sherain
  Harricharan}, \bibinfo{person}{Paul~A Frewen}, {and}
  \bibinfo{person}{Margaret~C McKinnon}.} \bibinfo{year}{2017}\natexlab{}.
\newblock \showarticletitle{The Innate Alarm System in {{PTSD}}: Conscious and
  Subconscious Processing of Threat}.
\newblock \bibinfo{journal}{\emph{Current Opinion in Psychology}}
  \bibinfo{volume}{14} (\bibinfo{year}{2017}), \bibinfo{pages}{109--115}.
\newblock
\showISSN{2352-250X}
\href{https://doi.org/10.1016/j.copsyc.2016.11.006}{doi:\nolinkurl{10.1016/j.copsyc.2016.11.006}}


\bibitem[LeDoux and Pine(2016)]%
        {ledoux2016using}
\bibfield{author}{\bibinfo{person}{Joseph~E LeDoux} {and}
  \bibinfo{person}{Daniel~S Pine}.} \bibinfo{year}{2016}\natexlab{}.
\newblock \showarticletitle{Using neuroscience to help understand fear and
  anxiety: a two-system framework}.
\newblock \bibinfo{journal}{\emph{American journal of psychiatry}}
  \bibinfo{volume}{173}, \bibinfo{number}{11} (\bibinfo{year}{2016}),
  \bibinfo{pages}{1083--1093}.
\newblock
\href{https://doi.org/10.1176/appi.ajp.2016.16030353}{doi:\nolinkurl{10.1176/appi.ajp.2016.16030353}}


\bibitem[Leichsenring et~al\mbox{.}(2024)]%
        {leichsenring2024Borderline}
\bibfield{author}{\bibinfo{person}{Falk Leichsenring}, \bibinfo{person}{Peter
  Fonagy}, \bibinfo{person}{Nikolas Heim}, \bibinfo{person}{Otto~F. Kernberg},
  \bibinfo{person}{Frank Leweke}, \bibinfo{person}{Patrick Luyten},
  \bibinfo{person}{Simone Salzer}, \bibinfo{person}{Carsten Spitzer}, {and}
  \bibinfo{person}{Christiane Steinert}.} \bibinfo{year}{2024}\natexlab{}.
\newblock \showarticletitle{Borderline personality disorder: a comprehensive
  review of diagnosis and clinical presentation, etiology, treatment, and
  current controversies}.
\newblock \bibinfo{journal}{\emph{World Psychiatry}} \bibinfo{volume}{23},
  \bibinfo{number}{1} (\bibinfo{year}{2024}), \bibinfo{pages}{4--25}.
\newblock
\showeprint{https://onlinelibrary.wiley.com/doi/pdf/10.1002/wps.21156}
\href{https://doi.org/10.1002/wps.21156}{doi:\nolinkurl{10.1002/wps.21156}}


\bibitem[Loucks et~al\mbox{.}(2019)]%
        {Loucks2019You}
\bibfield{author}{\bibinfo{person}{Laura Loucks}, \bibinfo{person}{Carly~W
  Yasinski}, \bibinfo{person}{S. Norrholm}, \bibinfo{person}{J. Maples-Keller},
  \bibinfo{person}{L. Post}, \bibinfo{person}{Liza~C Zwiebach},
  \bibinfo{person}{Devika Fiorillo}, \bibinfo{person}{Megan Goodlin},
  \bibinfo{person}{T. JovanoviÄ‡}, \bibinfo{person}{A. Rizzo}, {and}
  \bibinfo{person}{B. Rothbaum}.} \bibinfo{year}{2019}\natexlab{}.
\newblock \showarticletitle{You can do that?!: Feasibility of virtual reality
  exposure therapy in the treatment of PTSD due to military sexual trauma.}
\newblock \bibinfo{journal}{\emph{Journal of anxiety disorders}}
  \bibinfo{volume}{61} (\bibinfo{year}{2019}), \bibinfo{pages}{55--63}.
\newblock
\href{https://doi.org/10.1016/j.janxdis.2018.06.004}{doi:\nolinkurl{10.1016/j.janxdis.2018.06.004}}


\bibitem[Lyssenko et~al\mbox{.}(2018)]%
        {lyssenkoDissociationPsychiatricDisorders2018}
\bibfield{author}{\bibinfo{person}{Lisa Lyssenko}, \bibinfo{person}{Christian
  Schmahl}, \bibinfo{person}{Laura Bockhacker}, \bibinfo{person}{Ruben
  Vonderlin}, \bibinfo{person}{Martin Bohus}, {and} \bibinfo{person}{Nikolaus
  Kleindienst}.} \bibinfo{year}{2018}\natexlab{}.
\newblock \showarticletitle{Dissociation in {{Psychiatric Disorders}}: {{A
  Meta-Analysis}} of {{Studies Using}} the {{Dissociative Experiences Scale}}.}
\newblock \bibinfo{journal}{\emph{The American journal of psychiatry}}
  \bibinfo{volume}{175 1} (\bibinfo{year}{2018}), \bibinfo{pages}{37--46}.
\newblock
\href{https://doi.org/10.1176/appi.ajp.2017.17010025}{doi:\nolinkurl{10.1176/appi.ajp.2017.17010025}}


\bibitem[M.~Shae~Nester and Brand(2022)]%
        {Nester29072022}
\bibfield{author}{\bibinfo{person}{Sarah L.~Hawkins M.~Shae~Nester} {and}
  \bibinfo{person}{Bethany~L. Brand}.} \bibinfo{year}{2022}\natexlab{}.
\newblock \showarticletitle{Barriers to Accessing and Continuing Mental Health
  Treatment among Individuals with Dissociative Symptoms}.
\newblock \bibinfo{journal}{\emph{European Journal of Psychotraumatology}}
  \bibinfo{volume}{13}, \bibinfo{number}{1} (\bibinfo{year}{2022}),
  \bibinfo{pages}{2031594}.
\newblock
\showeprint{https://doi.org/10.1080/20008198.2022.2031594}
\href{https://doi.org/10.1080/20008198.2022.2031594}{doi:\nolinkurl{10.1080/20008198.2022.2031594}}


\bibitem[Maercker(2003)]%
        {maercker2003posttraumatische}
\bibfield{author}{\bibinfo{person}{Andreas Maercker}.}
  \bibinfo{year}{2003}\natexlab{}.
\newblock \showarticletitle{Posttraumatische-Stress-Skala-10 ({{PTSS-10}})}.
\newblock \bibinfo{journal}{\emph{Angstdiagnostik--Grundlagen und
  Testverfahren}} (\bibinfo{year}{2003}), \bibinfo{pages}{401--403}.
\newblock


\bibitem[Maercker et~al\mbox{.}(2022)]%
        {maercker2022complex}
\bibfield{author}{\bibinfo{person}{Andreas Maercker}, \bibinfo{person}{Marylene
  Cloitre}, \bibinfo{person}{Rahel Bachem}, \bibinfo{person}{Yolanda~R
  Schlumpf}, \bibinfo{person}{Brigitte Khoury}, \bibinfo{person}{Caitlin
  Hitchcock}, {and} \bibinfo{person}{Martin Bohus}.}
  \bibinfo{year}{2022}\natexlab{}.
\newblock \showarticletitle{Complex post-traumatic stress disorder}.
\newblock \bibinfo{journal}{\emph{The lancet}} \bibinfo{volume}{400},
  \bibinfo{number}{10345} (\bibinfo{year}{2022}), \bibinfo{pages}{60--72}.
\newblock


\bibitem[Maercker and Sch{\"u}tzwohl(1998)]%
        {maercker1998erfassung}
\bibfield{author}{\bibinfo{person}{Andreas Maercker} {and}
  \bibinfo{person}{Matthias Sch{\"u}tzwohl}.} \bibinfo{year}{1998}\natexlab{}.
\newblock \showarticletitle{Erfassung von Psychischen Belastungsfolgen: {{Die}}
  Impact of Event Skala-Revidierte Version ({{IES-r}}).}
\newblock \bibinfo{journal}{\emph{Diagnostica}} (\bibinfo{year}{1998}).
\newblock


\bibitem[McDonald et~al\mbox{.}(2019)]%
        {mcdonald_reliability_2019}
\bibfield{author}{\bibinfo{person}{Nora McDonald}, \bibinfo{person}{Sarita
  Schoenebeck}, {and} \bibinfo{person}{Andrea Forte}.}
  \bibinfo{year}{2019}\natexlab{}.
\newblock \showarticletitle{Reliability and {Inter}-rater {Reliability} in
  {Qualitative} {Research}: {Norms} and {Guidelines} for {CSCW} and {HCI}
  {Practice}}.
\newblock \bibinfo{journal}{\emph{Proceedings of the ACM on Human-Computer
  Interaction}} \bibinfo{volume}{3}, \bibinfo{number}{CSCW}
  (\bibinfo{date}{Nov.} \bibinfo{year}{2019}), \bibinfo{pages}{1--23}.
\newblock
\showISSN{2573-0142}
\href{https://doi.org/10.1145/3359174}{doi:\nolinkurl{10.1145/3359174}}


\bibitem[McKim(2023)]%
        {mckim2023-dy}
\bibfield{author}{\bibinfo{person}{Courtney McKim}.}
  \bibinfo{year}{2023}\natexlab{}.
\newblock \showarticletitle{Meaningful {Member-Checking}: A Structured Approach
  to {Member-Checking}}.
\newblock \bibinfo{journal}{\emph{American Journal of Qualitative Research}}
  \bibinfo{volume}{7}, \bibinfo{number}{2} (\bibinfo{year}{2023}),
  \bibinfo{pages}{41--52}.
\newblock


\bibitem[M.D(2015)]%
        {md_body_2015}
\bibfield{author}{\bibinfo{person}{Bessel van der~Kolk M.D}.}
  \bibinfo{year}{2015}\natexlab{}.
\newblock \bibinfo{booktitle}{\emph{The {Body} {Keeps} the {Score}: {Brain},
  {Mind}, and {Body} in the {Healing} of {Trauma}}}.
\newblock \bibinfo{publisher}{Penguin Books}, \bibinfo{address}{New York, NY}.
\newblock
\showISBNx{978-0-14-312774-1}


\bibitem[Meggelen et~al\mbox{.}(2022)]%
        {meggelenRandomizedControlledTrial2022}
\bibfield{author}{\bibinfo{person}{Marieke Meggelen},
  \bibinfo{person}{Nexhmedin Morina}, \bibinfo{person}{Colin {Van der Heiden}},
  \bibinfo{person}{Willem-Paul Brinkman}, \bibinfo{person}{Iris Yocarini},
  \bibinfo{person}{Myrthe Tielman}, \bibinfo{person}{Jan Rodenburg},
  \bibinfo{person}{Elisa Van~Ee}, \bibinfo{person}{Kevin Schie},
  \bibinfo{person}{Marijke Broekman}, {and} \bibinfo{person}{Ingmar Franken}.}
  \bibinfo{year}{2022}\natexlab{}.
\newblock \showarticletitle{A Randomized Controlled Trial to Pilot the Efficacy
  of a Computer-Based Intervention with Elements of Virtual Reality and Limited
  Therapist Assistance for the Treatment of Post-Traumatic Stress Disorder}.
\newblock \bibinfo{journal}{\emph{Frontiers in Digital Health}}
  \bibinfo{volume}{4} (\bibinfo{date}{Oct.} \bibinfo{year}{2022}),
  \bibinfo{pages}{974668}.
\newblock
\href{https://doi.org/10.3389/fdgth.2022.974668}{doi:\nolinkurl{10.3389/fdgth.2022.974668}}


\bibitem[Messman and Long(2003)]%
        {messman2003Childhood}
\bibfield{author}{\bibinfo{person}{Terri Messman} {and}
  \bibinfo{person}{Patricia Long}.} \bibinfo{year}{2003}\natexlab{}.
\newblock \showarticletitle{The role of childhood sexual abuse sequelae in the
  sexual revictimization of women: An empirical review and theoretical
  reformulation}.
\newblock \bibinfo{journal}{\emph{Clinical psychology review}}
  \bibinfo{volume}{23} (\bibinfo{date}{08} \bibinfo{year}{2003}),
  \bibinfo{pages}{537--71}.
\newblock
\href{https://doi.org/10.1016/S0272-7358(02)00203-9}{doi:\nolinkurl{10.1016/S0272-7358(02)00203-9}}


\bibitem[Moncur(2013)]%
        {moncur_emotional_2013}
\bibfield{author}{\bibinfo{person}{Wendy Moncur}.}
  \bibinfo{year}{2013}\natexlab{}.
\newblock \showarticletitle{The emotional wellbeing of researchers:
  considerations for practice}. In \bibinfo{booktitle}{\emph{Proceedings of the
  {SIGCHI} {Conference} on {Human} {Factors} in {Computing} {Systems}}}
  \emph{(\bibinfo{series}{{CHI} '13})}. \bibinfo{publisher}{Association for
  Computing Machinery}, \bibinfo{address}{New York, NY, USA},
  \bibinfo{pages}{1883--1890}.
\newblock
\showISBNx{978-1-4503-1899-0}
\href{https://doi.org/10.1145/2470654.2466248}{doi:\nolinkurl{10.1145/2470654.2466248}}


\bibitem[M{\"u}hlberger et~al\mbox{.}(2001)]%
        {muhlbergerRepeatedExposureFlight2001}
\bibfield{author}{\bibinfo{person}{Andreas M{\"u}hlberger},
  \bibinfo{person}{Martin Herrmann}, \bibinfo{person}{Georg Wiedemann},
  \bibinfo{person}{Heiner Ellgring}, {and} \bibinfo{person}{Paul Pauli}.}
  \bibinfo{year}{2001}\natexlab{}.
\newblock \showarticletitle{Repeated Exposure of Flight Phobics to Flights in
  Virtual Reality}.
\newblock \bibinfo{journal}{\emph{Behaviour Research and Therapy}}
  \bibinfo{volume}{39} (\bibinfo{date}{Oct.} \bibinfo{year}{2001}),
  \bibinfo{pages}{1033--1050}.
\newblock
\href{https://doi.org/10.1016/S0005-7967(00)00076-0}{doi:\nolinkurl{10.1016/S0005-7967(00)00076-0}}


\bibitem[Mullins et~al\mbox{.}(2014)]%
        {mullins_patient-centeredness_2014}
\bibfield{author}{\bibinfo{person}{C.~Daniel Mullins},
  \bibinfo{person}{Joseph~E. Vandigo}, \bibinfo{person}{Jason Zheng}, {and}
  \bibinfo{person}{Paul Wicks}.} \bibinfo{year}{2014}\natexlab{}.
\newblock \showarticletitle{Patient-centeredness in the design of clinical
  trials}.
\newblock \bibinfo{journal}{\emph{Value in health : the journal of the
  International Society for Pharmacoeconomics and Outcomes Research}}
  \bibinfo{volume}{17}, \bibinfo{number}{4} (\bibinfo{date}{June}
  \bibinfo{year}{2014}), \bibinfo{pages}{471--475}.
\newblock
\showISSN{1098-3015}
\href{https://doi.org/10.1016/j.jval.2014.02.012}{doi:\nolinkurl{10.1016/j.jval.2014.02.012}}


\bibitem[Murray et~al\mbox{.}(2007)]%
        {murrayAbsorptionDissociationLocus2007}
\bibfield{author}{\bibinfo{person}{Craig~D. Murray}, \bibinfo{person}{Jezz
  Fox}, {and} \bibinfo{person}{Steve Pettifer}.}
  \bibinfo{year}{2007}\natexlab{}.
\newblock \showarticletitle{Absorption, Dissociation, Locus of Control and
  Presence in Virtual Reality}.
\newblock \bibinfo{journal}{\emph{Computers in Human Behavior}}
  \bibinfo{volume}{23}, \bibinfo{number}{3} (\bibinfo{date}{May}
  \bibinfo{year}{2007}), \bibinfo{pages}{1347--1354}.
\newblock
\showISSN{0747-5632}
\href{https://doi.org/10.1016/j.chb.2004.12.010}{doi:\nolinkurl{10.1016/j.chb.2004.12.010}}


\bibitem[Neuner et~al\mbox{.}(2021)]%
        {neunerNarrativeExpositionstherapieNET2021}
\bibfield{author}{\bibinfo{person}{Frank Neuner}, \bibinfo{person}{Claudia
  Catani}, {and} \bibinfo{person}{Maggie Schauer}.}
  \bibinfo{year}{2021}\natexlab{}.
\newblock \bibinfo{booktitle}{\emph{{Narrative Expositionstherapie (NET)}}
  (\bibinfo{edition}{1} ed.)}. Vol.~\bibinfo{volume}{83}.
\newblock \bibinfo{publisher}{Hogrefe Verlag GmbH \& Co. KG}.
\newblock
\showISBNx{978-3-8444-3097-4}
\href{https://doi.org/10.1026/03097-000}{doi:\nolinkurl{10.1026/03097-000}}


\bibitem[Organization({[}n.\,d.{]}a)]%
        {worldhealthorganizationICD10Version2019}
\bibfield{author}{\bibinfo{person}{World~Health Organization}.}
  \bibinfo{year}{[n.\,d.]}\natexlab{a}.
\newblock \bibinfo{title}{{{ICD-10 Version}}:2019}.
\newblock \bibinfo{howpublished}{https://icd.who.int/browse10/2019/en}.
\newblock


\bibitem[Organization({[}n.\,d.{]}b)]%
        {worldhealthorganizationICD11}
\bibfield{author}{\bibinfo{person}{World~Health Organization}.}
  \bibinfo{year}{[n.\,d.]}\natexlab{b}.
\newblock \bibinfo{title}{{{ICD-11}}}.
\newblock \bibinfo{howpublished}{https://icd.who.int/en}.
\newblock


\bibitem[Pataranutaporn et~al\mbox{.}(2023)]%
        {pataranutaporn_living_2023}
\bibfield{author}{\bibinfo{person}{Pat Pataranutaporn},
  \bibinfo{person}{{https://orcid.org/0000-0002-1879-7340}},
  \bibinfo{person}{{View Profile}}, \bibinfo{person}{Valdemar Danry},
  \bibinfo{person}{{https://orcid.org/0000-0001-5225-0077}},
  \bibinfo{person}{{View Profile}}, \bibinfo{person}{Lancelot Blanchard},
  \bibinfo{person}{{https://orcid.org/0000-0003-1580-3116}},
  \bibinfo{person}{{View Profile}}, \bibinfo{person}{Lavanay Thakral},
  \bibinfo{person}{{https://orcid.org/0000-0001-7078-5331}},
  \bibinfo{person}{{View Profile}}, \bibinfo{person}{Naoki Ohsugi},
  \bibinfo{person}{{https://orcid.org/0000-0002-4489-4013}},
  \bibinfo{person}{{View Profile}}, \bibinfo{person}{Pattie Maes},
  \bibinfo{person}{{https://orcid.org/0000-0002-7722-6038}},
  \bibinfo{person}{{View Profile}}, \bibinfo{person}{Misha Sra},
  \bibinfo{person}{{https://orcid.org/0000-0001-8154-8518}}, {and}
  \bibinfo{person}{{View Profile}}.} \bibinfo{year}{2023}\natexlab{}.
\newblock \showarticletitle{Living {Memories}: {AI}-{Generated} {Characters} as
  {Digital} {Mementos}}.
\newblock In \bibinfo{booktitle}{\emph{Proceedings of the 28th {International}
  {Conference} on {Intelligent} {User} {Interfaces}}}.
  \bibinfo{pages}{889--901}.
\newblock
\showISBNx{9798400701061}
\href{https://doi.org/10.1145/3581641.3584065}{doi:\nolinkurl{10.1145/3581641.3584065}}


\bibitem[Posada-Quintero and Chon(2020)]%
        {posadaquintero2020eda}
\bibfield{author}{\bibinfo{person}{Hugo~F. Posada-Quintero} {and}
  \bibinfo{person}{Ki~H. Chon}.} \bibinfo{year}{2020}\natexlab{}.
\newblock \showarticletitle{Innovations in Electrodermal Activity Data
  Collection and Signal Processing: A Systematic Review}.
\newblock \bibinfo{journal}{\emph{Sensors}} \bibinfo{volume}{20},
  \bibinfo{number}{2} (\bibinfo{year}{2020}).
\newblock
\showISSN{1424-8220}
\href{https://doi.org/10.3390/s20020479}{doi:\nolinkurl{10.3390/s20020479}}


\bibitem[Ppali et~al\mbox{.}(2025a)]%
        {ppali_creating_2025}
\bibfield{author}{\bibinfo{person}{Sophia Ppali}, \bibinfo{person}{Ethan
  Cheung}, \bibinfo{person}{Alexandra Covaci}, \bibinfo{person}{Wan-Jou She},
  {and} \bibinfo{person}{Chee~Siang Ang}.} \bibinfo{year}{2025}\natexlab{a}.
\newblock \showarticletitle{Creating with {Care}: {Co}-{Designing} {Immersive}
  {Experiences} through {Art}-{Making} with {People} {Living} with {Dementia}}.
  In \bibinfo{booktitle}{\emph{Proceedings of the 2025 {CHI} {Conference} on
  {Human} {Factors} in {Computing} {Systems}}} \emph{(\bibinfo{series}{{CHI}
  '25})}. \bibinfo{publisher}{Association for Computing Machinery},
  \bibinfo{address}{New York, NY, USA}, \bibinfo{pages}{1--18}.
\newblock
\showISBNx{9798400713941}
\href{https://doi.org/10.1145/3706598.3714101}{doi:\nolinkurl{10.1145/3706598.3714101}}


\bibitem[Ppali et~al\mbox{.}(2025b)]%
        {ppali_cite_2025}
\bibfield{author}{\bibinfo{person}{Sophia Ppali}, \bibinfo{person}{Marios
  Constantinides}, \bibinfo{person}{Fotis Liarokapis}, \bibinfo{person}{Jaydon
  Farao}, \bibinfo{person}{Soraya~S. Anvari}, \bibinfo{person}{MinYoung Yoo},
  \bibinfo{person}{Ferran Altarriba~Bertran}, \bibinfo{person}{Shannon
  Rodgers}, \bibinfo{person}{Jihae Han}, \bibinfo{person}{Rina~R. Wehbe},
  \bibinfo{person}{Margot Brereton}, {and} \bibinfo{person}{Alexandra Covaci}.}
  \bibinfo{year}{2025}\natexlab{b}.
\newblock \showarticletitle{Cite {Your} {Well}-being {First}: {What} {Happens}
  {When} {Personal} {Life}, {Mental} {Health}, and {HCI} {Research} {Become}
  {Entangled}?}. In \bibinfo{booktitle}{\emph{Companion {Publication} of the
  2025 {ACM} {Designing} {Interactive} {Systems} {Conference}}}
  \emph{(\bibinfo{series}{{DIS} '25 {Companion}})}.
  \bibinfo{publisher}{Association for Computing Machinery},
  \bibinfo{address}{New York, NY, USA}, \bibinfo{pages}{52--56}.
\newblock
\showISBNx{9798400714863}
\href{https://doi.org/10.1145/3715668.3734181}{doi:\nolinkurl{10.1145/3715668.3734181}}


\bibitem[Randazzo and Ammari(2023)]%
        {randazzoIfSomeoneDownvoted2023}
\bibfield{author}{\bibinfo{person}{Casey Randazzo} {and}
  \bibinfo{person}{Tawfiq Ammari}.} \bibinfo{year}{2023}\natexlab{}.
\newblock \showarticletitle{``{{If Someone Downvoted My Posts}}---{{That}}'d
  {{Be}} the {{End}} of the {{World}}'': {{Designing Safer Online Spaces}} for
  {{Trauma Survivors}}}. In \bibinfo{booktitle}{\emph{Proceedings of the 2023
  {{CHI Conference}} on {{Human Factors}} in {{Computing Systems}}}}
  \emph{(\bibinfo{series}{{{CHI}} '23})}. \bibinfo{publisher}{Association for
  Computing Machinery}, \bibinfo{address}{New York, NY, USA}.
\newblock
\showISBNx{978-1-4503-9421-5}
\href{https://doi.org/10.1145/3544548.3581453}{doi:\nolinkurl{10.1145/3544548.3581453}}


\bibitem[Randazzo et~al\mbox{.}(2023)]%
        {randazzo2023TraumaInformedDesign}
\bibfield{author}{\bibinfo{person}{Casey Randazzo}, \bibinfo{person}{Carol~F.
  Scott}, \bibinfo{person}{Rosanna Bellini}, \bibinfo{person}{Tawfiq Ammari},
  \bibinfo{person}{Michael~Ann Devito}, \bibinfo{person}{Bryan Semaan}, {and}
  \bibinfo{person}{Nazanin Andalibi}.} \bibinfo{year}{2023}\natexlab{}.
\newblock \showarticletitle{Trauma-Informed Design: A Collaborative Approach to
  Building Safer Online Spaces}. In \bibinfo{booktitle}{\emph{Companion
  Publication of the 2023 Conference on Computer Supported Cooperative Work and
  Social Computing}} (Minneapolis, MN, USA) \emph{(\bibinfo{series}{CSCW '23
  Companion})}. \bibinfo{publisher}{Association for Computing Machinery},
  \bibinfo{address}{New York, NY, USA}, \bibinfo{pages}{470â€“475}.
\newblock
\showISBNx{9798400701290}
\href{https://doi.org/10.1145/3584931.3611277}{doi:\nolinkurl{10.1145/3584931.3611277}}


\bibitem[Rizzo et~al\mbox{.}(2006)]%
        {rizzo2006usercentered}
\bibfield{author}{\bibinfo{person}{Albert Rizzo}, \bibinfo{person}{Ken Graap},
  \bibinfo{person}{Jarrell Pair}, \bibinfo{person}{Greg Reger},
  \bibinfo{person}{Anton Treskunov}, {and} \bibinfo{person}{Thomas Parsons}.}
  \bibinfo{year}{2006}\natexlab{}.
\newblock \showarticletitle{User-Centered Design Driven Development of a VR
  Therapy Application for Iraq War Combat-Related Post Traumatic Stress
  Disorder}.
\newblock \bibinfo{journal}{\emph{Proceedings of the 2006 International
  Conference on Disability, Virtual Reality and Associated Technology}}.
\newblock
\href{https://doi.org/10.13140/RG.2.1.2757.9368}{doi:\nolinkurl{10.13140/RG.2.1.2757.9368}}


\bibitem[Rizzo et~al\mbox{.}(2014)]%
        {rizzoVirtualRealityExposure2014}
\bibfield{author}{\bibinfo{person}{Albert Rizzo}, \bibinfo{person}{Arno
  Hartholt}, \bibinfo{person}{Mario Grimani}, \bibinfo{person}{Andrew Leeds},
  {and} \bibinfo{person}{Matt Liewer}.} \bibinfo{year}{2014}\natexlab{}.
\newblock \showarticletitle{Virtual {{Reality Exposure Therapy}} for
  {{Combat-Related Posttraumatic Stress Disorder}}}.
\newblock \bibinfo{journal}{\emph{Computer}} \bibinfo{volume}{47},
  \bibinfo{number}{7} (\bibinfo{date}{July} \bibinfo{year}{2014}),
  \bibinfo{pages}{31--37}.
\newblock
\showISSN{1558-0814}
\href{https://doi.org/10.1109/MC.2014.199}{doi:\nolinkurl{10.1109/MC.2014.199}}


\bibitem[Rizzo et~al\mbox{.}(2021)]%
        {rizzo2021fromcombat}
\bibfield{author}{\bibinfo{person}{Albert Rizzo}, \bibinfo{person}{Arno
  Hartholt}, {and} \bibinfo{person}{Sharon Mozgai}.}
  \bibinfo{year}{2021}\natexlab{}.
\newblock \showarticletitle{From Combat to {COVID}-19â€“Managing the Impact of
  Trauma Using Virtual Reality}.
\newblock \bibinfo{journal}{\emph{Journal of Technology in Human Services}}
  \bibinfo{volume}{10.} (\bibinfo{year}{2021}).
\newblock


\bibitem[Rizzo and Shilling(2017)]%
        {rizzo_clinical_2017}
\bibfield{author}{\bibinfo{person}{Albert~â€˜Skipâ€™ Rizzo} {and}
  \bibinfo{person}{Russell Shilling}.} \bibinfo{year}{2017}\natexlab{}.
\newblock \showarticletitle{Clinical {Virtual} {Reality} tools to advance the
  prevention, assessment, and treatment of {PTSD}}.
\newblock \bibinfo{journal}{\emph{European Journal of Psychotraumatology}}
  \bibinfo{volume}{8}, \bibinfo{number}{sup5} (\bibinfo{date}{Jan.}
  \bibinfo{year}{2017}), \bibinfo{pages}{1414560}.
\newblock
\showISSN{2000-8066}
\href{https://doi.org/10.1080/20008198.2017.1414560}{doi:\nolinkurl{10.1080/20008198.2017.1414560}}


\bibitem[Rothbaum et~al\mbox{.}(1995)]%
        {rothbaumVirtualRealityGraded1995}
\bibfield{author}{\bibinfo{person}{Barbara~Olasov Rothbaum},
  \bibinfo{person}{Larry~F. Hodges}, \bibinfo{person}{Rob Kooper},
  \bibinfo{person}{Dan Opdyke}, \bibinfo{person}{James~S. Williford}, {and}
  \bibinfo{person}{Max North}.} \bibinfo{year}{1995}\natexlab{}.
\newblock \showarticletitle{Virtual Reality Graded Exposure in the Treatment of
  Acrophobia: {{A}} Case Report}.
\newblock \bibinfo{journal}{\emph{Behavior Therapy}} \bibinfo{volume}{26},
  \bibinfo{number}{3} (\bibinfo{date}{June} \bibinfo{year}{1995}),
  \bibinfo{pages}{547--554}.
\newblock
\showISSN{0005-7894}
\href{https://doi.org/10.1016/S0005-7894(05)80100-5}{doi:\nolinkurl{10.1016/S0005-7894(05)80100-5}}


\bibitem[Satinsky et~al\mbox{.}(2021)]%
        {satinsky_systematic_2021}
\bibfield{author}{\bibinfo{person}{Emily~N. Satinsky}, \bibinfo{person}{Tomoki
  Kimura}, \bibinfo{person}{Mathew~V. Kiang}, \bibinfo{person}{Rediet Abebe},
  \bibinfo{person}{Scott Cunningham}, \bibinfo{person}{Hedwig Lee},
  \bibinfo{person}{Xiaofei Lin}, \bibinfo{person}{Cindy~H. Liu},
  \bibinfo{person}{Igor Rudan}, \bibinfo{person}{Srijan Sen},
  \bibinfo{person}{Mark Tomlinson}, \bibinfo{person}{Miranda Yaver}, {and}
  \bibinfo{person}{Alexander~C. Tsai}.} \bibinfo{year}{2021}\natexlab{}.
\newblock \showarticletitle{Systematic review and meta-analysis of depression,
  anxiety, and suicidal ideation among {Ph}.{D}. students}.
\newblock \bibinfo{journal}{\emph{Scientific Reports}} \bibinfo{volume}{11},
  \bibinfo{number}{1} (\bibinfo{year}{2021}), \bibinfo{pages}{14370}.
\newblock


\bibitem[Sch{\"a}fer et~al\mbox{.}(2019)]%
        {12schafer2019s3}
\bibfield{author}{\bibinfo{person}{Ingo Sch{\"a}fer}, \bibinfo{person}{Ursula
  Gast}, \bibinfo{person}{Arne Hofmann}, \bibinfo{person}{Christine
  Knaevelsrud}, \bibinfo{person}{Astrid Lampe}, \bibinfo{person}{Peter
  Liebermann}, \bibinfo{person}{Annett Lotzin}, \bibinfo{person}{Andreas
  Maercker}, \bibinfo{person}{Rita Rosner}, {and} \bibinfo{person}{Wolfgang
  W{\"o}ller}.} \bibinfo{year}{2019}\natexlab{}.
\newblock \bibinfo{booktitle}{\emph{S3-leitlinie posttraumatische
  belastungsst{\"o}rung}}.
\newblock \bibinfo{publisher}{Springer}.
\newblock


\bibitem[Schubert et~al\mbox{.}(2001)]%
        {schubertExperiencePresenceFactor2001}
\bibfield{author}{\bibinfo{person}{Thomas Schubert}, \bibinfo{person}{Frank
  Friedmann}, {and} \bibinfo{person}{Holger Regenbrecht}.}
  \bibinfo{year}{2001}\natexlab{}.
\newblock \showarticletitle{The {{Experience}} of {{Presence}}: {{Factor
  Analytic Insights}}}.
\newblock \bibinfo{journal}{\emph{Presence}}  \bibinfo{volume}{10}
  (\bibinfo{date}{June} \bibinfo{year}{2001}), \bibinfo{pages}{266--281}.
\newblock
\href{https://doi.org/10.1162/105474601300343603}{doi:\nolinkurl{10.1162/105474601300343603}}


\bibitem[Scott et~al\mbox{.}(2023)]%
        {scott2023SocialMedia}
\bibfield{author}{\bibinfo{person}{Carol~F Scott}, \bibinfo{person}{Gabriela
  Marcu}, \bibinfo{person}{Riana~Elyse Anderson}, \bibinfo{person}{Mark~W
  Newman}, {and} \bibinfo{person}{Sarita Schoenebeck}.}
  \bibinfo{year}{2023}\natexlab{}.
\newblock \showarticletitle{Trauma-Informed Social Media: Towards Solutions for
  Reducing and Healing Online Harm}. In \bibinfo{booktitle}{\emph{Proceedings
  of the 2023 CHI Conference on Human Factors in Computing Systems}} (Hamburg,
  Germany) \emph{(\bibinfo{series}{CHI '23})}. \bibinfo{publisher}{Association
  for Computing Machinery}, \bibinfo{address}{New York, NY, USA}, Article
  \bibinfo{articleno}{341}, \bibinfo{numpages}{20}~pages.
\newblock
\showISBNx{9781450394215}
\href{https://doi.org/10.1145/3544548.3581512}{doi:\nolinkurl{10.1145/3544548.3581512}}


\bibitem[Shalev et~al\mbox{.}(2024)]%
        {shalev2024neurobiology}
\bibfield{author}{\bibinfo{person}{Arieh Shalev}, \bibinfo{person}{Dayeon Cho},
  {and} \bibinfo{person}{Charles~R Marmar}.} \bibinfo{year}{2024}\natexlab{}.
\newblock \showarticletitle{Neurobiology and Treatment of Posttraumatic Stress
  Disorder}.
\newblock \bibinfo{journal}{\emph{American Journal of Psychiatry}}
  \bibinfo{volume}{181}, \bibinfo{number}{8} (\bibinfo{year}{2024}),
  \bibinfo{pages}{705--719}.
\newblock
\href{https://doi.org/10.1176/appi.ajp.20240536}{doi:\nolinkurl{10.1176/appi.ajp.20240536}}


\bibitem[Sherrill et~al\mbox{.}(2020)]%
        {Sherrill2020-mn}
\bibfield{author}{\bibinfo{person}{A~M Sherrill}, \bibinfo{person}{J~R
  Goodnight}, \bibinfo{person}{M~S Burton}, {and} \bibinfo{person}{B~O
  Rothbaum}.} \bibinfo{year}{2020}\natexlab{}.
\newblock \showarticletitle{Use of virtual reality exposure therapy for trauma-
  and anxiety-related disorders}.
\newblock  (\bibinfo{year}{2020}), \bibinfo{pages}{75--91}.
\newblock


\bibitem[Sherrill et~al\mbox{.}(2025)]%
        {Sherrill2025-fo}
\bibfield{author}{\bibinfo{person}{Andrew~M Sherrill}, \bibinfo{person}{Natalie
  Hellman}, {and} \bibinfo{person}{Barbara~O Rothbaum}.}
  \bibinfo{year}{2025}\natexlab{}.
\newblock \showarticletitle{Virtual reality to enhance engagement in
  exposure-based treatments}.
\newblock In \bibinfo{booktitle}{\emph{{CBT}: Science Into Practice}}.
  \bibinfo{publisher}{Springer Nature Switzerland}, \bibinfo{address}{Cham},
  \bibinfo{pages}{129--147}.
\newblock


\bibitem[Sherrill et~al\mbox{.}(2019)]%
        {Sherrill2019-io}
\bibfield{author}{\bibinfo{person}{A~M Sherrill}, \bibinfo{person}{A~O
  Rothbaum}, \bibinfo{person}{L~B Mcsweeney}, {and} \bibinfo{person}{B~O
  Rothbaum}.} \bibinfo{year}{2019}\natexlab{}.
\newblock \showarticletitle{Virtual reality exposure therapy for posttraumatic
  stress disorder}.
\newblock \bibinfo{journal}{\emph{Psychiatric Annals}} \bibinfo{volume}{49},
  \bibinfo{number}{8} (\bibinfo{year}{2019}), \bibinfo{pages}{343--347}.
\newblock


\bibitem[Sherrill and Rothbaum(2022)]%
        {Sherrill2022-vl}
\bibfield{author}{\bibinfo{person}{A~M Sherrill} {and} \bibinfo{person}{B~O
  Rothbaum}.} \bibinfo{year}{2022}\natexlab{}.
\newblock \showarticletitle{Virtual reality exposure therapy}.
\newblock In \bibinfo{booktitle}{\emph{Encyclopedia of mental health
  interventions}}, \bibfield{editor}{\bibinfo{person}{E~H Taylor}} (Ed.).
  \bibinfo{publisher}{Springer}, \bibinfo{address}{Nature},
  \bibinfo{pages}{1--12}.
\newblock


\bibitem[Simmons(2017)]%
        {simmons_critical_2017}
\bibfield{author}{\bibinfo{person}{Nathaniel Simmons}.}
  \bibinfo{year}{2017}\natexlab{}.
\newblock \showarticletitle{Critical {Incident} {Method}}.
\newblock In \bibinfo{booktitle}{\emph{The {SAGE} {Encyclopedia} of
  {Communication} {Research} {Methods}}}. \bibinfo{publisher}{SAGE
  Publications, Inc}, \bibinfo{pages}{300--302}.
\newblock
\showISBNx{978-1-4833-8141-1}
\href{https://doi.org/10.4135/9781483381411}{doi:\nolinkurl{10.4135/9781483381411}}


\bibitem[Singh et~al\mbox{.}(2025)]%
        {singh_exploring_2025}
\bibfield{author}{\bibinfo{person}{Aneesha Singh},
  \bibinfo{person}{Martin~Johannes Dechant}, \bibinfo{person}{Dilisha Patel},
  \bibinfo{person}{Ewan Soubutts}, \bibinfo{person}{Giulia Barbareschi},
  \bibinfo{person}{Amid Ayobi}, {and} \bibinfo{person}{Nikki Newhouse}.}
  \bibinfo{year}{2025}\natexlab{}.
\newblock \showarticletitle{Exploring {Positionality} in {HCI}: {Perspectives},
  {Trends}, and {Challenges}}. In \bibinfo{booktitle}{\emph{Proceedings of the
  2025 {CHI} {Conference} on {Human} {Factors} in {Computing} {Systems}}}.
  \bibinfo{publisher}{ACM}, \bibinfo{address}{Yokohama Japan},
  \bibinfo{pages}{1--18}.
\newblock
\showISBNx{9798400713941}
\href{https://doi.org/10.1145/3706598.3713280}{doi:\nolinkurl{10.1145/3706598.3713280}}


\bibitem[Slater(2018)]%
        {slater_immersion_2018}
\bibfield{author}{\bibinfo{person}{Mel Slater}.}
  \bibinfo{year}{2018}\natexlab{}.
\newblock \showarticletitle{Immersion and the illusion of presence in virtual
  reality}.
\newblock \bibinfo{journal}{\emph{British Journal of Psychology}}
  \bibinfo{volume}{109}, \bibinfo{number}{3} (\bibinfo{date}{Aug.}
  \bibinfo{year}{2018}), \bibinfo{pages}{431--433}.
\newblock
\showISSN{0007-1269, 2044-8295}
\href{https://doi.org/10.1111/bjop.12305}{doi:\nolinkurl{10.1111/bjop.12305}}


\bibitem[Smeets et~al\mbox{.}(2010)]%
        {smeets2010Autobiographical}
\bibfield{author}{\bibinfo{person}{Tom Smeets}, \bibinfo{person}{Timo
  Giesbrecht}, \bibinfo{person}{Linsey Raymaekers}, \bibinfo{person}{Julia
  Shaw}, {and} \bibinfo{person}{Harald Merckelbach}.}
  \bibinfo{year}{2010}\natexlab{}.
\newblock \showarticletitle{Autobiographical integration of trauma memories and
  repressive coping predict post-traumatic stress symptoms in undergraduate
  students}.
\newblock \bibinfo{journal}{\emph{Clinical Psychology \& Psychotherapy}}
  \bibinfo{volume}{17}, \bibinfo{number}{3} (\bibinfo{year}{2010}),
  \bibinfo{pages}{211--218}.
\newblock
\showeprint{https://onlinelibrary.wiley.com/doi/pdf/10.1002/cpp.644}
\href{https://doi.org/10.1002/cpp.644}{doi:\nolinkurl{10.1002/cpp.644}}


\bibitem[Stamm(2010)]%
        {stamm2010concise}
\bibfield{author}{\bibinfo{person}{Beth Stamm}.}
  \bibinfo{year}{2010}\natexlab{}.
\newblock \showarticletitle{The concise manual for the professional quality of
  life scale}.
\newblock  (\bibinfo{year}{2010}).
\newblock


\bibitem[Stevens and Sherrill(2021)]%
        {stevens2021clinician}
\bibfield{author}{\bibinfo{person}{Trevor~M Stevens} {and}
  \bibinfo{person}{Andrew~M Sherrill}.} \bibinfo{year}{2021}\natexlab{}.
\newblock \showarticletitle{A clinicianâ€™s introduction to 360Â° video for
  exposure therapy.}
\newblock \bibinfo{journal}{\emph{Translational Issues in Psychological
  Science}} \bibinfo{volume}{7}, \bibinfo{number}{3} (\bibinfo{year}{2021}),
  \bibinfo{pages}{261}.
\newblock


\bibitem[Torrisi(2013)]%
        {torrisi_academic_2013}
\bibfield{author}{\bibinfo{person}{Benedetto Torrisi}.}
  \bibinfo{year}{2013}\natexlab{}.
\newblock \showarticletitle{Academic productivity correlated with well-being at
  work}.
\newblock \bibinfo{journal}{\emph{Scientometrics}} \bibinfo{volume}{94},
  \bibinfo{number}{2} (\bibinfo{year}{2013}), \bibinfo{pages}{801--815}.
\newblock


\bibitem[{van der Kolk} and Fisler(1995)]%
        {vanderkolkDissociationFragmentaryNature1995}
\bibfield{author}{\bibinfo{person}{Bessel~A. {van der Kolk}} {and}
  \bibinfo{person}{Rita Fisler}.} \bibinfo{year}{1995}\natexlab{}.
\newblock \showarticletitle{Dissociation and the Fragmentary Nature of
  Traumatic Memories: {{Overview}} and Exploratory Study}.
\newblock \bibinfo{journal}{\emph{Journal of Traumatic Stress}}
  \bibinfo{volume}{8}, \bibinfo{number}{4} (\bibinfo{date}{Oct.}
  \bibinfo{year}{1995}), \bibinfo{pages}{505--525}.
\newblock
\showISSN{1573-6598}
\href{https://doi.org/10.1007/BF02102887}{doi:\nolinkurl{10.1007/BF02102887}}


\bibitem[{van Gelderen} et~al\mbox{.}(2018)]%
        {gelderen2018Innovative}
\bibfield{author}{\bibinfo{person}{Marieke~J. {van Gelderen}},
  \bibinfo{person}{Mirjam~J. Nijdam}, {and} \bibinfo{person}{Eric Vermetten}.}
  \bibinfo{year}{2018}\natexlab{}.
\newblock \showarticletitle{An Innovative Framework for Delivering
  Psychotherapy to Patients with Treatment-Resistant Posttraumatic Stress
  Disorder: {{Rationale}} for Interactive Motion-Assisted Therapy}.
\newblock \bibinfo{journal}{\emph{Frontiers in Psychiatry}}
  \bibinfo{volume}{Volume 9 - 2018} (\bibinfo{year}{2018}).
\newblock
\showISSN{1664-0640}
\href{https://doi.org/10.3389/fpsyt.2018.00176}{doi:\nolinkurl{10.3389/fpsyt.2018.00176}}


\bibitem[van Meggelen(2019)]%
        {meggelenComputerBasedInterventionElements2019}
\bibfield{author}{\bibinfo{person}{Marieke van Meggelen}.}
  \bibinfo{year}{2019}\natexlab{}.
\newblock \showarticletitle{A {{Computer-Based Intervention}} with {{Elements}}
  of {{Virtual Reality}} and {{Limited Therapist Assistance}} for the
  {{Treatment}} of {{PTSD}} : {{Efficacy}}, {{Acceptance}} and {{Future
  Implications}}}. \bibinfo{publisher}{Erasmus Universiteit Rotterdam (EUR)},
  \bibinfo{address}{Rotterdam}.
\newblock
\showISBNx{978-94-6375-491-0}


\bibitem[Vermetten et~al\mbox{.}(2025)]%
        {Vermetten2025zu}
\bibfield{author}{\bibinfo{person}{Eric Vermetten}, \bibinfo{person}{Lisa
  Burback}, \bibinfo{person}{Phillip~R Sevigny}, \bibinfo{person}{Mirjam~J
  Nijdam}, \bibinfo{person}{Olga Winkler}, \bibinfo{person}{Emmanuel Espejo},
  \bibinfo{person}{Pinata Sessoms}, \bibinfo{person}{Katherine Bright},
  \bibinfo{person}{Michael~J Roy}, {and} \bibinfo{person}{Suzette
  Br{\'e}mault-Phillip}.} \bibinfo{year}{2025}\natexlab{}.
\newblock \showarticletitle{Brief manual for multi-Modal {Motion-Assisted}
  Memory Desensitization and reconsolidation therapy for the treatment of
  post-traumatic stress disorder}.
\newblock \bibinfo{journal}{\emph{Psyc.. Clin. Psychopharmacol.}}
  \bibinfo{volume}{35}, \bibinfo{number}{Suppl 1} (\bibinfo{date}{Aug.}
  \bibinfo{year}{2025}), \bibinfo{pages}{S122--S134}.
\newblock


\bibitem[Vervliet et~al\mbox{.}(2013)]%
        {vervliet2013fear}
\bibfield{author}{\bibinfo{person}{Bram Vervliet}, \bibinfo{person}{Michelle~G
  Craske}, {and} \bibinfo{person}{Dirk Hermans}.}
  \bibinfo{year}{2013}\natexlab{}.
\newblock \showarticletitle{Fear extinction and relapse: state of the art}.
\newblock \bibinfo{journal}{\emph{Annual review of clinical psychology}}
  \bibinfo{volume}{9}, \bibinfo{number}{1} (\bibinfo{year}{2013}),
  \bibinfo{pages}{215--248}.
\newblock


\bibitem[Viergever(2019)]%
        {viergever_critical_2019}
\bibfield{author}{\bibinfo{person}{Roderik~F. Viergever}.}
  \bibinfo{year}{2019}\natexlab{}.
\newblock \showarticletitle{The {Critical} {Incident} {Technique}: {Method} or
  {Methodology}?}
\newblock \bibinfo{journal}{\emph{Qualitative Health Research}}
  \bibinfo{volume}{29}, \bibinfo{number}{7} (\bibinfo{date}{June}
  \bibinfo{year}{2019}), \bibinfo{pages}{1065--1079}.
\newblock
\showISSN{1049-7323}
\href{https://doi.org/10.1177/1049732318813112}{doi:\nolinkurl{10.1177/1049732318813112}}


\bibitem[Vincelli(1999)]%
        {vincelli1999imagination}
\bibfield{author}{\bibinfo{person}{Francesco Vincelli}.}
  \bibinfo{year}{1999}\natexlab{}.
\newblock \showarticletitle{From imagination to virtual reality: the future of
  clinical psychology}.
\newblock \bibinfo{journal}{\emph{CyberPsychology and Behavior}}
  \bibinfo{volume}{2}, \bibinfo{number}{3} (\bibinfo{year}{1999}),
  \bibinfo{pages}{241--248}.
\newblock


\bibitem[Williamson et~al\mbox{.}(2020)]%
        {williamson_secondary_2020}
\bibfield{author}{\bibinfo{person}{Emma Williamson}, \bibinfo{person}{Alison
  Gregory}, \bibinfo{person}{Hilary Abrahams}, \bibinfo{person}{Nadia Aghtaie},
  \bibinfo{person}{Sarah-Jane Walker}, {and} \bibinfo{person}{Marianne
  Hester}.} \bibinfo{year}{2020}\natexlab{}.
\newblock \showarticletitle{Secondary {Trauma}: {Emotional} {Safety} in
  {Sensitive} {Research}}.
\newblock \bibinfo{journal}{\emph{Journal of Academic Ethics}}
  \bibinfo{volume}{18}, \bibinfo{number}{1} (\bibinfo{date}{March}
  \bibinfo{year}{2020}), \bibinfo{pages}{55--70}.
\newblock
\showISSN{1572-8544}
\href{https://doi.org/10.1007/s10805-019-09348-y}{doi:\nolinkurl{10.1007/s10805-019-09348-y}}


\bibitem[Wilson(2004)]%
        {wilsonPTSDComplexPTSD2004}
\bibfield{author}{\bibinfo{person}{John~P. Wilson}.}
  \bibinfo{year}{2004}\natexlab{}.
\newblock \showarticletitle{{{PTSD}} and {{Complex PTSD}}: {{Symptoms}},
  {{Syndromes}}, and {{Diagnoses}}.}
\newblock In \bibinfo{booktitle}{\emph{Assessing Psychological Trauma and
  {{PTSD}}, 2nd Ed.}} \bibinfo{publisher}{The Guilford Press},
  \bibinfo{address}{New York, NY, US}, \bibinfo{pages}{7--44}.
\newblock
\showISBNx{1-59385-035-2 (Hardcover)}


\bibitem[Wobbrock and Kientz(2016)]%
        {wobbrock_research_2016}
\bibfield{author}{\bibinfo{person}{Jacob~O Wobbrock} {and}
  \bibinfo{person}{Julie~A Kientz}.} \bibinfo{year}{2016}\natexlab{}.
\newblock \showarticletitle{Research contributions in human-computer
  interaction}.
\newblock \bibinfo{journal}{\emph{interactions}} \bibinfo{volume}{23},
  \bibinfo{number}{3} (\bibinfo{year}{2016}), \bibinfo{pages}{38--44}.
\newblock
\href{https://doi.org/10.1145/2907069}{doi:\nolinkurl{10.1145/2907069}}
\newblock
\shownote{Publisher: ACM New York, NY, USA}.


\bibitem[Wolters et~al\mbox{.}(2017)]%
        {wolters_emotional_2017}
\bibfield{author}{\bibinfo{person}{Maria~K. Wolters},
  \bibinfo{person}{Zawadhafsa Mkulo}, {and} \bibinfo{person}{Petra~M.
  Boynton}.} \bibinfo{year}{2017}\natexlab{}.
\newblock \showarticletitle{The {Emotional} {Work} of {Doing} {eHealth}
  {Research}}. In \bibinfo{booktitle}{\emph{Proceedings of the 2017 {CHI}
  {Conference} {Extended} {Abstracts} on {Human} {Factors} in {Computing}
  {Systems}}} \emph{(\bibinfo{series}{{CHI} {EA} '17})}.
  \bibinfo{publisher}{Association for Computing Machinery},
  \bibinfo{address}{New York, NY, USA}, \bibinfo{pages}{816--826}.
\newblock
\showISBNx{978-1-4503-4656-6}
\href{https://doi.org/10.1145/3027063.3052764}{doi:\nolinkurl{10.1145/3027063.3052764}}


\bibitem[Wright and Mccarthy(2008)]%
        {wright_empathy_2008}
\bibfield{author}{\bibinfo{person}{Peter Wright} {and} \bibinfo{person}{J.
  Mccarthy}.} \bibinfo{year}{2008}\natexlab{}.
\newblock \bibinfo{booktitle}{\emph{Empathy and experience in {HCI}}}.
\newblock
\href{https://doi.org/10.1145/1357054.1357156}{doi:\nolinkurl{10.1145/1357054.1357156}}
\newblock
\shownote{Pages: 646}.


\bibitem[Ye et~al\mbox{.}(2024)]%
        {ye_generative_2024}
\bibfield{author}{\bibinfo{person}{Yilin Ye}, \bibinfo{person}{Jianing Hao},
  \bibinfo{person}{Yihan Hou}, \bibinfo{person}{Zhan Wang},
  \bibinfo{person}{Shishi Xiao}, \bibinfo{person}{Yuyu Luo}, {and}
  \bibinfo{person}{Wei Zeng}.} \bibinfo{year}{2024}\natexlab{}.
\newblock \showarticletitle{Generative {AI} for visualization: {State} of the
  art and future directions}.
\newblock \bibinfo{journal}{\emph{Visual Informatics}} \bibinfo{volume}{8},
  \bibinfo{number}{2} (\bibinfo{date}{June} \bibinfo{year}{2024}),
  \bibinfo{pages}{43--66}.
\newblock
\showISSN{2468-502X}
\href{https://doi.org/10.1016/j.visinf.2024.04.003}{doi:\nolinkurl{10.1016/j.visinf.2024.04.003}}


\end{thebibliography}

\appendix
\section*{}
\begin{table*}[t]
        \centering
        \begin{tabular}{ c p{1cm} p{8cm} l }\hline
            \multicolumn{3}{l}{\textbf{Questionnaire on VR experience}}\\
            Question & Language & Description & Item Type \\ \hline\hline
             \hypertarget{VRQ1}{1} & G & Wie oft haben Sie bisher eine VR-Brille getragen (Vorerfahrung
                    Nutzungshäufigkeit)? & <3x, 3-10x, >10x \\
            & E & \textit{How often have you worn VR glasses so far? (Previous experience/ Frequency of use)?} &\\
             \hypertarget{VRQ2}{2} & G & In welchem Maße waren Sie in der VR-Situation mental vertieft? & 7-point Likert\\
             & E & \textit{To what extent were you mentally immersed in the VR situation?} & \\
             \hypertarget{VRQ3}{3} & G & Wie sehr hat Sie die VR-Simulation vereinnahmt? & 7-point Likert \\
             & E & \textit{How much has the VR simulation absorbed you?} & \\
             \hypertarget{VRQ4}{4} & G & Wie aufregend war die VR-Simulation?  & 7-point Likert \\
             & E & \textit{How exciting was the VR simulation?} & \\
             \hypertarget{VRQ5}{5} & G & Die Verwendung der VR-Brille hatte für die Konfrontation mit meinen traumatischen Erinnerungen einen bedeutenden Mehrwert gegenüber imaginativer Konfrontation. & 7-point Likert \\
             & E & \textit{The use of VR glasses had a significant added value for the confrontation with my traumatic memories compared to imaginative confrontation.} & \\
            \hypertarget{VRQ6}{6} & G & Wie wahrscheinlich würden Sie anderen eine Konfrontation in VR im Rahmen einer Therapie weiterempfehlen? & 7-point Likert \\
            & E & \textit{How likely would you be to recommend a confrontation in VR to others as part of a therapy?} & \\
             \hypertarget{VRQ7}{7}& G & Wie sehr hat Ihnen die cooperative Erstellung von Expositionsszenarien bzw. -triggern am Bildschirm geholfen Ihre Angst vor der Konfrontation in VR zu reduzieren. & 7-point Likert \\
             & E & \textit{How much has the cooperative creation of exposure scenarios or triggers on the screen helped you to reduce your fear of confrontation in VR?} & \\
             \hypertarget{VRQ8}{8}& G & Was hat Ihnen an der Umsetzung der Therapie mit VR gut gefallen? & Open-ended \\
             & E & \textit{What did you like about the implementation of therapy with VR?} & \\
            \hypertarget{VRQ9}{9}& G & Was hat Ihnen an der Umsetzung der Therapie mit VR nicht gefallen? & Open-ended \\
            & E& \textit{What didn't you like about the implementation of therapy with VR?} & \\ \hline        
        \end{tabular}
        \caption{Questionnaire on VR experience assessed after each exposure session ($t_1$) in original version (german (G)) and translated into english (E)\label{tab:questionnaire_vrexperience}}
        \Description{Questionnaire items overview. The table's columns from left to right are item number, language, summary, and type of questionnaire item.}
    \end{table*}

       \begin{table*}[t]
        \centering
        \begin{tabular}{ p{6cm} p{6cm} }\hline
            \multicolumn{2}{l}{\textbf{Inclusion \& Exclusion Criteria for Patient Recruitment}}\\
            Inclusion criteria & Exclusion criteria\\ \hline\hline 
        1. Diagnosed PTSD $\pm$ borderline personality disorder comorbidity & 1. Presence of a disorder from the schizophrenia spectrum\\
        2. At least $18$ years old& 2. Severe affective disorder\\
        & 3. Substance dependence, at least 6 months abstinent\\
        & 4. Organic brain impairment\\
        & 5. Neurological disorder\\
        & 6. Intellectual disability\\
        & 7. Insufficient command of the German language\\
        & 8. Acute suicidality\\
        & 9. Pregnancy or breastfeeding\\
        & 10. Absence of informed consent\\ \hline 
    \end{tabular}
        \caption{Inclusion and exclusion criteria for the recruitment of patients of this study.}\label{tab:exclusioncritera}
        \Description{Criteria overview. The table's columns from left to right are inclusion and exclusion criteria.}
    \end{table*}

       \begin{table*}[t]
        \centering
        \begin{tabular}{p{1cm} p{3cm} p{6cm}  p{2cm}}\hline
        ID & Name & Description & Trauma type\\ \hline\hline 
        \hypertarget{GT1}{GT1} & Distance to observer & Content was initially visualized at a distance of $60 meters$ and could be brought closer to the observer via keyboard or controller input. Content could further be rotated around the z-axis. & CT\\
        \hypertarget{GT2}{GT2} & Number of people & The amount of people and the distribution area & CT\\
        \hypertarget{GT3}{GT3} & People's dynamics & People could either look around, talk, walk or run. Combinations of different dynamics were used to increase complexity further. & MT\\
        \hypertarget{GT4}{GT4} & People's appearance & Gender, physical appearance, and ethnicity were altered to calibrate perceived hazards. & MT\\
        \hypertarget{GT5}{GT5} & Lighting conditions & Brightness of scenario was increased to increase exposure. & CT\\
        \hypertarget{GT6}{GT6} & Sensory diversity & Audio-only, visuals-only and a combination were used to alter complexity. & MT\\
        \hypertarget{GT7}{GT7} & Location of observer & Gradation through observer's proximity to the action. In agoraphobic scenarios, exposure level was lower if the observer could rule out the possibility of something happening behind them. & CT, MT\\
        \hypertarget{GT8}{GT8} & Similarity with feared stimulus & The usage of another type of the actual feared content, e.g. a rescue helicopter instead of a military helicopter, to reduce the exposure level. & MT\\
        \hypertarget{GT9}{GT9} & Level of detail & Content was partially hidden, and objects' materials were exchanged with a neutral white material to lower the exposure level. & CT\\ \hline 
    \end{tabular}
        \caption{Techniques used to calibrate the exposure level, including the trauma types that they have been used for (CT = childhood trauma, MT = military trauma).}\label{tab:gradation}
        \Description{Overview of used gradation techniques. The table's columns from left to right are ID, name, and description.}
    \end{table*}

   \begin{table*}[t]
        \centering
        \begin{tabular}{ l p{1.5cm} p{2cm} p{1.75cm} p{1.4cm} p{3cm} p{3cm}}\hline
            ID & Environment & Objects & Events & Interaction & Gradation & Assets \\ \hline\hline 
            S1 & White, neutral & People & Walking, talking & Observe & Number, dynamics and appearance of people, observer distance (\hyperlink{GT2}{GT2}, \hyperlink{GT3}{GT3}, \hyperlink{GT4}{GT4}, \hyperlink{GT7}{GT7})& \href{https://assetstore.unity.com/packages/3d/characters/citizens-pro-2024-143604}{Citizens Pro}\\ \hline
            S2 & White, neutral & Helicopter & Landing  & Observe & Type of helicopter sound, audio-only (\hyperlink{GT6}{GT6}, \hyperlink{GT8}{GT8}) & \href{https://assetstore.unity.com/packages/3d/vehicles/air/20-helicopters-pack-62654}{20 Helicopters Pack}\\ \hline
            S3 & White, neutral & Doll House & None & Observe & Distance, Rotation on z-axis (\hyperlink{GT1}{GT1}) & \href{https://www.turbosquid.com/3d-models/free-house-toy-3d-model/891246}{Toy house}\\ \hline
            S4 & White, neutral & Bed with white bedsheet & None  & Observe & Distance, Rotation on z-axis (\hyperlink{GT1}{GT1}) & Asset no longer available\\ \hline
            S5 & White, neutral & Bed with red bedsheet & None  & Observe & Distance, Rotation on z-axis (\hyperlink{GT1}{GT1})& \href{https://www.blenderkit.com/asset-gallery-detail/7a580762-d059-441e-853d-de08052e0fa4/}{Bed For Relaxing}\\ \hline
            S6 & White, neutral & Chest & None  & Observe & Distance, Rotation on z-axis (\hyperlink{GT1}{GT1})&  \href{https://www.blenderkit.com/asset-gallery-detail/1294ce74-8310-4f8f-9006-01329f3968d7/?query=category_subtree:model+Wooden+chest+order:_score}{Wooden Box}\\ \hline
            S7 & White, neutral & Corset & None  & Observe & Distance, Rotation on z-axis (\hyperlink{GT1}{GT1})&  Custom 3D model\\ \hline
            S8 & White, neutral & Funeral images & None  & Observe & Number of visible images (\hyperlink{GT9}{GT9})& \href{https://reliefweb.int/report/iran-islamic-republic/afghan-refugees-reach-iran-violence-escalates}{Example image}\\ \hline
            S9 & White, neutral & Train, Trees, above-ground train station & Train runs, arrives, or leaves & Observe & Number, appearance and dynamics of people (\hyperlink{GT2}{GT2}, \hyperlink{GT3}{GT3}, \hyperlink{GT4}{GT4})& \href{https://assetstore.unity.com/packages/3d/characters/citizens-pro-2024-143604}{Citizens Pro}, \href{https://assetstore.unity.com/packages/3d/vehicles/land/train-high-speed-66612}{Train - High Speed}, \href{https://assetstore.unity.com/packages/3d/environments/landscapes/railway-station-sunshade-173266}{Railway Station Sunshade}\\ \hline
            S10 & Underground train station & Train, train station & Train runs, train arrives & Observe, leave train & Number, appearance and dynamics of people (\hyperlink{GT2}{GT2}, \hyperlink{GT3}{GT3}, \hyperlink{GT4}{GT4})&\href{https://assetstore.unity.com/packages/3d/characters/citizens-pro-2024-143604}{Citizens Pro}, \href{https://assetstore.unity.com/packages/3d/environments/urban/tokyo-metro-station-02-278375}{Tokyo Metro Station 02}, \href{https://assetstore.unity.com/packages/3d/vehicles/land/train-high-speed-66612}{Train - High Speed}\\ \hline
            S11 & Town & People, Café outdoor area & Talking, walking, sitting in café & Observe, teleportation & Number, appearance and dynamics of people, location of observer (\hyperlink{GT2}{GT2}, \hyperlink{GT3}{GT3}, \hyperlink{GT4}{GT4}, \hyperlink{GT7}{GT7})&\href{https://assetstore.unity.com/packages/3d/characters/citizens-pro-2024-143604}{Citizens Pro}, \href{https://assetstore.unity.com/packages/3d/environments/high-detail-city-2-243252}{High Detail City 2}\\ \hline
            S12 & Café & People & People coming, ordering and leaving & Observe & Number and dynamics of people (\hyperlink{GT2}{GT2}, \hyperlink{GT3}{GT3})& \href{https://assetstore.unity.com/packages/3d/props/interior/coffee-shop-interior-209622}{Coffee Shop interior}\\ \hline
            S13 & Street & People, bench & People walking by & Observe, get up & Number, appearance and dynamics of people (\hyperlink{GT2}{GT2}, \hyperlink{GT3}{GT3}, \hyperlink{GT4}{GT4})& \href{https://assetstore.unity.com/packages/3d/characters/citizens-pro-2024-143604}{Citizens Pro}\\ \hline
            S14 & Cellar & Washing machine, chest, shelves & None & Observe & Brightness of the light (\hyperlink{GT5}{GT5})& \href{https://www.blenderkit.com/asset-gallery-detail/eedcbf8c-b380-45d5-abc9-76a4ae800307/}{Shelf}\\ \hline
            S15 & Brick wall room & Blue carpet & None & Observe & None & None\\ \hline
            S16 & Childhood room & Bed, commode, lamp, boxes & None & Observe & Perspective, material shading (\hyperlink{GT7}{GT7}, \hyperlink{GT9}{GT9})& \href{https://www.blenderkit.com/asset-gallery-detail/4b66dd0e-d3c5-488d-919c-e8d6dd684eca/}{Chest Of Drawers Ikea}\\ \hline
            S17 & Living room & Couches, lamp, wall cupboard, couch table & None & Observe & Perspective (\hyperlink{GT7}{GT7})& \href{https://www.blenderkit.com/asset-gallery-detail/02d6f727-d945-4e4e-894c-a3b533af7a4e/}{Casting couch}, \href{https://www.blenderkit.com/asset-gallery-detail/131ca093-28c6-4dc5-8e16-a3ca11322b81/}{Pendant Hanging Lamp}, \href{https://www.blenderkit.com/asset-gallery-detail/8ce55dc8-575f-4c35-baa3-0bdd3a501d39/}{Armchair}\\ \hline
            S18 & Childhood room & Bed, desk, lamp, doll house, cupboard, rug & None & Observe & Perspective, material shading (\hyperlink{GT7}{GT7}, \hyperlink{GT9}{GT9})&\href{https://www.turbosquid.com/de/3d-models/3d-cama-malm-ikea-virtualizer-model-1902701}{Cama MALM}, \href{https://www.turbosquid.com/de/3d-models/desk-blender-3d-model-1232007}{Desk} \href{https://www.turbosquid.com/3d-models/free-house-toy-3d-model/891246}{Toy house}, \href{https://www.turbosquid.com/de/3d-models/radiator-max-free/719020}{Radiator}, \href{https://www.turbosquid.com/de/3d-models/free-max-mode-measuring-rulers/786272}{Meausring rulers}\\ \hline
        \end{tabular}
        \caption{Exposure scenarios used in this study, including environmental design, content, events, interaction possibilities for patient, gradation techniques, and used assets for this scenario.}\label{tab:scenarios}
        \Description{Exposure scenario overview. The table's columns from left to right are scenario ID, environment, objects, events, interaction technique, gradation, and assets.}
    \end{table*}

    \begin{table*}[t]
        \centering
        \begin{tabular}{p{3cm} p{1cm} p{1cm} p{0.5cm} p{4cm}}\hline
            \multicolumn{3}{l}{\textbf{Demographics of Recruited Patients}}\\
            Trauma Type & Pseudonym & Gender & Age & Visited VR Exposure Scenarios \\ \hline\hline 
            \multirow{6}{*}{Childhood trauma}&Lisa & Female & 23 & S4*\\
            &Lydia & Female & 23 & S16*\\
            &Andrea & Female & 25 & S3*, S18*\\
            &Natalie & Female & 34 & S6*, S14*, S15*\\
            &Sabrina & Female & 23 & S7*\\
            &Charlotte & Female & 38 & S5*, S17*\\\hline
            \multirow{6}{*}{Military trauma} &Robert & Male & 55 & S1*, S11*, S12*\\
            &John & Male & 51 & S2*\\
            &Marc & Male & 44 & S10*, S9*\\
            &Peter & Male & 42 & S1, S8*, S11*, S13\\
            &Stephan & Male & 52 & S2*\\\hline
        \end{tabular}
        \caption{Demographics of the eleven recruited patients as well as list of Virtual Reality (VR) exposure scenarios that they visited. For each scenario \textbf{*} indicates that a skin conductance response was measured ($delta>0.005$, \cite{posadaquintero2020eda}).}\label{tab:demographics}
        \Description{Demographics overview. The table's columns from left to right are pseudonym, gender, age, and trauma type.}
    \end{table*}

\begin{table*}[t]
    \centering
    \begin{tabular}{p{4cm} c c c c c c}\hline
        Construct & \multicolumn{2}{c}{Childhood/ Sexual} & \multicolumn{2}{c}{Military} & \multicolumn{2}{c}{Overall}\\ \cline{2-7}
        & Mean & SD & Mean & SD & Mean & SD \\ \hline
            IPQ& $0.03$ & $1.10$ & $0.18$ & $0.91$ & $0.09$ & $1.03$\\
            \hspace{0.5cm}Spatial Presence& $-0.19$ & $0.76$ &$0.78$ & $1.15$ & $0.17$ & $1.04$\\
            \hspace{0.5cm}Involvement& $-0.20$ & $1.40$ & $-0.13$ & $1.46$ & $-0.17$ & $1.42$\\
            \hspace{0.5cm}Realism& $0.07$ & $1.03$ & $-0.56$ & $0.85$ & $-0.16$ & $1.02$\\ \hline
            VR Experience &&&&&&\\
            \hspace{0.5cm} \hyperlink{VRQ2}{Q2 - Involvement} & $4.91$ & $1.20$ & $5.56$ & $1.69$ & $5.18$ & $1.48$\\
            \hspace{0.5cm} \hyperlink{VRQ3}{Q3 - Absorption} & $5.14$ & $1.32$ & $6.06$ & $1.37$ & $5.53$ & $1.52$\\
            \hspace{0.5cm} \hyperlink{VRQ4}{Q4 - Excitement} & $5.05$ & $1.64$ & $5.75$ & $1.54$ & $5.34$ & $1.74$\\
            \hspace{0.5cm} \hyperlink{VRQ5}{Q5 - Visual imagery} & $5.18$ & $1.44$ & $4.43$ & $1.67$ & $4.89$ & $1.52$\\
            \hspace{0.5cm} \hyperlink{VRQ6}{Q6 - Recommendation} & $5.09$ & $0.93$ & $5.86$ & $0.74$ & $5.39$ & $0.98$\\
            \hspace{0.5cm} \hyperlink{VRQ7}{Q7 - Collaborative design} & $3.81$ & $1.10$ & $3.86$ & $1.92$ & $3.83$ & $1.48$\\\hline
        \end{tabular}
        \caption{Mean and standard deviation of the presence scores (assessed using the igroup presence questionnaire (IPQ)) and virtual reality experience scores (assessed using a custom questionnaire).}\label{tab:presence}
        \Description{Summary of the presence and acceptance scores. The table shows the overall presence scores and the scores of the virtual reality experience questionnaire of the patients with childhood trauma, the patients with military trauma, and overall.}
    \end{table*}


\end{document}